\documentclass[11pt,a4paper]{article}
\pdfoutput=1
\usepackage{jheppub}

\usepackage{multirow, graphicx,amssymb,url,mathrsfs,amsmath}
\usepackage{wrapfig,boxedminipage,setspace,epsfig}
\usepackage{amsxtra,amstext,latexsym,dsfont,amsfonts}
\usepackage{color}
\usepackage[dvipsnames]{xcolor}
\usepackage{float}
\usepackage{slashed,comment}
\usepackage{contour}
\usepackage{tikz}
\usetikzlibrary{calc,patterns,angles,quotes}
\usetikzlibrary{decorations.pathreplacing,decorations.markings,snakes}
\usepackage{tabularx, array}
\usepackage{pgfplots}
\pgfplotsset{compat=1.18}
\usepackage{float}

 \newcommand{\w}{\omega}

\newcommand{\calc}{\mbox{${\cal C}$}}

\newcommand{\LA}[1]{\label{#1}}
\newcommand{\half}{\frac{1}{2}}

\newcommand{\dd}{\mathrm{d}}

\newcommand{\bra}[1]{\mbox{$\langle #1 |$}}
\newcommand{\ket}[1]{\mbox{$| #1 \rangle$}}

\newcolumntype{L}[1]{>{\raggedright\arraybackslash}p{#1}}
\newcolumntype{C}[1]{>{\centering\arraybackslash}p{#1}}
\newcolumntype{R}[1]{>{\raggedleft\arraybackslash}p{#1}}

\usepackage{xparse}

\newcommand{\hypr}{{}_2F_1}
\usepackage{subcaption}
\newcommand{\Arg}{\text{Arg}}

\title{Cosmological brick walls \& quantum chaotic dynamics of de Sitter horizons}

\author[a]{Jos\'e M. Begines,}
\author[b]{Suman Das,}
\author[c,d]{Hyun-Sik Jeong,}
\author[b]{and Juan F. Pedraza}

\emailAdd{jmbsphysics@outlook.es}
\emailAdd{suman.das@ift.csic.es}
\emailAdd{hyunsik.jeong@apctp.org}
\emailAdd{j.pedraza@csic.es}

\preprint{\texttt{IFT-UAM/CSIC-26-40, APCTP Pre2026-002}}

\affiliation[a]{Departamento de Física, Universidad de Oviedo, 33007 Oviedo, Spain}
\affiliation[b]{Instituto de F\'isica Te\'orica UAM/CSIC, 28049 Madrid, Spain}
\affiliation[c]{Asia Pacific Center for Theoretical Physics, 37673 Pohang, Korea}
\affiliation[d]{Department of Physics, Pohang University of Science and Technology, 37673 Pohang, Korea}

\abstract{
Originally proposed by ’t~Hooft, the brick wall model has recently reemerged as a useful framework for probing quantum aspects of horizon physics, particularly in the context of holography. In this paper, we apply it to asymptotically de~Sitter spacetimes. We compute the normal modes of a massless scalar field in pure de~Sitter space and in the Schwarzschild--de~Sitter black hole, and analyze the resulting single-particle spectra using the level-spacing distribution, the spectral form factor, and Krylov complexity. In pure de~Sitter, the spectrum exhibits clear long-range signatures of chaos despite not obeying a conventional Wigner--Dyson level-spacing distribution. The Schwarzschild--de~Sitter case is qualitatively richer: in the WKB regime, where tunneling between the two classically allowed regions is exponentially suppressed, the presence of both an event horizon and a cosmological horizon gives rise to two independent near-horizon sectors, so that the full spectrum is the superposition of two subsequences. As a result, the combined level-spacing distribution develops a nonzero value at $s=0$ even when spectral correlations remain. Nevertheless, for sufficiently small stretched-horizon fluctuations, the superposed spectrum still exhibits an approximately linear ramp in the spectral form factor and a pronounced peak in Krylov complexity. Our results show that the absence of strict level repulsion should not, by itself, be taken as evidence against chaos, and that the spectral form factor and Krylov complexity provide sharper diagnostics of the underlying chaotic dynamics.}

\begin{document}
\maketitle


\section{Introduction}\label{intro}

The study of black hole horizons remains central to the ongoing effort to reconcile general relativity with quantum mechanics within a consistent framework of quantum gravity. Since the seminal works of Bekenstein and Hawking~\cite{Bekenstein1972,Hawking_1975,Strominger:1996sh}, it has been clear that horizons are not merely causal boundaries: they are thermodynamic objects, endowed with a temperature and an entropy proportional to their area. A basic challenge is therefore to identify the microscopic degrees of freedom underlying this entropy and, more broadly, to determine which observables are sensitive to their underlying dynamics.

One of the simplest and most fruitful frameworks for probing these questions is the brick wall model, originally proposed by 't~Hooft~\cite{THOOFT1985727}. By introducing a stretched horizon~\cite{Susskind:1993if}, i.e., a regulator located a short proper distance outside the true horizon, the model makes it possible to quantize probe fields and compute the thermal free energy in a controlled semiclassical setting. In this construction, the familiar area law for entropy emerges from ordinary statistical mechanics applied to the normal modes of quantum fields propagating outside the horizon. This is already highly nontrivial, and strongly suggests that the corresponding mode spectrum captures meaningful information about horizon physics.

This observation raises a broader question: if these modes already encode the thermodynamic entropy, might they also contain more refined information about the horizon, such as its spectral correlations or chaotic properties? Recent work suggests that the answer is yes. In asymptotically AdS spacetimes, the brick wall model has become a useful framework for exploring quantum spectral properties of horizons, particularly in the context of holography~\cite{Maldacena:1997re,Gubser:1998bc,Witten:1998qj}. Early discussions of brick walls in AdS appeared in~\cite{Kay:2011np,Iizuka:2013kma}, while more recent developments have shown that probe-field normal modes provide a natural arena for investigating thermality and chaos-related features through correlators, spectral statistics, spectral form factors, stretched-horizon fluctuations, and, more recently, Krylov complexity~\cite{Das:2022evy,Das:2023ulz,Das:2023xjr,Krishnan:2023jqn, Burman:2023kko,Banerjee:2024dpl,Das:2024fwg,Burman:2024egy,Jeong:2024jjn, Banerjee:2024ivh,Ageev:2024gem,Jeong:2025jyx, Caceres:2025qlh} (see also \cite{Gayari:2026jwn, Krishnan:2026mpa} for recent progress along these lines). 

A particularly striking lesson of these developments is that no single spectral diagnostic should be overinterpreted in isolation. In the original BTZ brick wall setup~\cite{Das:2022evy}, the single-particle spectral form factor (SFF) constructed from the probe-field normal modes exhibits a clear dip--ramp--plateau structure, even though the corresponding level-spacing distribution is not of the conventional Wigner--Dyson type. More generally, it has been emphasized that a linear ramp in the SFF, while often associated with random-matrix behavior, can also arise in deterministic spectra~\cite{Das:2023yfj}. This means that the SFF captures robust long-range information about the spectrum, whereas nearest-neighbor spacing statistics are much more sensitive to short-range structure. Once fluctuations of the stretched-horizon boundary condition are introduced---motivated in part by the profile functions of known fuzzball solutions~\cite{Lunin:2001jy,Kanitscheider:2007wq,Bena:2015bea}---the same brick wall framework can also produce Wigner--Dyson-like statistics and interpolate between Poissonian and correlated spectra~\cite{Das:2023ulz,Jeong:2024jjn}.

These results fit naturally with the broader expectation that black holes are among the fastest scramblers in nature~\cite{Cotler:2016fpe,Chen:2024oqv}. In holography, this expectation is tied to the near-horizon dynamics of black holes~\cite{tHooft:1993gx,Susskind:1994vu} and is sharply quantified by the Maldacena--Shenker--Stanford bound on the quantum Lyapunov exponent~\cite{Maldacena:2015waa}. Gravitationally, chaotic dynamics manifests itself through the exponential growth of out-of-time-ordered correlators (OTOCs) associated with near-horizon scattering~\cite{Roberts:2014isa,Shenker:2013pqa,Shenker:2013yza,Shenker:2014cwa}, as well as through pole-skipping~\cite{Grozdanov:2017ajz,Blake:2017ris,Blake:2018leo,Jeong:2021zhz}, which encodes universal horizon constraints in the analytic structure of thermal Green's functions. The brick wall model provides a complementary probe-side window into this physics: the chaotic signatures arise in the spectrum of probe excitations rather than directly in the microscopic black hole spectrum, but the fact that the same probe sector also reproduces the area law suggests that it may still capture nontrivial indirect information about the underlying horizon dynamics.

It is therefore natural to ask whether the same picture extends to spacetimes with a positive cosmological constant. De~Sitter space is of obvious interest both for quantum gravity and for cosmology, yet it remains much less understood than AdS. See~\cite{Spradlin:2001pw,Anninos:2012qw,Galante:2023uyf} for reviews. Its defining feature is the presence of a cosmological horizon, which limits the observable region of any inertial observer. Much like a black hole horizon, this cosmological horizon carries a temperature and an entropy~\cite{Gibbons:1977mu}. Various approaches to de~Sitter holography have been proposed, including dS/CFT~\cite{Strominger:2001pn} and static-patch holography~\cite{Susskind:2021esx}, but a fully satisfactory microscopic description remains elusive.

At the same time, several recent developments indicate that scrambling and chaos remain meaningful notions in de~Sitter settings as well. The presence of a cosmological horizon strongly suggests that de~Sitter space should behave as a fast scrambler~\cite{Susskind:2011ap}, and recent studies based on gravitational probes such as OTOCs and pole-skipping support the idea that both black hole and cosmological horizons in de~Sitter backgrounds exhibit chaotic features closely analogous to those familiar from AdS black holes~\cite{Anninos:2018svg,Aalsma:2020aib,Kolchmeyer:2024fly,Ahn:2025exp}. This makes the brick wall model a particularly attractive tool: it provides a simple, geometric, and tractable framework in which one can ask whether de~Sitter horizons organize their probe spectra in a way that mirrors the AdS story.

In this paper, we apply the brick wall model to asymptotically de~Sitter spacetimes, focusing on the normal modes of a probe scalar field and on three complementary diagnostics of spectral correlations: the level-spacing distribution, the spectral form factor, and Krylov (state/spread) complexity. We first study pure de~Sitter space, where one can place a stretched horizon near the cosmological horizon and ask whether the resulting normal modes exhibit the same slow angular-momentum dependence and the same chaotic diagnostics found previously in AdS black holes. We then turn to the Schwarzschild--de~Sitter black hole, where the static patch is bounded by both an event horizon and a cosmological horizon, and therefore naturally admits two stretched horizons.

The presence of two horizons leads to a qualitatively new feature that is absent in the AdS examples studied so far. In the WKB regime relevant for this work, tunneling between the two classically allowed regions is exponentially suppressed, so the quantization problem factorizes into two independent sectors, one associated with the black hole horizon and the other with the cosmological horizon. The full spectrum is therefore not a single sequence, but rather the superposition of two subsequences localized near the two horizons.\footnote{This structure is also consistent with a broader semiclassical picture in which, at leading order in Newton's constant, the degrees of freedom associated with the two horizons organize into approximately independent sectors, $
\mathcal{H}_{\text{total}} \simeq \mathcal{H}_{\mathrm{bh}} \otimes \mathcal{H}_{\mathrm{c}}$. A closely related holographic interpretation was recently proposed in~\cite{Ahn:2025exp}, where Schwarzschild--de~Sitter is described in terms of two entangled sectors associated with the black hole and cosmological horizons. In our setup, the WKB factorization of the quantization conditions suggests that these sectors effectively decouple at leading order, with interactions between them suppressed.} This simple observation has important consequences for the interpretation of spectral diagnostics. In particular, once two statistically independent or weakly coupled subsequences are superposed, the combined nearest-neighbor spacing distribution can acquire a nonzero value at the origin simply because levels from the two subsequences interleave. Thus, the appearance of \(p(s=0)\neq 0\) in the full spectrum does not by itself imply the absence of chaos; it may instead be a kinematical consequence of spectral superposition, in the same general spirit as Berry--Robnik-type statistics in mixed systems~\cite{Berry:1984, Prosen:1993, Prosen:1994, Prosen:1998, Prosen:1999}.\footnote{Related lessons have also emerged in recent studies of quantum chaos in mixed systems, where spectral statistics were analyzed through Brody-type interpolating distributions~\cite{Amore:2024ihm,Huh:2024ytz}. There, the mixed character was encoded statistically, without a clean decomposition into independent spectral subsectors. By contrast, the Schwarzschild--de~Sitter setup studied here provides a more geometric realization of mixed spectral behavior: in the WKB regime, the full spectrum is the explicit superposition of two near-horizon spectra.} Schwarzschild--de~Sitter black holes therefore provide a clean setting in which to sharpen the question: which signatures of chaos survive the superposition of two near-horizon spectra, and which do not? In particular, what becomes of the ramp in the SFF and of the peak in Krylov complexity when the full level-spacing distribution is no longer strictly Wigner--Dyson-like?

A second issue concerns the role of stretched-horizon fluctuations. As in the AdS case, we allow for Gaussian-distributed boundary conditions, which provide an effective way to model fluctuations of the stretched horizon and generalize the original Dirichlet wall. Mathematically, increasing the variance can drive the spectrum toward a more Poisson-like regime. Physically, however, that limit should be interpreted with care. Since horizons are expected on general grounds to be efficient scramblers, and since sufficiently large boundary fluctuations can wash out correlations already visible in the controlled probe sector, the large-variance Poissonian regime is best viewed as a formal extrapolation rather than as the generic expectation for gravitational horizons. From this perspective, the more interesting question is not whether arbitrarily strong noise can destroy spectral correlations, but rather how robust the ramp and the Krylov peak remain within the regime where the probe approximation is still trustworthy.

Our goal is therefore twofold. First, we want to determine whether pure de~Sitter space and Schwarzschild--de~Sitter black holes exhibit the same qualitative probe-sector signatures of quantum chaos that have emerged in the brick wall description of AdS black holes. Second, and more importantly, we want to understand how those signatures are modified in a genuine two-horizon geometry, where the relevant spectrum is naturally a superposition of two near-horizon spectra. In this sense, de~Sitter black holes offer the simplest horizon system in which one can study, within a controlled semiclassical setup, the interplay between spectral superposition, spectral ramps, and Krylov complexity in the search for universal signatures of horizon chaos.

This manuscript is organized as follows. In section~\ref{sec:definitions}, we establish our notation and briefly review the main diagnostics used throughout the paper: the level-spacing distribution, the spectral form factor, and Krylov complexity. In section~\ref{emptydS}, we study pure de~Sitter space, first deriving the exact normal modes of a probe scalar field and then analyzing their chaotic properties for both standard and fluctuating brick wall boundary conditions. In section~\ref{sec:sds}, we turn to the Schwarzschild--de~Sitter black hole. There, we derive the normal modes in the WKB regime, emphasize the factorization into two near-horizon sectors, and investigate the spectral properties of the resulting superposed spectrum in various regimes. We conclude in section~\ref{sec:conclusions} with a discussion of the main lessons and several directions for future work. Technical details of the analysis are collected in the appendices.

\section{Useful definitions and conventions}\label{sec:definitions}

In this section, we collect a few useful definitions and conventions that will be used throughout the paper. We also briefly review the main diagnostics employed in our analysis in order to fix notation and make the discussion self-contained.

In this work, we study the normal modes of a probe scalar field in de~Sitter geometry. These modes are labeled by two quantum numbers: the principal quantum number \(n\) and the angular momentum quantum number \(l\). Since our goal is to search for signatures of quantum chaos in the spectrum, we focus on three standard diagnostics: the level-spacing distribution (LSD), the spectral form factor (SFF), and the Krylov state/spread complexity (KC). In all three cases, we primarily analyze sectors with fixed \(n\).\footnote{For fixed \(l\), the spectrum turns out to be linear.} Although the definitions of these quantities are standard and well documented in the literature, we briefly review them here for completeness and to establish our notation.

\paragraph{Level-spacing distribution.}
The level-spacing distribution (LSD) is defined from the ordered energy eigenvalues
$E_1 \le E_2 \le \cdots$, after unfolding the spectrum so that the local mean level spacing is unity.
Denoting the unfolded eigenvalues by $\{\tilde E_n\}$, the nearest-neighbor spacings are therefore
\begin{equation}
    s_n = \tilde E_{n+1}-\tilde E_n \, .
\end{equation}
The level-spacing distribution $p(s)$ is then defined as the normalized probability density of the spacings $s_n$.
For chaotic systems, the general expectation is that $p(s)$ should exhibit level repulsion, implying $p(s\to 0)\to 0$, whereas for integrable systems, $p(s)$ is maximal at $s=0$. As we will see, however, this expectation need not hold in the presence of spectral superposition, where nontrivial chaotic correlations may persist even when $p(s)$ develops a nonzero value at the origin.

\paragraph{Spectral form factor.}
The spectral form factor (SFF) of a quantum system with discrete energy eigenvalues $\{E_n\}$ is defined as
\begin{equation}
    g(t)
    \;=\;
    \frac{Z(\beta,t)\,Z^{\ast}(\beta,t)}{Z(\beta)^2}\,,
    \qquad
    Z(\beta,t)=\sum_n e^{-(\beta+it)E_n}\,,
    \qquad
    Z(\beta)=\sum_n e^{-\beta E_n}\, ,
\end{equation}
or, equivalently,
\begin{equation}
    g(t)
    \;=\;
    \frac{1}{Z(\beta)^2}
    \sum_{m,n}
    e^{-\beta(E_m+E_n)}\,e^{-i(E_m-E_n)t}\, .
\end{equation}

This quantity depends only on the spectrum of the Hamiltonian and is well defined for arbitrary inverse temperature $\beta$.
At early times, the SFF typically exhibits a decay, often referred to as the dip or slope, which arises from coarse-graining of the spectrum.
At late times, it saturates to a constant value, known as the plateau, reflecting the discreteness of the spectrum.
These features are generic to finite-dimensional unitary quantum systems.
The nontrivial information is encoded in how the dip connects to the plateau: for chaotic systems, this interpolation is characterized by a linear ramp.

When the Hamiltonian depends on a random parameter $\mathcal{J}$, it is often useful to consider an averaged spectral form factor~\cite{Cotler:2016fpe},
\begin{equation}
    g_{\text{avg}}(t)
    \;=\;
    \frac{\big\langle Z(\beta,t)\,Z^{\ast}(\beta,t)\big\rangle_{\mathcal{J}}}
    {\big\langle Z(\beta)\big\rangle_{\mathcal{J}}^{\,2}} \, ,
\end{equation}
which exhibits significantly reduced fluctuations compared to the unaveraged quantity.

\paragraph{Krylov complexity.}
Beyond traditional probes, Krylov complexity (KC) has emerged as a modern diagnostic of quantum chaos~\cite{Balasubramanian:2022tpr,Caputa:2024vrn}. It effectively captures the transition from integrability to chaos in a manner consistent with standard spectral diagnostics~\cite{Baggioli:2024wbz,Baggioli:2025knt}. While originally developed to characterize operator growth in the Heisenberg picture~\cite{Parker:2018yvk}, the framework has been extended to the Schr\"odinger picture to quantify the spread of states within the Krylov subspace~\cite{Balasubramanian:2022tpr}. In this manuscript, we focus specifically on the Krylov complexity of states. Comprehensive overviews of Krylov complexity in diverse physical contexts, including both operator and state perspectives, are provided in~\cite{Nandy:2024evd,Rabinovici:2025otw,wipREVIEW}.\footnote{Recent developments in holography have seen a surge of interest in both Krylov state and operator complexity (see, e.g., \cite{Rabinovici:2023yex,Lin:2022rbf,Heller:2024ldz,Ambrosini:2024sre,Jian:2020qpp,Caputa:2024sux,Fu:2025kkh,Jeong:2026iac,Balasubramanian:2024lqk}). These studies have related Krylov complexity to various geometric observables, such as proper momentum in AdS black holes and wormhole lengths in two-dimensional gravity.}

Krylov state/spread complexity is defined as follows. Starting from an initial state $\ket{\psi(0)}$, the time-evolved state is
\begin{equation}
    \ket{\psi(t)} = e^{-iHt}\ket{\psi(0)}
    = \sum_{n=0}^{\infty}\frac{(-it)^n}{n!} H^n \ket{\psi(0)} .
\end{equation}
This naturally introduces the set $\{\ket{\psi_n}=H^n\ket{\psi(0)}\}$, which does not form an orthonormal basis.
An orthonormal basis
\begin{equation}
    \mathcal{K}=\{\ket{K_n},\; n=0,1,2,\ldots,K\},
\end{equation}
with $\ket{K_0}=\ket{\psi(0)}$, can be constructed using the Lanczos algorithm~\cite{Lanczos:1950zz,ViswanathMuller1994}.
The span of $\mathcal{K}$ defines the Krylov subspace, whose dimension $K$ is finite for any finite-dimensional system.
In this basis, the Hamiltonian takes a tridiagonal form,
\begin{equation}
    \bra{K_m}H\ket{K_n}
    = a_n\,\delta_{m,n}
    + b_{n+1}\,\delta_{m,n+1}
    + b_n\,\delta_{m,n-1},
\end{equation}
where $a_n$ and $b_n$ are the so-called Lanczos coefficients.

The evolved state can be expanded as
\begin{equation}
    \ket{\psi(t)}=\sum_n \psi_n(t)\ket{K_n},
\end{equation}
and the Krylov complexity is then defined as
\begin{equation}\label{spread_K}
    \mathcal{C}(t)=\sum_n n\,|\psi_n(t)|^2 .
\end{equation}
Throughout this work, we take the initial state to be the thermofield double (TFD) state, so that the complexity depends only on the spectrum; for simplicity, we set $\beta=0$.

Krylov state complexity exhibits a characteristic growth--peak--plateau structure, closely paralleling the dip--ramp--plateau behavior of the spectral form factor~\cite{Baggioli:2024wbz,Baggioli:2025knt}. As shown in~\cite{Erdmenger:2023wjg}, the two quantities are related by an Ehrenfest-type argument. In particular, the emergence of a pronounced peak in $\mathcal{C}(t)$ is associated with chaotic dynamics, and its height has been observed to increase from GOE to GUE to GSE~\cite{Erdmenger:2023wjg,Balasubramanian:2022tpr}, making Krylov state complexity a useful diagnostic of random-matrix universality classes.

\section{Pure de~Sitter space}\label{emptydS}

In this section, we focus on pure de~Sitter space, which nevertheless possesses a cosmological horizon. In the static patch, this horizon appears as a coordinate singularity at a finite radius, reflecting the fact that an inertial observer can access only a finite causal region of the full spacetime due to the accelerated expansion. The metric of the static patch of de~Sitter spacetime is given by
\begin{equation}
    \dd s^2_{d+1}
    =
    -f(r)\,\dd t^2
    + \frac{\dd r^2}{f(r)}
    + r^2 \dd\Omega_{d-1}^2 \, ,
\end{equation}
with
\begin{equation}
    f(r) = 1 - r^2 ,
    \qquad
    0 \le r < 1 \,,
\end{equation}
where the cosmological horizon is at $r=1$. Here $\dd\Omega_{d-1}^2$ denotes the line element on \(S^{d-1}\), and the curvature radius of de~Sitter space has been set to unity.

\subsection{Computation of normal modes}

We are interested in computing the normal modes of massless scalar fluctuations subject to brick wall boundary conditions. To do so, we first solve the Klein--Gordon equation
\begin{equation}\label{KG_empty ds}
    \Box \Phi = 0 \, .
\end{equation}
We adopt the separation ansatz
\begin{equation}
    \Phi(r,\Omega,t)
    =
    \Psi(r)\,e^{-i\omega t} Y(\Omega) ,
\end{equation}
where $Y(\Omega)$ are the spherical harmonics on $S^{d-1}$. This leads to the radial equation
\begin{equation}\label{radial1}
    \begin{aligned}
        (r^2-1)\Psi''
        + \Bigl[\frac{1-d}{r} + r(1+d)\Bigr]\Psi'
        + \Bigl[\frac{l(l+d-2)}{r^2} - \frac{\omega^2}{1-r^2}\Bigr]\Psi
        = 0 \, .
    \end{aligned}
\end{equation}
This equation admits analytic solutions in terms of hypergeometric functions:
\begin{equation}\LA{SOL_STATIC}
    \begin{aligned}
        \Psi(r) = \calc_1(1-r^2)^{-i\w/2}r^l&\hypr\left[\half\left(l-i\w\right),\half\left(d+l-i\w\right),\frac{d}{2}+l,r^2\right] \\
        + \calc_2(1-r^2)^{-i\w/2}r^{2-d-l}&\hypr\left[1-\half\left(l+i\w\right),1-\half\left(d+l+i\w\right),2-\frac{d}{2}-l,r^2\right].
    \end{aligned}
\end{equation}
Expanding near the origin, $r \to 0$, one finds
\begin{equation}
    \Psi(r)
    =
    \mathcal{C}_1 r^l
    \left[
        1 + \frac{l(d+l)-\omega^2}{2(d+2l)} r^2
        + \mathcal{O}(r^3)
    \right]
    +
    \mathcal{C}_2 r^{-d-l}
    \left[
        r^2 + \mathcal{O}(r^3)
    \right] .
\end{equation}
Regularity at the origin therefore requires
\begin{equation}
    \mathcal{C}_2 = 0 \, .
\end{equation}
The regular solution, expanded near the horizon, behaves as
\begin{equation}
    \Psi_{\mathrm{hor}}(r)
    \sim
    \mathcal{C}_1
    \left[
        P_2 (2-2r)^{-i\omega/2}
        + Q_2 (2-2r)^{i\omega/2}
    \right] .
\end{equation}
The coefficients are given by
\begin{equation}\LA{STATIC_BOUNDARY}
    P_2
    =
    \frac{1}{
        \Gamma\!\left(\tfrac{l+i\omega}{2}\right)
        \Gamma\!\left(\tfrac{d+l+i\omega}{2}\right)
        \Gamma(-i\omega)
    } ,
    \qquad
    Q_2
    =
    \frac{1}{
        \Gamma\!\left(\tfrac{l-i\omega}{2}\right)
        \Gamma\!\left(\tfrac{d+l-i\omega}{2}\right)
        \Gamma(i\omega)
    } .
\end{equation}
For real $\omega$, one finds $|P_2| = |Q_2|$.

Imposing the conventional ingoing boundary condition at the horizon yields quasinormal modes. Since we are instead interested in normal modes, we introduce a brick wall at \(r=r_0\), close to the horizon, and impose the Dirichlet condition
\begin{equation}\label{brick0}
    \Psi_{\mathrm{hor}}(r_0) = 0 \, .
\end{equation}
This condition leads to a discrete set of real frequencies \(\omega\), labeled by \(l\) and a principal quantum number \(n\). The spectrum is degenerate in the magnetic quantum numbers, since they do not appear in the radial equation~\eqref{radial1}.

A more general boundary condition is
\begin{equation}\label{nonzero}
    \Psi_{\mathrm{hor}}(r_0) = \Psi_0 \, ,
\end{equation}
with \(\Psi_0\) a nonvanishing constant. A generic choice of \(\Psi_0\) leads to a reflection coefficient with nonunit modulus,
\begin{equation}
    \left| \frac{Q_2}{P_2} \right|^2 \neq 1 ,
\end{equation}
which in turn implies the presence of complex modes. To avoid this, we choose \(\Psi_0\) (following~\cite{Das:2023ulz}) in a specific way, such that Eq.~\eqref{nonzero} can be written as
\begin{equation}
    P_2 (2-2r_0)^{-i\omega/2}
    + Q_2 (2-2r_0)^{i\omega/2}
    =
    \frac{\Psi_0}{\mathcal{C}_1} .
\end{equation}
This condition can be recast as
\begin{equation}\label{QUANTIZATION_BTZ}
    e^{i(\theta_P-\theta_Q)}
    =
    \mu_l\, e^{i\left(\lambda_l \omega + \frac{\theta}{2}\right)}
    - e^{i\theta} ,
\end{equation}
where
\begin{equation}
    \theta_P = \Arg(P_2) ,
    \qquad
    \theta_Q = \Arg(Q_2) ,
    \qquad
    \theta = \Arg\!\left[(2-2r_0)^{i\omega}\right] ,
\end{equation}
and
\begin{equation}
    \mu_l
    =
    \left|
        \frac{\Psi_0}{\mathcal{C}_1 Q_2}
    \right| ,
    \qquad
    \lambda_l \omega
    =
    \Arg\!\left(
        \frac{\Psi_0}{\mathcal{C}_1 Q_2}
    \right) .
\end{equation}
The index \(l\) emphasizes that, for each angular momentum, we choose \(\lambda_l\) from a fixed normal distribution. Separating the real and imaginary parts yields
\begin{equation}
    \begin{gathered}
        \mu_l
        =
        2 \cos\!\left(
            \lambda_l \omega - \frac{\theta}{2}
        \right) ,
        \\
        \cos(\theta_P-\theta_Q)
        =
        \cos(2\lambda_l \omega) ,
        \qquad
        \sin(\theta_P-\theta_Q)
        =
        \sin(2\lambda_l \omega) ,
    \end{gathered}
\end{equation}
which implies
\begin{equation}
    \theta_Q = 2\lambda_l \omega + 2\pi n ,
    \qquad
    n \in \mathds{Z} .
\end{equation}
The standard Dirichlet condition~\eqref{brick0} is recovered by setting \(\mu_l=0\). The quantity controlling a nonzero \(\mu_l\) is therefore the difference between \(\lambda_l\) and \(\omega \log(1-r_0)\). As discussed in~\cite{Das:2023ulz}, fluctuations in \(\lambda_l\) can thus be interpreted as fluctuations in the position of the brick wall. We fix the mean value to be \(\langle \lambda_l \rangle = \tfrac{1}{2}\log(2-2r_0)\) and choose the variance \(\sigma^2=\sigma_0^2/l^2\). This yields the normal mode spectrum shown in Fig.~\ref{NORMAL_STATIC}; along the \(n\) direction, the modes remain linear (not shown in the figure).
\begin{figure}[t!]
    \centering
    \begin{subfigure}[b]{0.48\textwidth}
        \centering
        \includegraphics[width=\textwidth]{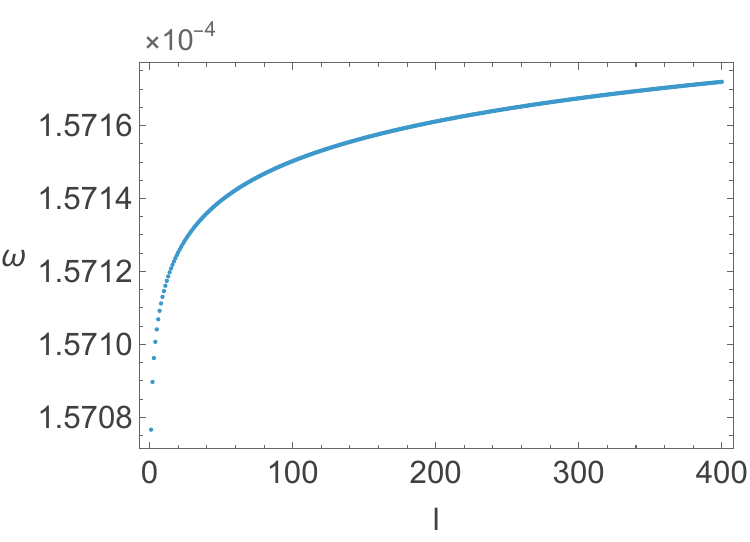}
        \caption{$\sigma_0^2=0$}
    \end{subfigure}
    \hfill
    \begin{subfigure}[b]{0.48\textwidth}
        \centering
        \includegraphics[width=\textwidth]{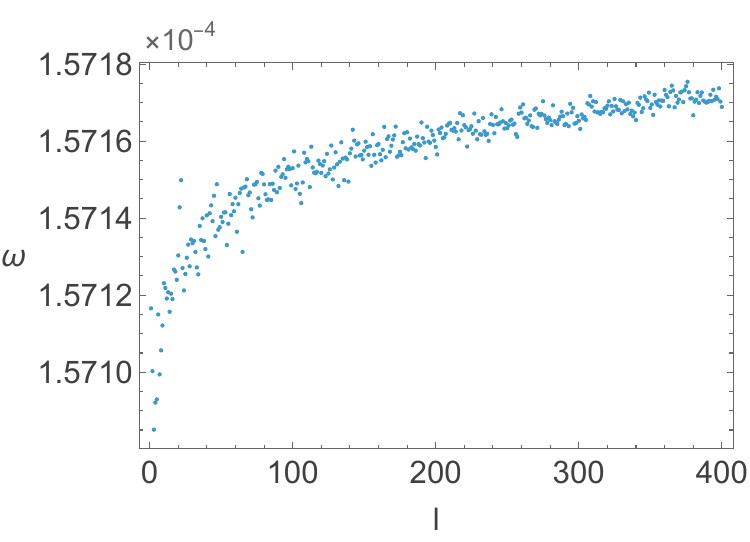}
        \caption{$\sigma_0^2=2$}
    \end{subfigure}
    \caption{Normal modes of a massless scalar field in the static patch of dS$_4$, for \(n=0\), with brick wall boundary condition \(\Psi(r_0)=\mu_l e^{i\lambda_l\omega}\). The parameter \(\lambda_l\) is drawn from a Gaussian distribution with mean $\langle\lambda_l\rangle=\half\log\left(2-2r_0\right)\approx 10^{-4}$ and variance \(\sigma^2=\sigma_0^2/l\).}\LA{NORMAL_STATIC}
\end{figure}
%

\subsection{Spectral properties}

As in the AdS case~\cite{Das:2022evy,Das:2023ulz}, the modes exhibit a nontrivial logarithmic growth with the angular quantum number \(l\). Since this is the same behavior that gives rise to the area-law entropy in the brick wall model, we focus on the spectral properties of the modes in the \(l\) sector. This suggests that the logarithmic dependence is a generic consequence of near-horizon redshift or, equivalently, of the universal Rindler kinematics of non-extremal horizons, rather than a peculiarity of AdS black holes.

\begin{figure}[t!]
    \centering
    \begin{subfigure}[b]{0.48\textwidth}
        \centering
        \includegraphics[width=\textwidth]{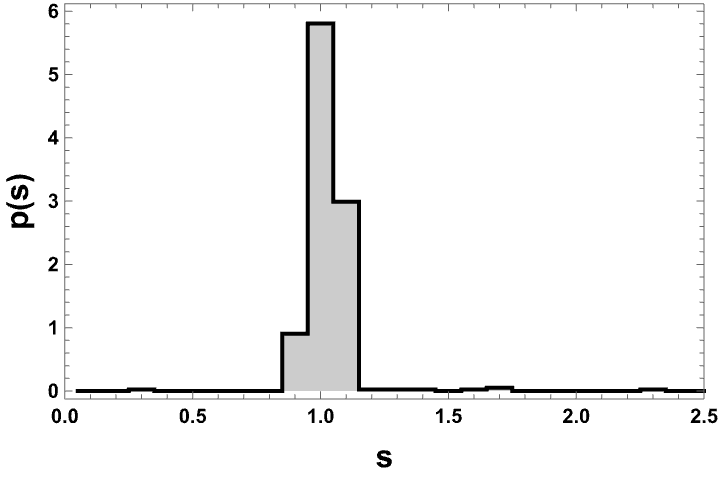}
        \caption{$\sigma_0^2=0$ (Delta-like)}
    \end{subfigure}
    \hfill
    \begin{subfigure}[b]{0.48\textwidth}
        \centering
        \includegraphics[width=\textwidth]{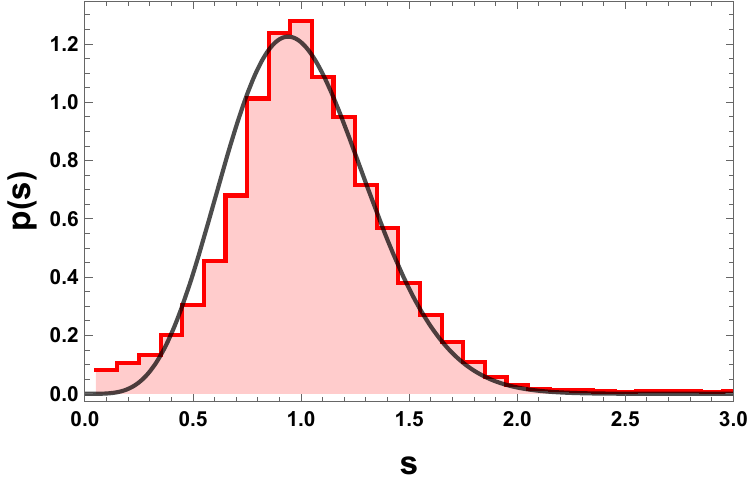}
        \caption{$\sigma_0^2=0.017$ (GSE)}
    \end{subfigure}
    \hfill
    \vspace{.3cm}
    \begin{subfigure}[b]{0.48\textwidth}
        \centering
        \includegraphics[width=\textwidth]{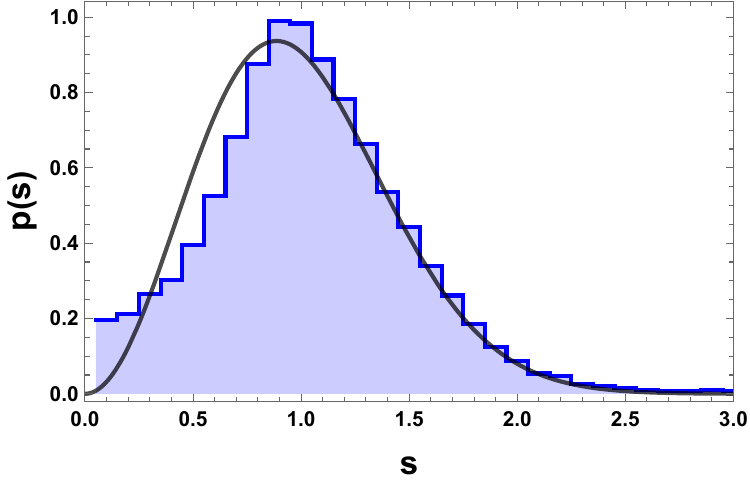}
        \caption{$\sigma_0^2=0.024$ (GUE)}
    \end{subfigure}
    \hfill
    \begin{subfigure}[b]{0.48\textwidth}
        \centering
        \includegraphics[width=\textwidth]{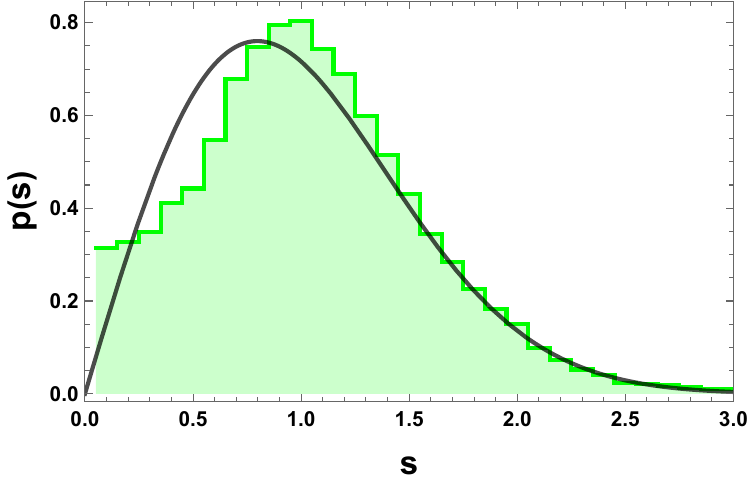}
        \caption{$\sigma_0^2=0.030$ (GOE)}
    \end{subfigure}
    \hfill
    \vspace{.3cm}
    \begin{subfigure}[b]{0.48\textwidth}
        \centering
        \includegraphics[width=\textwidth]{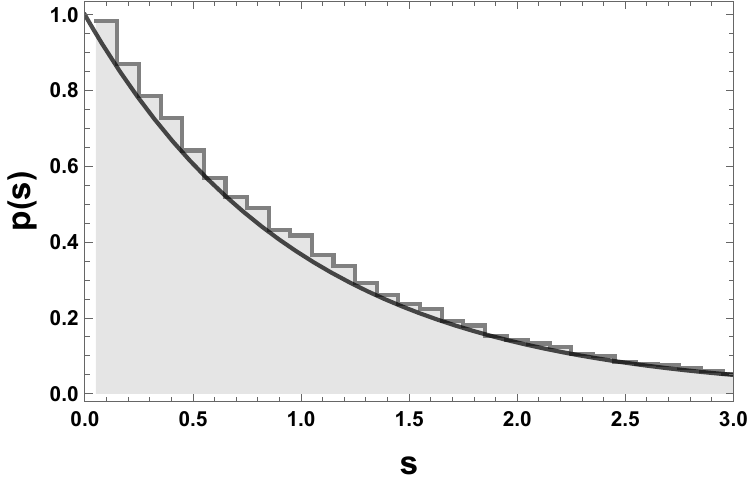}
        \caption{$\sigma_0^2=0.5$ (Poisson)}
    \end{subfigure}
    \vspace{-0.1cm}
    \caption{Behavior of the level-spacing distributions of the normal modes as a function of the variance \(\sigma_0^2\), averaged over 200 realizations. The solid lines correspond to the conventional Wigner--Dyson distributions of the standard ``\(\beta\)-ensemble,'' (GSE/GUE/GOE) as indicated in parentheses.}\LA{FIVE_CASES}
\end{figure}
In Fig.~\ref{FIVE_CASES}, we show the behavior of the level--spacing distribution (LSD) as a function of the variance \(\sigma_0^2\). As \(\sigma_0\) increases, the LSD evolves from a delta--function--like peak to a Poisson distribution, passing through a family of intermediate distributions that resemble Wigner--Dyson statistics. However, as in AdS brick wall spectra, the agreement is not exact: the Gaussian boundary fluctuations produce only effective Wigner--Dyson-like statistics, rather than an exact random-matrix ensemble, since they deform an underlying quasi-logarithmic spectrum in a structured way. As a result, finite-size effects, binning, or residual inhomogeneity in the noisy spectrum can leave a small but nonzero weight near the origin~\cite{Jeong:2024jjn}. A second difference is that the distribution consistently exhibits its maximum near \(s=1\), rather than at the smaller values characteristic of the standard Wigner--Dyson surmises (\(s_{\mathrm{peak}}^{\mathrm{GOE}}=\sqrt{2/\pi}\approx 0.80\), \(s_{\mathrm{peak}}^{\mathrm{GUE}}=\sqrt{\pi}/2\approx 0.89\), and \(s_{\mathrm{peak}}^{\mathrm{GSE}}=\sqrt{2}\,3^{1/4}/\pi^{1/4}\approx 0.97\)).

\begin{figure}[t!]
    \centering
    \begin{subfigure}[b]{0.48\textwidth}
        \centering
        \includegraphics[width=\textwidth]{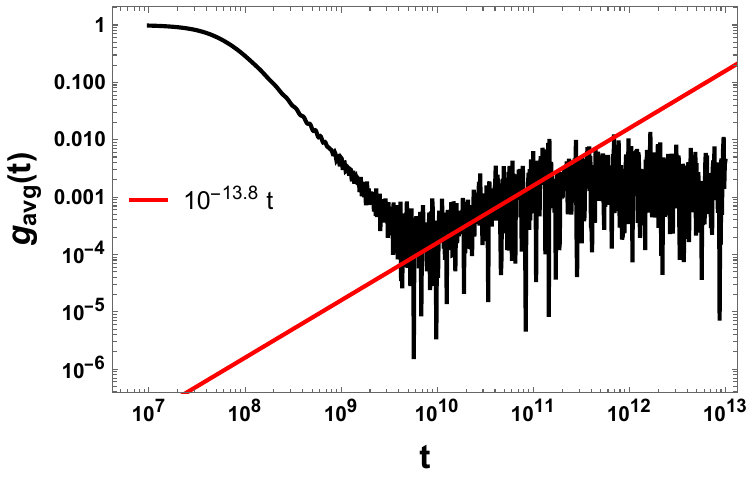}
        \caption{$\sigma_0^2=0$}
    \end{subfigure}
    \hfill
    \begin{subfigure}[b]{0.48\textwidth}
        \centering
        \includegraphics[width=\textwidth]{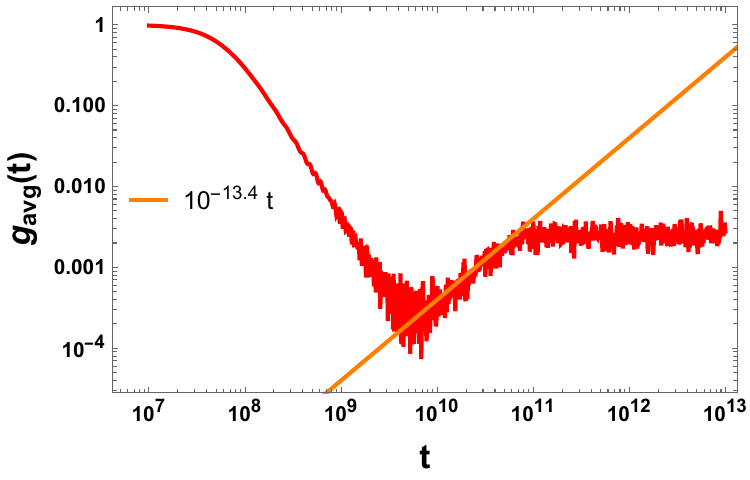}
    \caption{$\sigma_0^2=0.017$}
    \end{subfigure}
    \hfill
        \begin{subfigure}[b]{0.48\textwidth}
        \centering
        \includegraphics[width=\textwidth]{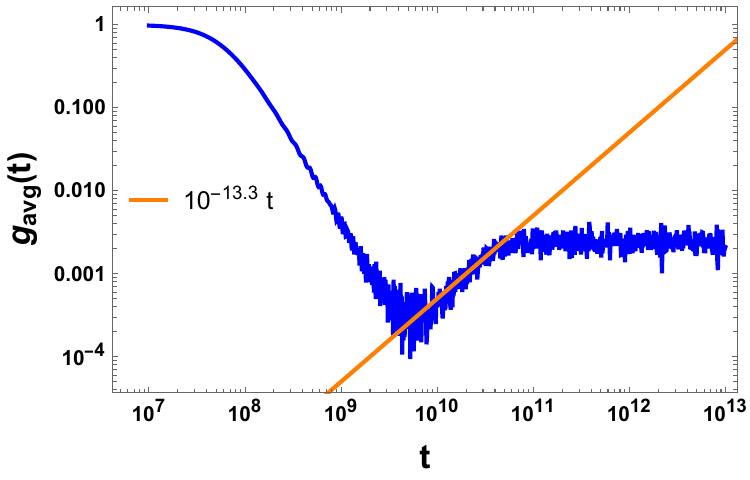}
        \caption{$\sigma_0^2=0.024$}
    \end{subfigure}
    \hfill
    \begin{subfigure}[b]{0.48\textwidth}
        \centering
        \includegraphics[width=\textwidth]{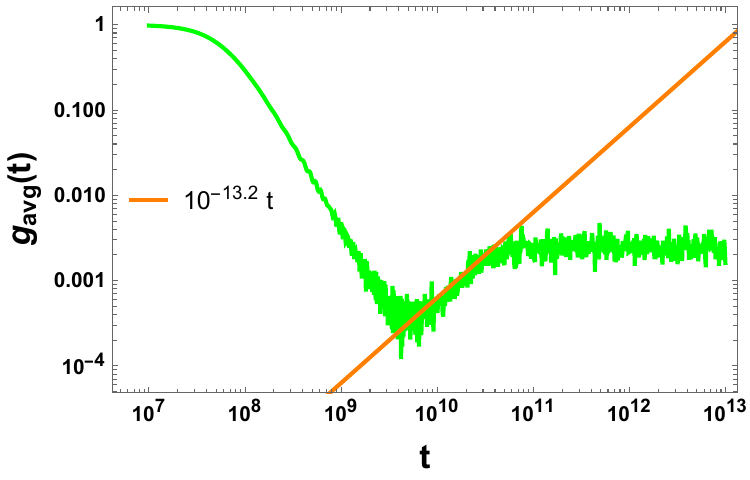}
        \caption{$\sigma_0^2=0.030$}
    \end{subfigure}
    \hfill
    \begin{subfigure}[b]{0.48\textwidth}
        \centering
        \includegraphics[width=\textwidth]{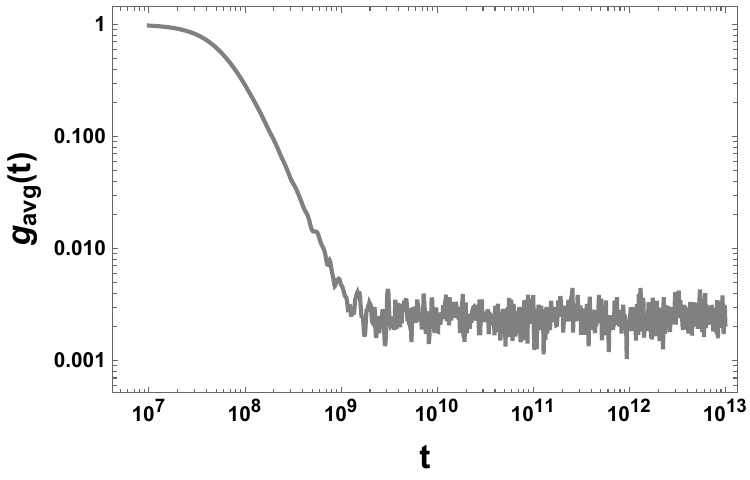}
        \caption{$\sigma_0^2=0.5$}
    \end{subfigure}
    \vspace{-0.1cm}
    \caption{Ensemble-averaged spectral form factor for the scalar field, with the orange lines indicating a unit-slope ramp on the log-log scale.}\LA{SFF_Averaged}
\end{figure}
In Fig.~\ref{SFF_Averaged}, we show the behavior of the SFF at \(\beta=0\), for various values of the variance. We focus on this infinite-temperature case because it provides the cleanest view of the spectral correlations and allows for a direct comparison with our analysis of Krylov complexity. As \(\sigma_0\) increases, the ramp becomes progressively shorter and eventually disappears. Whenever the ramp is present, its slope is equal to one, which is commonly regarded as an indicator of chaotic spectral correlations. However, it is worth emphasizing that this ramp is not exactly identical to the linear ramp observed in random matrix theory (RMT). In RMT, the slope of the ramp is always equal to one whenever a ramp appears in the SFF, whereas in the present case the slope is temperature-dependent: it is equal to one at \(\beta=0\) but decreases as \(\beta\) increases. For further discussion, see~\cite{Basu:2025zkp}.

Finally, in Fig.~\ref{KRYLOV_FIG}, we show the behavior of the Krylov complexity for the TFD state at \(\beta=0\) for various values of \(\sigma_0\). For small \(\sigma_0\), the complexity exhibits a growth--peak--plateau structure, which is commonly observed in chaotic systems. For \(\sigma_0=0\), the peak height is maximal; however, the complexity approaches its plateau value through oscillations, as shown by the black curve. This behavior closely resembles that of the simple harmonic oscillator, which is integrable. In contrast, for nonzero but small variance, the behavior is more akin to that of quantum chaotic systems, with a characteristic peak but without oscillations. As the variance increases further, the peak becomes less pronounced and eventually disappears, consistent with Poissonian behavior. All of these trends are consistent with our observations from the level-spacing distribution and the spectral form factor.
\begin{figure}[t!]
    \centering
    \includegraphics[width=0.6\textwidth]{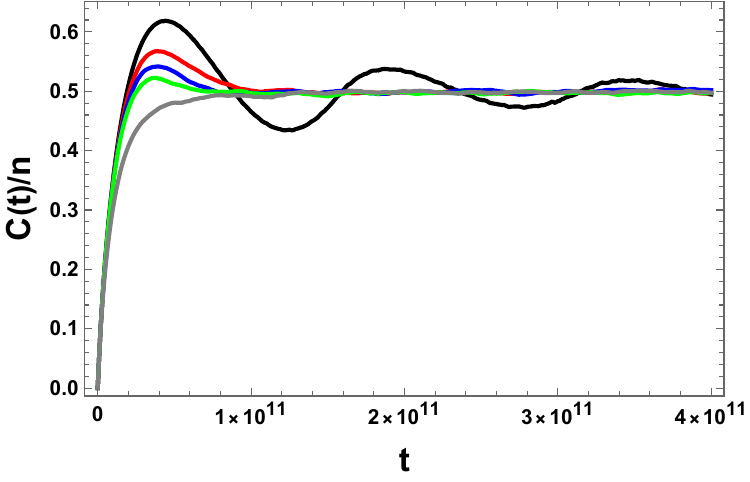}
    \caption{Ensemble-averaged Krylov complexity of the scalar field normal modes when $\sigma_0^2=\{0,~ 0.017,~ 0.024,~ 0.030,~ 0.5\}$  for black, red, blue, green and black, respectively.}\LA{KRYLOV_FIG}
\end{figure}

\section{Schwarzschild--de~Sitter black hole}\label{sec:sds}

In the previous section, we considered pure de~Sitter spacetime and showed that placing a brick wall near the cosmological horizon reproduces the same qualitative features obtained by placing a wall near the horizon of an AdS black hole. In this section, we extend the analysis to the Schwarzschild--de~Sitter black hole. For definiteness, we work in five spacetime dimensions, where the metric functions and horizon structure take a relatively simpler form, although we expect the general lessons to apply in all dimensions.

The metric of the 5-dimensional Schwarzschild--de~Sitter black hole is given by
\begin{equation}
    \dd s^2
    = -f(r)\,\dd t^2
    + \frac{\dd r^2}{f(r)}
    + r^2\,\dd\Omega_3^2 \, ,
\end{equation}
where the blackening function \(f(r)\) takes the form
\begin{equation}
    f(r)
    = \frac{\Lambda}{6r^2}\,
      \bigl(r^2 - r_e^2\bigr)\,
      \bigl(r_c^2 - r^2\bigr) \, .
\end{equation}
Here \(\Lambda>0\) is the cosmological constant. The constants \(r_e\) and \(r_c\) denote the radii of the black hole event horizon and the cosmological horizon, respectively, and are given by
\begin{equation}
    r_e
    = \left(
        \frac{3 - \sqrt{9 - 12 M \Lambda}}{\Lambda}
      \right)^{1/2},
    \qquad
    r_c
    = \left(
        \frac{3 + \sqrt{9 - 12 M \Lambda}}{\Lambda}
      \right)^{1/2},
\end{equation}
where $M$ is the mass parameter. The radial coordinate $r$ is restricted to the static patch
\begin{equation}
    r_e < r < r_c \, ,
\end{equation}
which corresponds to the region accessible to a static observer between the two horizons.

In this background, we are once again interested in computing the normal modes of massless scalar fluctuations and analyzing their properties. To this end, we first need to solve the Klein--Gordon equation using the standard separation ansatz
\begin{equation}
    \Phi
    = e^{-i\omega t}\,\Psi(r)\,Y(\Omega_3) \, ,
\end{equation}
where \(Y(\Omega_3)\) are spherical harmonics on \(S^3\). The radial equation then takes the form
\begin{equation}\label{rad_BH}
    \alpha(r)\,\Psi''(r)
    - \beta(r)\,\Psi'(r)
    + \gamma(r)\,\Psi(r)
    = 0 \, ,
\end{equation}
with
\begin{align}
    \alpha(r)
    &= \frac{\Lambda}{6r^2}\,
       (r^2 - r_e^2)(r_c^2 - r^2), \\
    \beta(r)
    &= \frac{\Lambda}{6r^3}
       \Bigl(
           5r^4 + r_c^2 r_e^2
           - 3r^2(r_c^2 + r_e^2)
       \Bigr), \\
    \gamma(r)
    &= \frac{6r^2\omega^2}
            {(r_c^2 - r^2)(r^2 - r_e^2)}
       - \frac{l(l+2)}{r^2} \, .
\end{align}

To compute the normal modes, we introduce two brick walls placed close to the event horizon and the cosmological horizon, respectively, so that no flux leaks through either horizon. At this stage, an important qualitative difference from the previous cases arises. In the geometries studied earlier, there was only a single horizon. In the present case, however, the spacetime contains two horizons. A natural dimensionless parameter is the ratio of the proper distances from the two horizons to their respective brick walls, namely \(s_e/s_c\), where \(s_e\) and \(s_c\) denote the proper distances from the event horizon and the cosmological horizon, respectively. In what follows, we study the behavior of the normal modes as a function of this parameter. Placing the brick walls at \(r = r_e + \epsilon\) and \(r = r_c - \delta\), respectively, we find that the corresponding proper distances are given by
\begin{equation}
    s_e(\epsilon)=
    \sqrt{\frac{6}{\Lambda}}\,
    \arcsin\!\left(
    \sqrt{\frac{(r_e+\epsilon)^2 - r_e^2}{\,r_c^2 - r_e^2\,}}
    \right), 
    \qquad
    s_c(\delta)
    =\sqrt{\frac{6}{\Lambda}}\,
    \arccos\!\left(
    \sqrt{\frac{(r_c-\delta)^2 - r_e^2}{\,r_c^2 - r_e^2\,}}
    \right).
\end{equation}
These relations can be inverted to give
\begin{align}
    \epsilon(s_e) &=
    \sqrt{r_e^2+\left(r_c^2-r_e^2\right)
    \sin^2\!\left(\sqrt{\frac{\Lambda}{6}}\;s_e\right)}-r_e , \\
    \delta(s_c) &=
    r_c-\sqrt{r_c^2-\left(r_c^2-r_e^2\right)
    \sin^2\!\left(\sqrt{\frac{\Lambda}{6}}\;s_c\right)} .
\end{align}
This implies
\begin{equation}
    \frac{\epsilon}{\delta}
    = \frac{r_c}{r_e} \left( \frac{s_e}{s_c} \right)^2 + \ldots
\end{equation}
Since \(r_c>r_e\), setting \(s_e=s_c\) implies \(\epsilon > \delta\), which in turn gives
\begin{equation}
    g^{\text{event}}_{tt}(s_e) > g^{\text{cosmo}}_{tt}(s_c) .
\end{equation}
Therefore, the scalar field experiences a smaller redshift at the brick wall near the event horizon than at the brick wall near the cosmological horizon. Since the very slow growth of the modes with angular momentum relies on a large redshift, we expect slower growth for modes localized near the cosmological horizon than for those localized near the event horizon. Such behavior can be reversed if the ratio \(s_e/s_c\) is chosen such that\footnote{At first sight, it may appear that for \(\frac{s_e}{s_c} \leq \sqrt{\frac{r_e}{r_c}}\) one would have \(g^{\text{event}}_{tt}(s_e) < g^{\text{cosmo}}_{tt}(s_c)\). However, this is not the case because the redshift factors near the two horizons differ. In particular, one can check that placing the walls at the same \emph{coordinate} distance \(\epsilon_0\) from their corresponding horizons,
\(
\frac{g^{\text{event}}_{tt}(\epsilon_0)}{g^{\text{cosmo}}_{tt}(\epsilon_0)}=\frac{r_c}{r_e}>1 .
\)
}
\begin{equation}
     g^{\text{event}}_{tt}(s_e) < g^{\text{cosmo}}_{tt}(s_c) .
\end{equation}
We will demonstrate this explicitly by computing the normal modes.

\subsection{Computation of normal modes}

\begin{figure}[t!]
    \centering
    \includegraphics[width=0.6\textwidth]{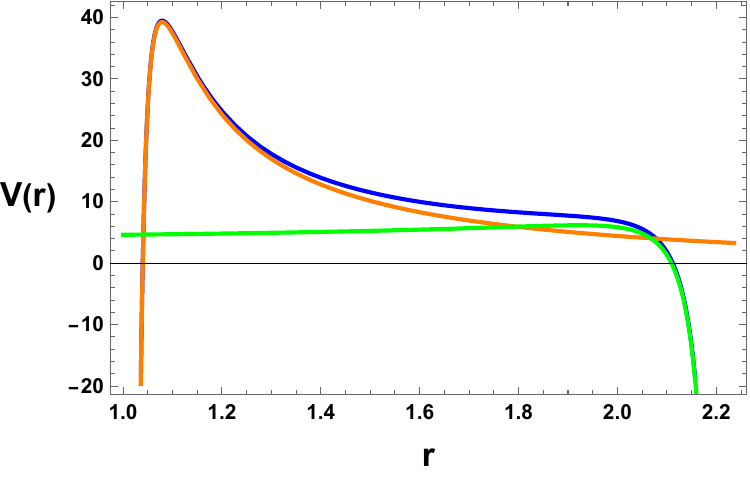}
    \caption{Potential for the Schwarzschild--de~Sitter black hole with \(r_e=1\), \(r_c=\sqrt{5}\), \(\omega=0.1\), and \(l=2\). The blue curve represents the full potential, while the orange and green curves show the leading-order expansions around the two horizons.}
    \label{SchPotential}
\end{figure}

For the sake of concreteness, in what follows we fix
\begin{equation}
    \Lambda=1 \,, \qquad M=\frac{5}{12}
    \quad \Longrightarrow \quad
    r_e=1 \,, \qquad r_c=\sqrt{5} .
\end{equation}
With these choices, the radial equation~\eqref{rad_BH} takes the form
\begin{equation}\label{KG_Schwarzschild}
    \left(1-\frac{5+r^4}{6 r^2}\right)\Psi''
    +\left(\frac{3}{r}-\frac{5r}{6}-\frac{5}{6r^3}\right)\Psi'
    +\left(\frac{6r^2\omega^2}{5-6r^2+r^4}
    -\frac{l(l+2)}{r^2}\right)\Psi=0 .
\end{equation}
It is worth emphasizing that, unlike the pure de~Sitter case, this equation cannot be solved analytically in terms of known functions. We therefore resort to the WKB approximation, which we review in Appendix~\ref{WKBapp}.\footnote{As a consistency check, we have benchmarked the method by repeating the analysis for pure de~Sitter within the WKB approximation, finding excellent agreement. See Appendix~\ref{WKBpuredS} for details.} Although Eq.~\eqref{KG_Schwarzschild} is not in Schr\"odinger form, it can be recast as a zero-energy Schr\"odinger equation for a new function \(\chi(r)\), leading to
\begin{equation}
    \chi''(r)-V(r)\,\chi(r)=0 \, ,
\end{equation}
where the effective potential is given by
\begin{equation}
    V(r)
    =
    \frac{-25-60\!\left(3+2l(2+l)\right)r^2
    +6\!\left(43+24l(l+2)\right)r^4
    -12\!\left(11+2l(l+2)+12\omega^2\right)r^6
    +15r^8}
    {4r^2\!\left(5-6r^2+r^4\right)^2} \, .
\end{equation}
A representative form of the potential is shown in Fig.~\ref{SchPotential}, illustrating that the effective potential possesses two turning points as a result of the two-horizon structure of the geometry. We denote them by \(r_e^*\) and \(r_c^*\), corresponding to the turning points on the event horizon and cosmological horizon sides, respectively.

The brick walls are placed at the stretched horizons, \(r_{0e}\) and \(r_{0c}\), respectively. They lie between the true horizons and their corresponding turning points, allowing for the existence of bound states. As usual, we impose Dirichlet boundary conditions at the brick walls,
\begin{equation}
    \Psi(r_{0e})=\Psi(r_{0c})=0 \, ,
\end{equation}
which amounts to introducing infinite potential barriers at \(r_{0e}\) and \(r_{0c}\). The corresponding WKB quantization conditions, derived in detail in Appendix~\ref{WKBapp}, take the form
\begin{equation}\label{Event_Condition}
    \int_{r_{0e}}^{r_e^*} \sqrt{-V(r)}\,\dd r
    =
    \left(n-\frac{1}{4}\right)\pi \, ,
\end{equation}
\begin{equation}\label{cosmo_Condition}
    \int_{r_c^*}^{r_{0c}} \sqrt{-V(r)}\,\dd r
    =
    \left(n-\frac{1}{4}\right)\pi \, ,
\end{equation}
where \(n \in \mathbb{Z}^+\). In writing these conditions, we have assumed that there is no tunneling between the two classically allowed regions, namely
\begin{equation}
    \exp(-S)
    =
    \exp\!\left(
    -\int_{r_e^*}^{r_c^*} \sqrt{V(r)}\,\dd r
    \right)
    \;\rightarrow\; 0 \, ,
\end{equation}
which occurs when the two turning points are sufficiently far apart, when the potential barrier between them is sufficiently high, or due to a combination of both effects. Although for our chosen parameters \(r_c\) is not parametrically larger than \(r_e\), we restrict to a regime in which the maximal cutoff in \(\omega\) is much smaller than the minimum of the potential, \(V_{\text{min}}(r)\), in the region \(r_e^*<r<r_c^*\). In this regime, the mixing between modes localized near the two horizons can be consistently neglected. Furthermore, in deriving Eqs.~\eqref{Event_Condition} and~\eqref{cosmo_Condition}, we have neglected corrections of order
\(\mathcal{O}\!\left(1/\!\int_{r_*}^{r_0}\sqrt{|V(r)|}\,\dd r\right)\).
Therefore, the quantization conditions are valid only for \(n\gg1\); throughout this work we take \(n\geq6\).

If we consider the full potential, the left-hand sides of~\eqref{Event_Condition} and~\eqref{cosmo_Condition} do not admit closed analytic expressions. Our strategy is therefore as follows. First, we assume that the turning points lie near their respective horizons. Second, we perform a Taylor expansion of the potential near each horizon and retain only the first three terms, which allows us to evaluate the integrals analytically. The regime of validity of this approximation is evaluated and justified in Appendix~\ref{appenC}. The Taylor-expanded potentials take the form
\begin{align}
    \lim_{r \to r_e + \epsilon}V(r)
    &\approx
    -\frac{4+9\omega^2}{16\epsilon^2}
    +\frac{2+6l+3l^2-9\omega^2}{4\epsilon}
    -\frac{7}{16}(4+9\omega^2)
    +\mathcal{O}(\epsilon) \, ,  \\
    \lim_{r \to r_c -\delta} V(r)
    &\approx
    -\frac{4+45\omega^2}{16\delta^2}
    -\frac{16-12l-6l^2+45\omega^2}{8\sqrt{5}\,\delta}
    +\frac{1}{20}\!\left(18l-4+9l^2-45\omega^2\right)
    +\mathcal{O}(\delta) \, .
\end{align}
We first focus on normal modes localized near the event horizon, for which
\begin{equation}
     \lim_{r \to r_e + \epsilon} V(r)
     \approx
     A_0 + \frac{A_1}{\epsilon} - \frac{A_2}{\epsilon^2} \, ,
\end{equation}
with
\begin{align}
    A_0 = -\frac{7}{16}(4+9\omega^2), \qquad
    A_1 = \frac{1}{4}(2+6l+3l^2-9\omega^2), \qquad
    A_2 = \frac{4+9\omega^2}{16} .
\end{align}
Using this expansion, the integral on the left-hand side of~\eqref{Event_Condition} can be written as
\begin{equation}
    \int_{r_*}^{r_0=r_e+\epsilon_0} \!\sqrt{|V(r)|}\,\dd r
    =
    -\Biggl(
    \sqrt{A_2-\epsilon_0 (A_0 \epsilon_0+A_1)}
    +\frac{A_1}{2 \sqrt{A_0}} \tan^{-1}\!\frac{\tilde{b}}{\tilde{a}}
    + \frac{\sqrt{A_2}}{2}
    \log \!\frac{(a+1)^2+b^2}{(1-a)^2+b^2}
    \Biggr) .
\end{equation}
Here \(\tilde{a}\), \(\tilde{b}\), \(a\), and \(b\) are defined as
\begin{align}
    \tilde{a} &= 1+2\frac{A_0}{A_1}\,\epsilon_0, \hspace{3cm}
    \tilde{b}=\frac{2 \sqrt{A_0 (A_2-\epsilon_0 (A_0 \epsilon_0+A_1))}}{A_1}, \nonumber \\
    a &=-\frac{\sqrt{A_2-\epsilon_0 (A_0 \epsilon_0+A_1)}}{\sqrt{A_2}}, \qquad
    b=\sqrt{\frac{A_0}{A_2}}\,\epsilon_0 .
\end{align}
The resulting quantization condition is therefore
\begin{equation}\label{quanthor}
    -\Biggl(
    \sqrt{A_2-\epsilon_0 (A_0 \epsilon_0+A_1)}
    +\frac{A_1}{2 \sqrt{A_0}} \tan^{-1}\!\frac{\tilde{b}}{\tilde{a}}
    + \frac{\sqrt{A_2}}{2}
    \log \!\frac{(a+1)^2+b^2}{(1-a)^2+b^2}
    \Biggr)
    =
    \left(n - \frac{1}{4} \right) \pi .
\end{equation}
It is worth emphasizing that the allowed values of \(\omega\) are labeled by two quantum numbers, \(n\) and \(l\), with the \(l\)-dependence entering through the left-hand side of~\eqref{quanthor}.

Following a similar procedure, we find that the quantization condition for modes localized near the cosmological horizon is
\begin{equation}\label{quantcosmo}
    -\Biggl(
    \sqrt{B_2-\delta_0 (B_0 \delta_0+B_1)}
    +\frac{B_1}{2 \sqrt{B_0}} \tan^{-1}\!\frac{\tilde{d}}{\tilde{c}}
    + \frac{\sqrt{B_2}}{2}
    \log \!\frac{(c+1)^2+d^2}{(1-c)^2+d^2}
    \Biggr)
    =
    \left(n - \frac{1}{4} \right) \pi ,
\end{equation}
where
\begin{align}
    B_0 &= \frac{-4 + 18\,l + 9\,l^2 - 45\,\omega^2}{20}, \qquad
    B_1 = -\,\frac{16 - 12\,l - 6\,l^2 + 45\,\omega^2}{8\sqrt{5}}, \qquad
    B_2 = \frac{4 + 45\,\omega^2}{16} ,
\end{align}
and
\begin{align}
    \tilde{c} &= 1 + 2\frac{B_0}{B_1}\,\delta_0, \hspace{3cm}
    \tilde{d} = \frac{2 \sqrt{ B_0 \bigl( B_2 - \delta_0 ( B_0 \delta_0 + B_1 ) \bigr) }}{B_1}, \nonumber \\
    c &= -\frac{\sqrt{ B_2 - \delta_0 ( B_0 \delta_0 + B_1 ) }}{\sqrt{B_2}}, \qquad
    d = \sqrt{\frac{B_0}{B_2}}\, \delta_0 .
\end{align}

\begin{figure}[t!]
     \centering
     \begin{subfigure}[b]{0.45\textwidth}
         \centering
         \includegraphics[width=\textwidth]{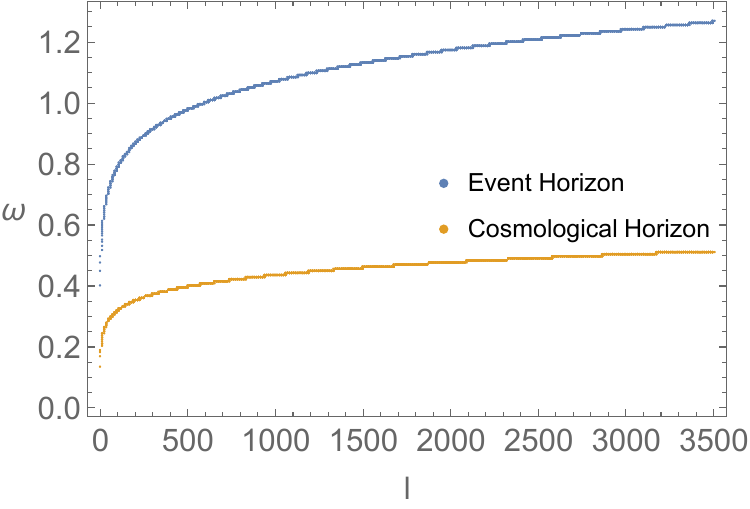}
         \caption{$s_e/s_c=1$}
         \label{SdScomp1}
     \end{subfigure}
     \hfill
      \begin{subfigure}[b]{0.45\textwidth}
         \centering
         \includegraphics[width=\textwidth]{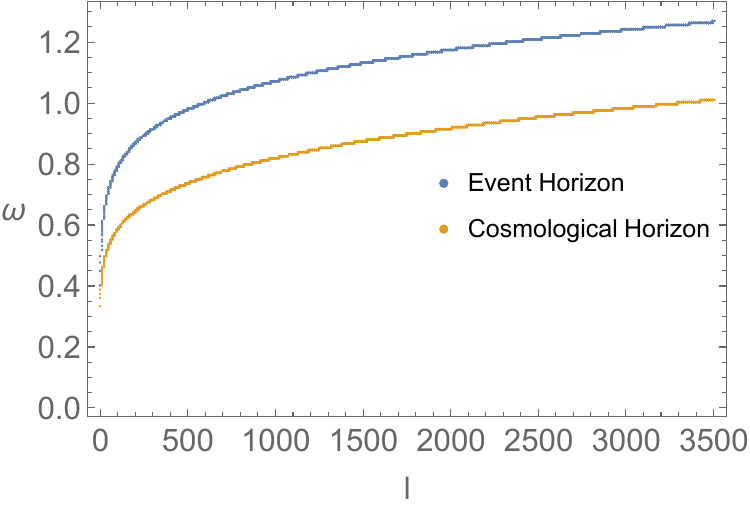}
         \caption{$s_e/s_c=10^{-2}$}
         \label{SdScomp2}
     \end{subfigure}
     \hfill
     \begin{subfigure}[b]{0.45\textwidth}
         \centering
         \includegraphics[width=\textwidth]{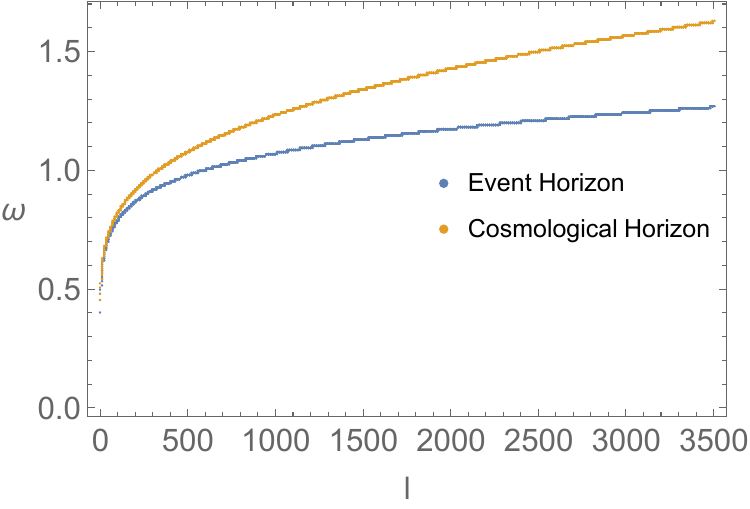}
         \caption{$s_e/s_c=10^{-3}$}
         \label{SdScomp3}
     \end{subfigure}
     \hfill
  \begin{subfigure}[b]{0.45\textwidth}
         \centering
         \includegraphics[width=\textwidth]{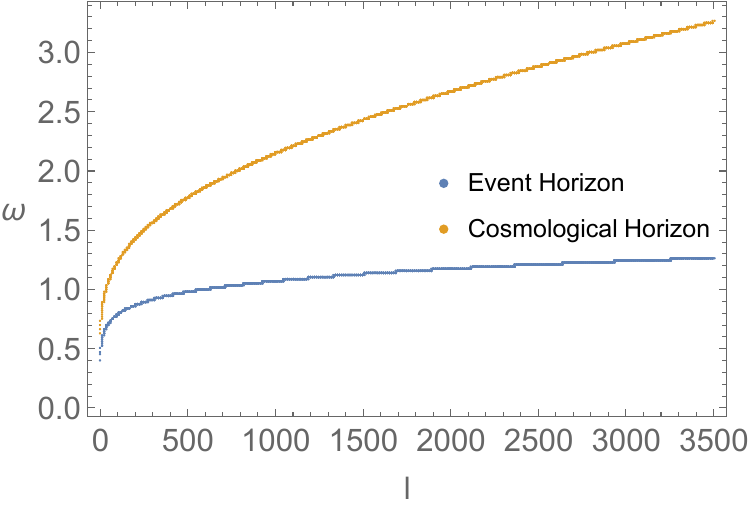}
         \caption{$s_e/s_c=10^{-4}$}
         \label{SdScomp4}
     \end{subfigure}
        \caption{Normal modes of the Schwarzschild--de~Sitter black hole as functions of the angular quantum number \(l\) (\(n=6\)). Modes localized near the event and cosmological horizons are shown in blue and yellow, respectively. While the panels illustrate the dependence on the ratio \(s_e/s_c\), the absolute values of the proper distances are also important. In all panels, \(s_e\) is kept fixed at \(10^{-7}\) (in units of $1/{\sqrt{\Lambda}}$), and \(s_c\) is varied to change the ratio.}
        \label{modes_BH}
\end{figure}

We solve these quantization conditions numerically using \texttt{Mathematica}. The resulting normal modes are shown in Fig.~\ref{modes_BH} for various values of \(s_e/s_c\). As discussed earlier, for equal proper distances between the brick walls and their respective horizons, we have \(\lvert g^{\text{event}}_{tt}(s_e)/g^{\text{cosmo}}_{tt}(s_c) \rvert \sim 4.17\), implying a larger redshift near the cosmological horizon. Accordingly, Fig.~\ref{SdScomp1} shows that modes localized near \(r_c\) are smaller in magnitude and grow much more slowly with \(l\) than those localized near \(r_e\). In contrast, in the last panel, Fig.~\ref{SdScomp4}, we find \(\lvert g^{\text{event}}_{tt}(s_e)/g^{\text{cosmo}}_{tt}(s_c) \rvert \sim 10^{-8}\), leading to the opposite behavior.

Next, we turn to the case with brick wall random fluctuations, following~\cite{Das:2023ulz,Jeong:2024jjn}. Instead of imposing vanishing Dirichlet boundary conditions at the brick walls, we allow for
\begin{equation}
    \Psi(r_{0e}) = C_1 , \qquad \Psi(r_{0c}) = C_2 .
\end{equation}
For arbitrary choices of \(C_1\) and \(C_2\), several difficulties arise. First, the resulting quantization conditions are not easily solvable. More importantly, generic choices do not lead to real normal modes. Following~\cite{Das:2023ulz,Jeong:2024jjn}, we therefore adopt a specific choice, which in the two-horizon case is given by~\eqref{eqC1C2}. With this choice, the WKB quantization conditions become
\begin{align}
    \int_{r_{0e}}^{r_e^*} \sqrt{-V(r)} \,\dd r
    &= \left(n-\frac{1}{4}\right)\pi - \lambda_1 ,
    \label{quantbh1} \\
    \int_{r_c^*}^{r_{0c}} \sqrt{-V(r)} \,\dd r
    &= \left(n-\frac{1}{4}\right)\pi - \lambda_2 .
    \label{quantbh2}
\end{align}
For each \(l\), we draw \(\lambda_{1,2}\) from a Gaussian distribution with zero mean. The zero-variance limit reproduces the standard brick wall boundary conditions discussed earlier. Since the geometry fixes \(V(r)\) and the turning points \(r_e^*\) and \(r_c^*\), introducing fluctuations in \(\lambda_{1,2}\) effectively induces fluctuations in the positions of the brick walls, in analogy with~\cite{Das:2023ulz,Jeong:2024jjn}.

\begin{figure}[t!]
     \centering
     \begin{subfigure}[b]{0.45\textwidth}
         \centering
         \includegraphics[width=\textwidth]{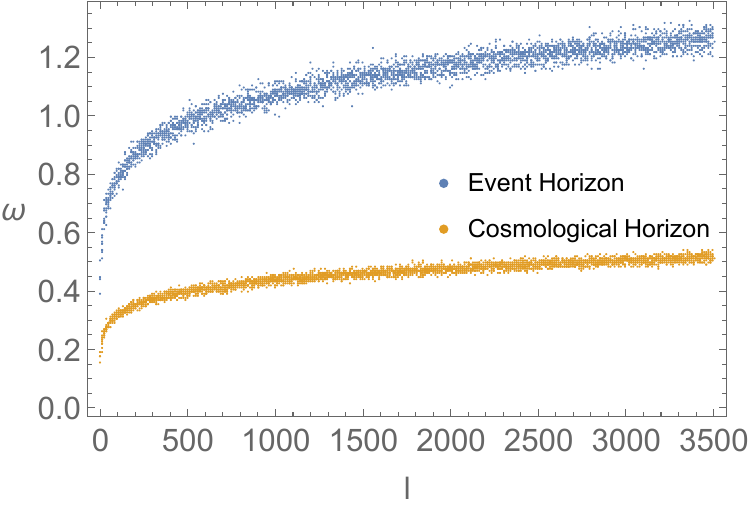}
         \caption{$s_e/s_c=1$}
         \label{comp1}
     \end{subfigure}
     \hfill
      \begin{subfigure}[b]{0.45\textwidth}
         \centering
         \includegraphics[width=\textwidth]{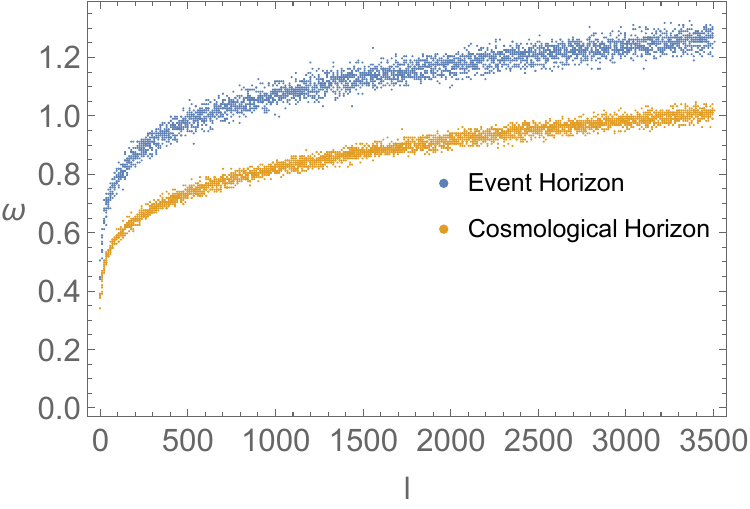}
         \caption{$s_e/s_c=10^{-2}$}
         \label{comp2}
     \end{subfigure}
     \hfill
     \begin{subfigure}[b]{0.45\textwidth}
         \centering
         \includegraphics[width=\textwidth]{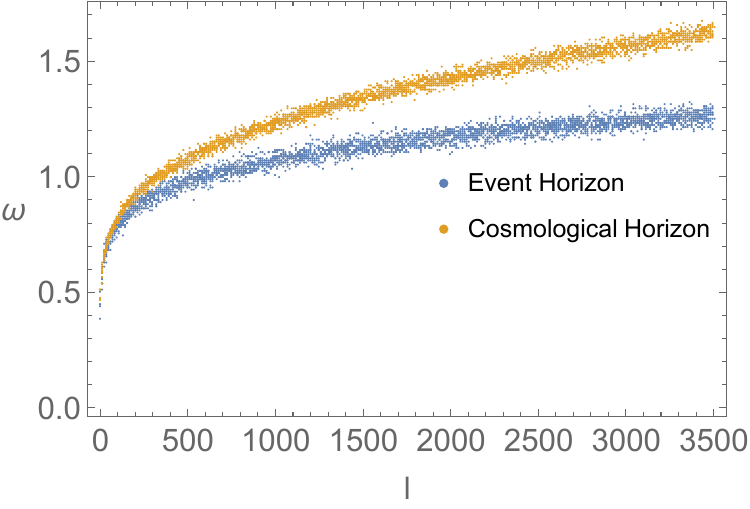}
         \caption{$s_e/s_c=10^{-3}$}
         \label{comp3}
     \end{subfigure}
     \hfill
  \begin{subfigure}[b]{0.45\textwidth}
         \centering
         \includegraphics[width=\textwidth]{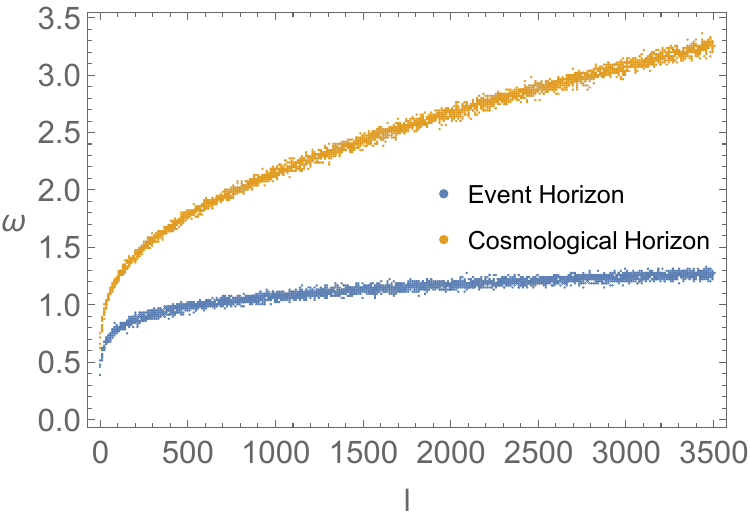}
         \caption{$s_e/s_c=10^{-4}$}
         \label{comp4}
     \end{subfigure}
        \caption{Normal modes as functions of the angular quantum number \(l\), obtained by solving~\eqref{quantbh1}-\eqref{quantbh2}. The parameters \(\lambda_{1,2}\) are drawn from a normal distribution with zero mean and variance \(0.1\). In all panels, \(n=6\) and \(s_e\) is fixed to \(10^{-7}\) (in units of \(1/\sqrt{\Lambda}\)), while \(s_c\) is varied to adjust the ratio.}
        \label{modes_noise}
\end{figure}

As in the previous case, we solve~\eqref{quantbh1}-\eqref{quantbh2} numerically using \texttt{Mathematica} for nonzero variance of \(\lambda_{1,2}\). The resulting quantized modes as functions of the angular quantum number \(l\) are shown in Fig.~\ref{modes_noise}. To avoid introducing an excessive number of parameters, we present results for the case in which both \(\lambda_1\) and \(\lambda_2\) have the same variance; a more general analysis is deferred to the next subsection. It is worth noting that, although both brick walls fluctuate with the same variance, their different locations in the geometry lead to different redshift factors, and consequently the resulting modes are affected differently. In particular, modes localized near the event horizon are influenced more strongly than those localized near the cosmological horizon.
%
%

\subsection{Spectral properties}

The goal of this section is to understand how the presence of two decoupled near-horizon regions and fluctuations of the brick walls manifest in the spectral properties of the system.  The spectrum is obtained in a regime where tunneling between the two potential wells can be neglected. Consequently, the system admits two independent sets of energy levels, \(\{\omega_{\text{event}}\}\) and \(\{\omega_{\text{cosmo}}\}\), associated with the event- and cosmological-horizon wells, respectively. For the spectral analysis, we focus on the single-particle sector. The full single-particle spectrum is therefore given by the union
\begin{equation}
    \{\omega_{\text{full}}\}
    =
    \{\omega_{\text{event}}\}
    \cup
    \{\omega_{\text{cosmo}}\} .
\end{equation}
Since the spectrum contains infinitely many energy levels, we introduce a cutoff \(\omega_{\text{cut}}\) and retain all modes associated with both brick walls whose frequencies lie below it. We then construct the union of the two spectra. Throughout, we work in the regime \(\omega_{\text{cut}} \ll V_{\text{min}}(r)\) in the region \(r_e^*<r<r_c^*\), so that tunneling between the two wells remains suppressed and can be safely neglected. To simplify the analysis, we consider the following two cases:
\begin{enumerate}
    \item Fix the variances \(\lambda_{1,2}\) and vary the relative positions of the brick walls, studying how the spectral properties change as a function of their ratio.
    \item Fix the ratio $s_e/s_c$ and vary the variances, analyzing the resulting spectral properties.
\end{enumerate}

\subsubsection{Case 1: varying the brick wall positions}

We first consider the case in which the two fluctuating brick walls have the same variance. 
Concretely, we take an \(l\)-dependent variance of the form \(\sigma^2=(\sigma_0/l)^2\), with \(\sigma_0^2=0.08\).

\begin{figure}[t!]
     \centering
     \begin{subfigure}[b]{0.45\textwidth}
         \centering
         \includegraphics[width=\textwidth]{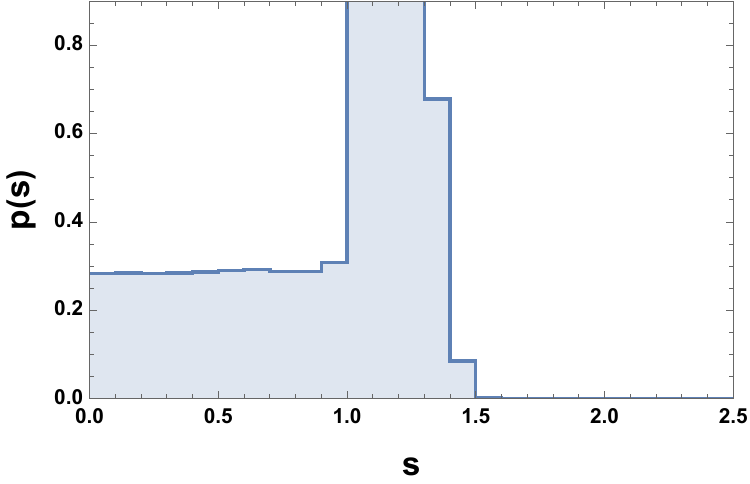}
         \caption{$s_e/s_c=10^{-2}$}
         \label{comp01}
     \end{subfigure}
     \hfill
     \begin{subfigure}[b]{0.45\textwidth}
         \centering
         \includegraphics[width=\textwidth]{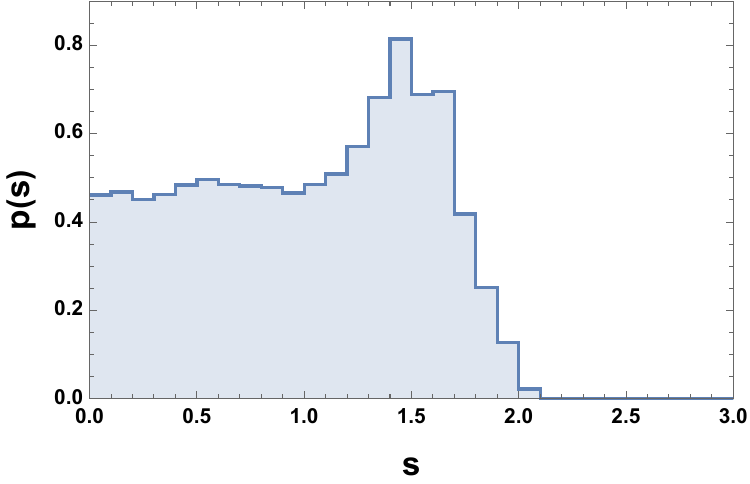}
         \caption{$s_e/s_c=10^{-2.5}$}
         \label{comp02}
     \end{subfigure}
     \hfill
     \begin{subfigure}[b]{0.45\textwidth}
         \centering
         \includegraphics[width=\textwidth]{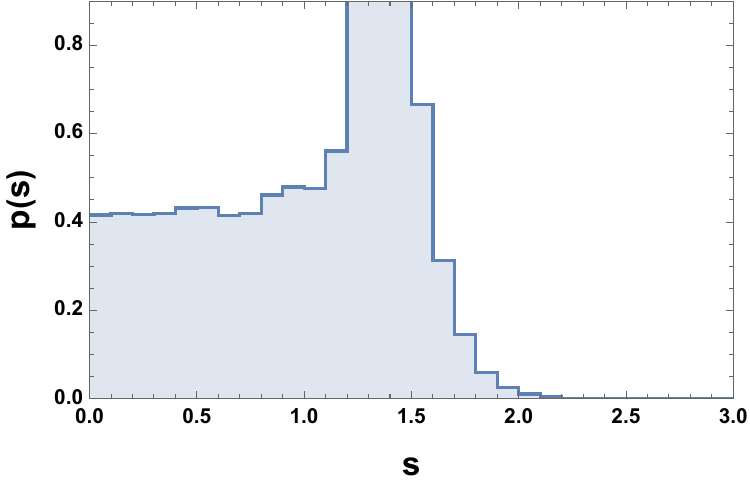}
         \caption{$s_e/s_c=10^{-3}$}
         \label{comp03}
     \end{subfigure}
     \hfill
     \begin{subfigure}[b]{0.45\textwidth}
         \centering
         \includegraphics[width=\textwidth]{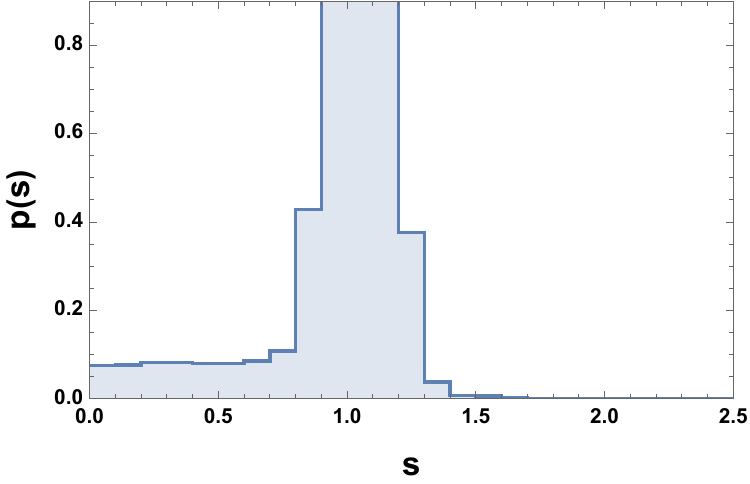}
         \caption{$s_e/s_c=10^{-4}$}
         \label{comp04}
     \end{subfigure}
     \caption{Level-spacing distribution as a function of \(s_e/s_c\), with \(\omega_{\text{cut}} = 1.09\) and \(\sigma_0^2 = 0.08\). Here \(s_e\) is fixed at \(10^{-7}\) (in units of \(1/\sqrt{\Lambda}\)), while \(s_c\) is varied to adjust the ratio. The profiles are consistent with a Berry--Robnik-type distribution. The panels show a zoomed-in view for clarity.}
     \label{avg_LSD_1}
\end{figure}
\paragraph{Level-spacing distribution.} In Fig.~\ref{avg_LSD_1}, we present the LSD as a function of the ratio \(s_e/s_c\), namely the ratio of the proper distances between the brick walls and their respective horizons. In all panels, we fix the cutoff to \(\omega_{\text{cut}} = 1.09\) and include all modes with \(\omega \leq \omega_{\text{cut}}\). It is important to emphasize that, although fluctuations modify the spectrum of each individual potential well in a manner very similar to what we found in pure de~Sitter space, it is far less obvious how the noise should organize the LSD of the full spectrum, which is given by the union of the two nearly independent subsectors. Nevertheless, the two-horizon structure already suggests that Berry--Robnik-type statistics should arise naturally: once the full spectrum is formed by superposing two subsequences, a nonzero value of \(p(s)\) at the origin is no longer in tension with the persistence of nontrivial spectral correlations.

Indeed, the LSD shown in Fig.~\ref{avg_LSD_1} exhibits qualitative features reminiscent of the Berry--Robnik distribution~\cite{Berry:1984,Prosen:1993,Prosen:1994,Prosen:1998,Prosen:1999}, which arises in mixed systems containing regular and chaotic components. In the present setup, however, the interpretation is somewhat different. Rather than modeling mixed behavior statistically from the outset, the Schwarzschild--de~Sitter geometry itself leads, in the WKB regime, to a decomposition of the spectrum into two nearly independent near-horizon subsectors, in the regime of suppressed tunneling. To the best of our knowledge, this provides a particularly clean gravitational realization of a setting in which Berry--Robnik-type statistics are expected on general grounds.

A useful diagnostic in this context is the value of \(p(s)\) at \(s=0\), which is known to encode information about the relative sizes of the two spectral components whose union is being considered. In the standard case where one subsector follows Wigner--Dyson statistics and the other Poisson statistics, one has
\begin{equation}
\label{eqBR}
p(0) = 1 - \rho^2 \, ,
\end{equation}
where \(\rho\) denotes the fractional density of the component for which \(p(0)\) is smaller. In our case, neither subsector is \emph{exactly} Wigner--Dyson nor Poisson, so the exact relation~\eqref{eqBR} need not hold. Still, we expect the same qualitative trend to survive. For the data shown in the four panels Fig.~\ref{avg_LSD_1}, we obtain
\[
1 - \rho^2 = \{0.321,\; 0.644,\; 0.534,\; 0.085 \}\, ,
\]
which is indeed reflected in the corresponding values of \(p(0)\) in the LSD. The associated \(r\)-ratios for these four examples are
\[
\langle r \rangle =\{0.693,\; 0.420,\; 0.495,\; 0.856 \}\, ,
\]
and these are likewise consistent with the observed behavior of \(p(0)\), in the same spirit as the results reported in~\cite{Yan_2025}.

\begin{figure}[t!]
     \centering
     \begin{subfigure}[b]{0.45\textwidth}
         \centering
         \includegraphics[width=\textwidth]{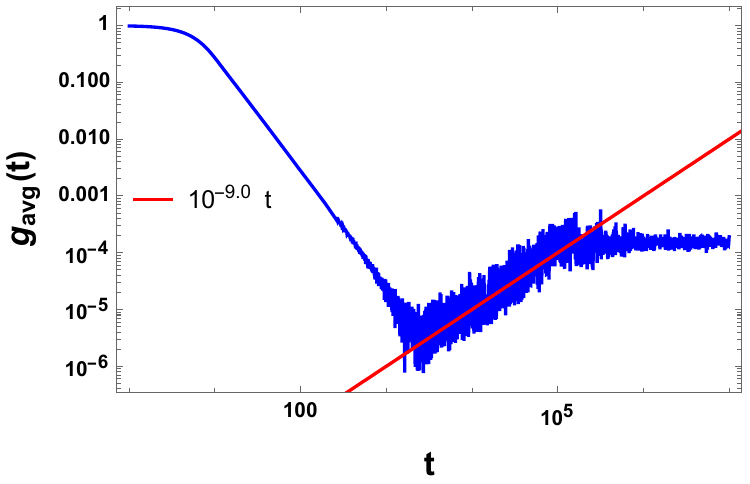}
         \caption{$s_e/s_c=10^{-2}$}
         \label{comp11}
     \end{subfigure}
     \hfill
     \begin{subfigure}[b]{0.45\textwidth}
         \centering
         \includegraphics[width=\textwidth]{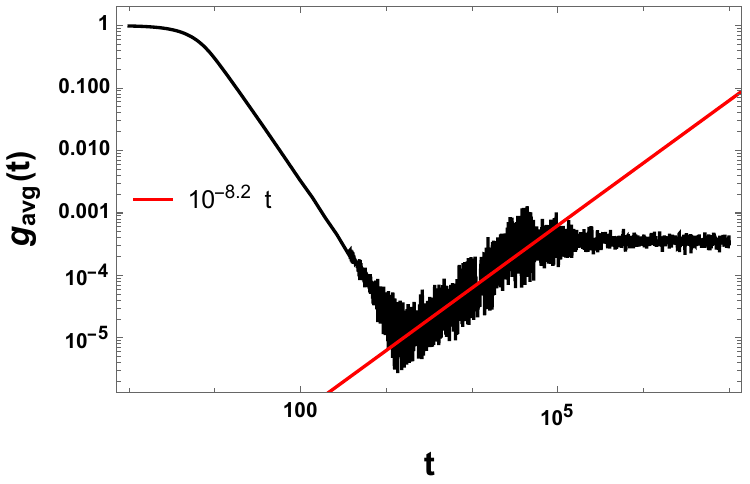}
         \caption{$s_e/s_c=10^{-2.5}$}
         \label{comp22}
     \end{subfigure}
     \hfill
     \begin{subfigure}[b]{0.45\textwidth}
         \centering
         \includegraphics[width=\textwidth]{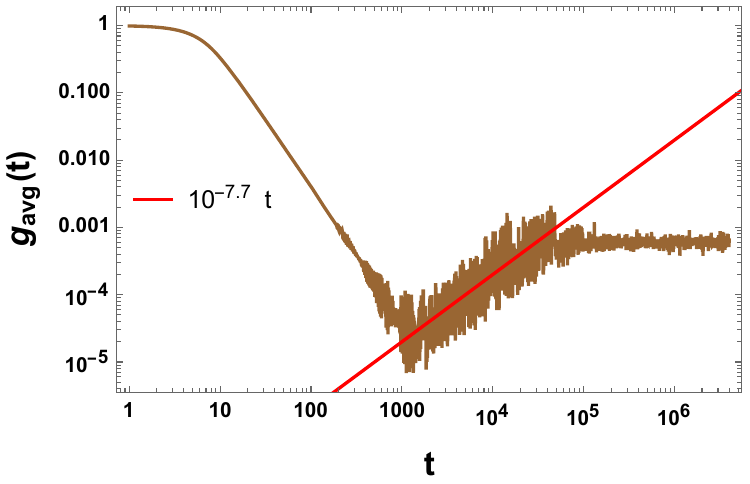}
         \caption{$s_e/s_c=10^{-3}$}
         \label{comp33}
     \end{subfigure}
     \hfill
     \begin{subfigure}[b]{0.45\textwidth}
         \centering
         \includegraphics[width=\textwidth]{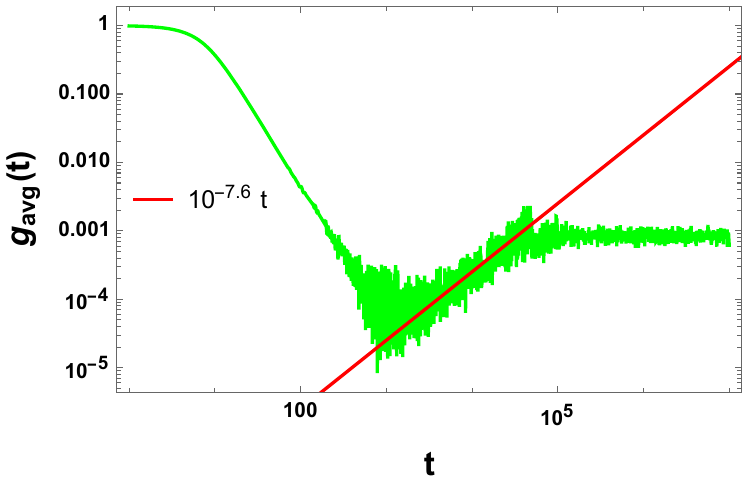}
         \caption{$s_e/s_c=10^{-4}$}
         \label{comp44}
     \end{subfigure}
     \caption{Ensemble-averaged spectral form factor as a function of \(s_e/s_c\), computed from the same set of modes used for the LSD. All plots correspond to \(\beta=0\).}
     \label{avg_SFF_1}
\end{figure}

\paragraph{Spectral form factor.}
In Fig.~\ref{avg_SFF_1}, we present the behavior of the spectral form factor at \(\beta=0\) as a function of the ratio \(s_e/s_c\). In all cases, the SFF exhibits a clear dip--ramp--plateau structure, with a ramp whose slope is very close to one. This is particularly noteworthy because the ramp persists even though the corresponding level-spacing distribution is mixed and has \(p(0)\neq 0\), reinforcing the point that the absence of strict level repulsion in the full spectrum does not by itself preclude nontrivial chaotic correlations. For comparison, we also plot a reference line with unit slope in red.

\paragraph{Krylov complexity.}
In Fig.~\ref{KRYLOV_SCHW_FIG}, we present the Krylov complexity, defined in~\eqref{spread_K}, for the TFD state at \(\beta=0\) in the four cases discussed above. In all cases, the complexity exhibits the characteristic growth--peak--plateau structure: it initially grows, reaches a peak, and then oscillates around \(0.5\) with decreasing amplitude before eventually saturating at that value. Although the spectrum in the present setup is mixed, the peak height may still be used as a useful indicator of chaos in the spectrum~\cite{Erdmenger:2023wjg,Balasubramanian:2022tpr}. From this perspective, a clear pattern emerges: as the \(r\)-ratio increases, the peak becomes progressively higher, consistent with the well-known trend that the Krylov peak height increases as one moves from GOE to GUE to GSE statistics. At the same time, the persistence of oscillations around the plateau suggests that a Poissonian component remains imprinted in the dynamics, as expected for a mixed spectrum. It is worth emphasizing that this behavior is also consistent with the corresponding values of \(1-\rho^2\) obtained above, further supporting the interpretation of these spectra in terms of Berry--Robnik-type statistics.
\begin{figure}[t!]
    \centering
    \includegraphics[width=0.6\textwidth]{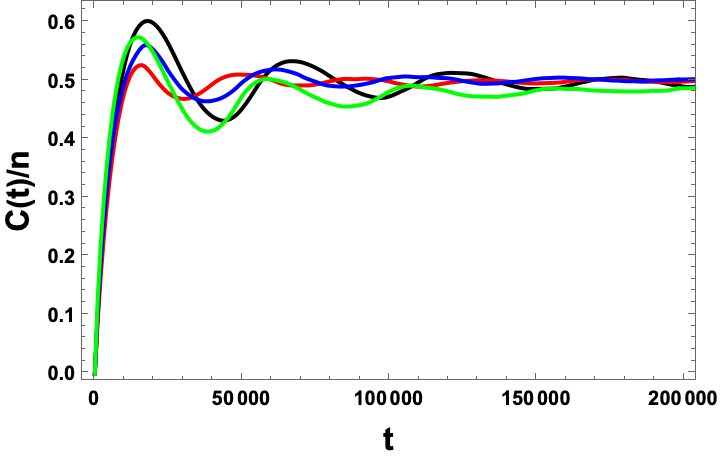}
    \caption{Ensemble-averaged Krylov complexity for the TFD state at \(\beta=0\), shown for \(s_e/s_c = 10^{-2},\,10^{-2.5},\,10^{-3},\,10^{-4}\) in black, red, blue, and green, respectively.}\LA{KRYLOV_SCHW_FIG}
\end{figure}

\subsubsection{Case 2: varying the brick wall variances}

In the second case, we fix the proper distances and vary the relative size of the two variances. This can be done either by changing one variance while keeping the other fixed, or by varying both simultaneously. To avoid clutter, we choose to keep the variance near the event horizon fixed and vary the variance near the cosmological horizon. Other interesting examples are discussed in Appendix~\ref{appenD}. Throughout, the ratio of the proper distances is chosen so that the two horizons contribute approximately equally to the full spectrum.

\begin{figure}[t!]
     \centering
     \begin{subfigure}[b]{0.45\textwidth}
         \centering
         \includegraphics[width=\textwidth]{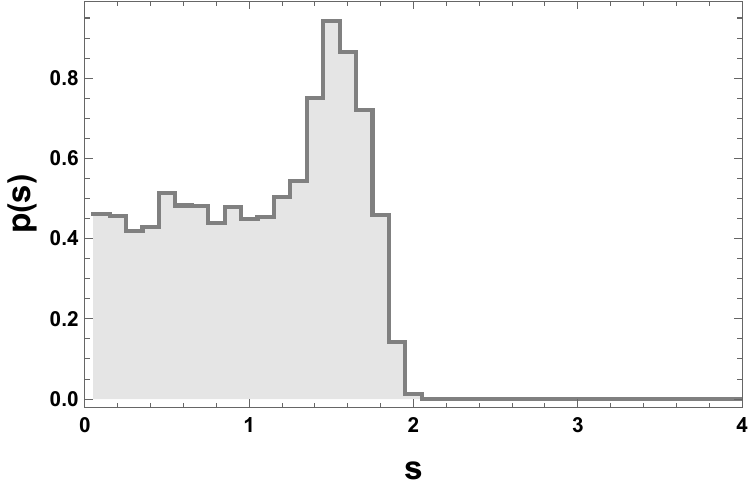}
         \caption{$\sigma_0^2=0.1$}
         \label{comp11}
     \end{subfigure}
     \hfill
     \begin{subfigure}[b]{0.45\textwidth}
         \centering
         \includegraphics[width=\textwidth]{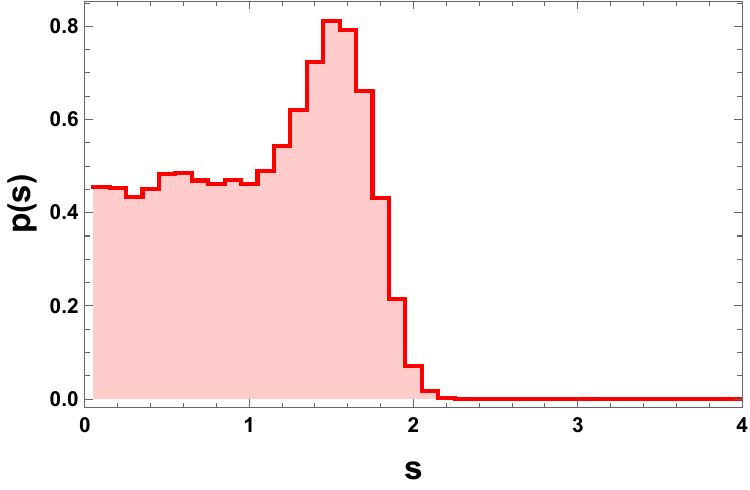}
         \caption{$\sigma_0^2=0.2$}
         \label{comp21}
     \end{subfigure}
     \hfill
     \begin{subfigure}[b]{0.45\textwidth}
         \centering
         \includegraphics[width=\textwidth]{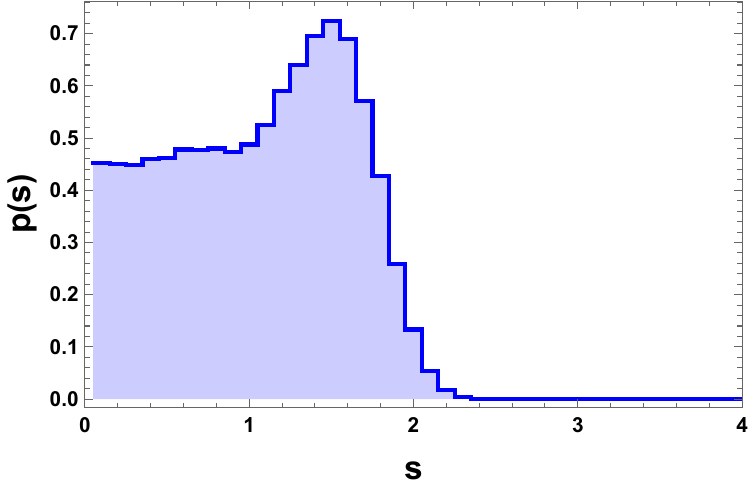}
         \caption{$\sigma_0^2=0.3$}
         \label{comp31}
     \end{subfigure}
     \hfill
     \begin{subfigure}[b]{0.45\textwidth}
         \centering
         \includegraphics[width=\textwidth]{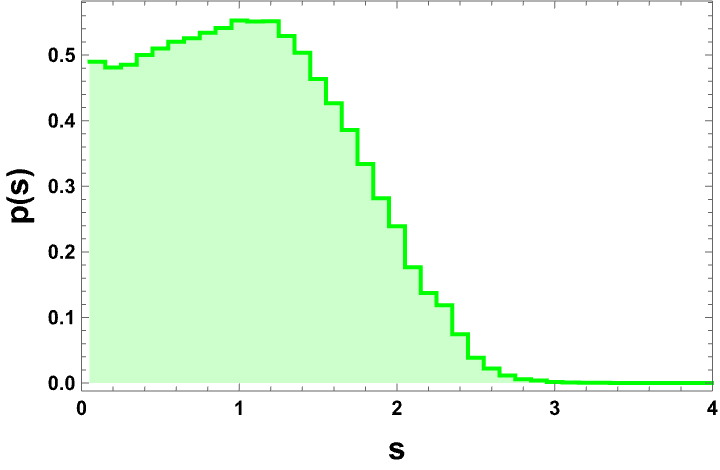}
         \caption{$\sigma_0^2=0.8$}
         \label{comp41}
     \end{subfigure}
     \hfill
     \begin{subfigure}[b]{0.45\textwidth}
         \centering
         \includegraphics[width=\textwidth]{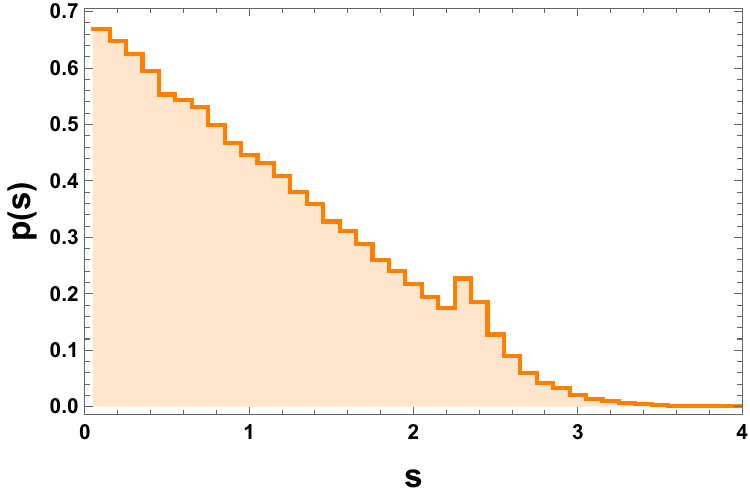}
         \caption{$\sigma_0^2=2$}
         \label{comp51}
     \end{subfigure}
     \hfill
     \begin{subfigure}[b]{0.45\textwidth}
         \centering
         \includegraphics[width=\textwidth]{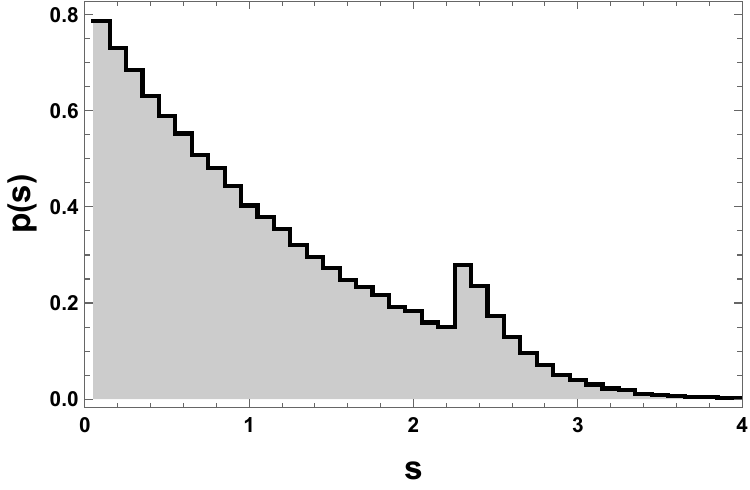}
         \caption{$\sigma_0^2=10$}
         \label{comp61}
     \end{subfigure}
     \caption{Level-spacing distribution as a function of the variances. For concreteness, the variance at the black hole wall is fixed at \(\sigma_0^2=10^{-4}\), while the variance at the cosmological wall is varied. The other parameters are fixed at \(\omega_{\text{cut}}=1.09\) and \(s_e/s_c=10^{-2.5}\).}
     \label{avg_LSD_22_COSMO}
\end{figure}

In Fig.~\ref{avg_LSD_22_COSMO} we show the level-spacing distribution of the combined spectrum. For small variance, the LSD is similar to that shown in panels~\ref{comp02} and~\ref{comp03}, as expected. As \(\sigma_0\) increases, the LSD of the modes localized near the cosmological horizon becomes strongly Poissonian. Consequently, the combined LSD approaches a Poisson distribution with a small bump inherited from the modes localized near the event horizon.

\begin{figure}[t!]
     \centering
     \begin{subfigure}[b]{0.45\textwidth}
         \centering
         \includegraphics[width=\textwidth]{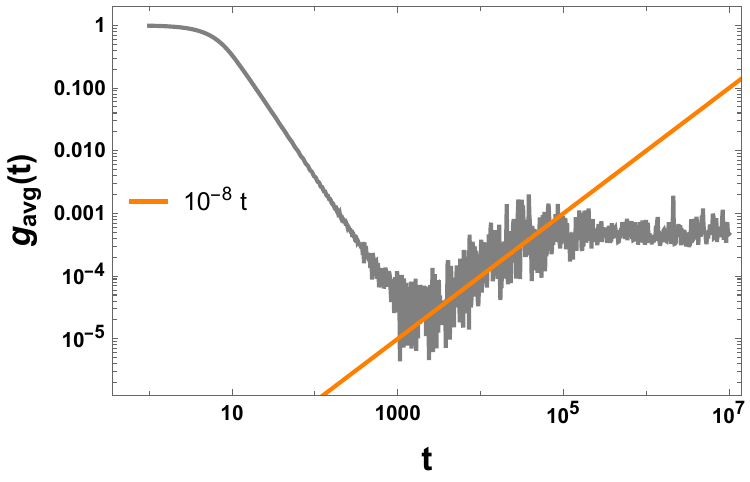}
         \caption{$\sigma_0^2=0.1$}
         \label{comp1}
     \end{subfigure}
     \hfill
     \begin{subfigure}[b]{0.45\textwidth}
         \centering
         \includegraphics[width=\textwidth]{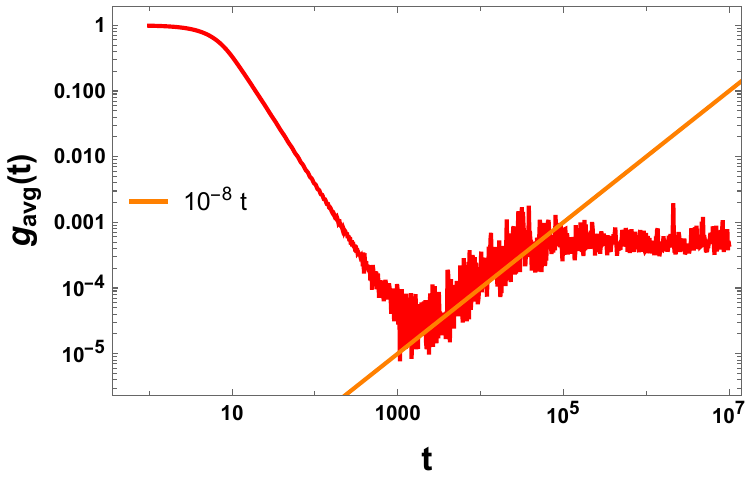}
         \caption{$\sigma_0^2=0.2$}
         \label{comp2}
     \end{subfigure}
     \hfill
     \begin{subfigure}[b]{0.45\textwidth}
         \centering
         \includegraphics[width=\textwidth]{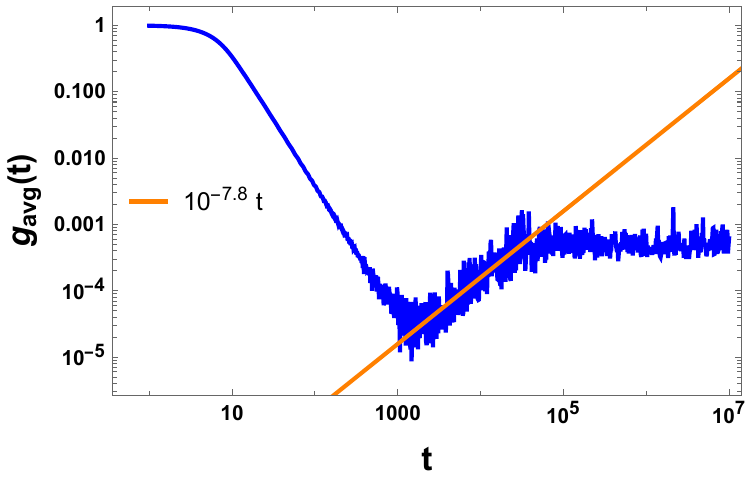}
         \caption{$\sigma_0^2=0.3$}
         \label{comp3}
     \end{subfigure}
     \hfill
     \begin{subfigure}[b]{0.45\textwidth}
         \centering
         \includegraphics[width=\textwidth]{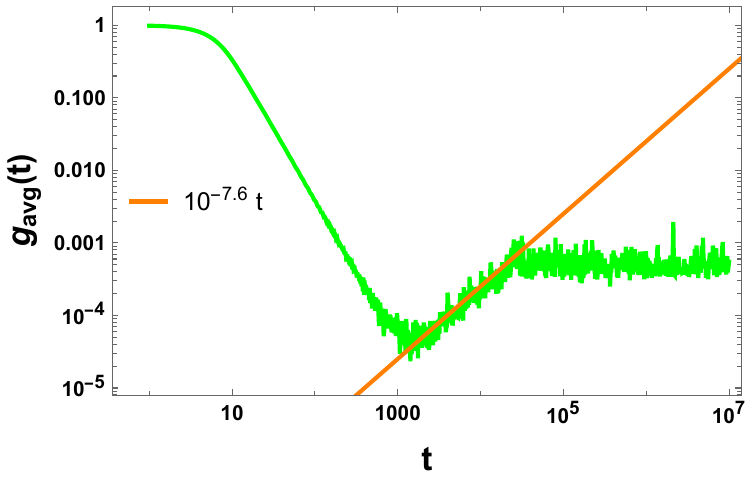}
         \caption{$\sigma_0^2=0.8$}
         \label{comp4}
     \end{subfigure}
     \hfill
     \begin{subfigure}[b]{0.45\textwidth}
         \centering
         \includegraphics[width=\textwidth]{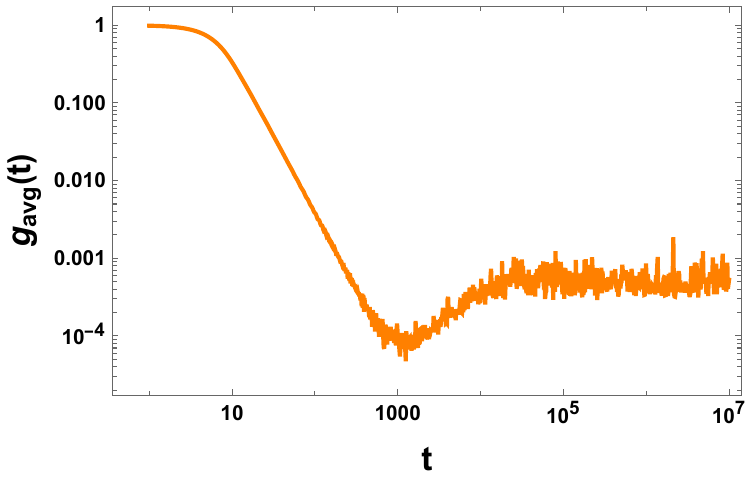}
         \caption{$\sigma_0^2=2$}
         \label{comp4}
     \end{subfigure}
     \hfill
     \begin{subfigure}[b]{0.45\textwidth}
         \centering
         \includegraphics[width=\textwidth]{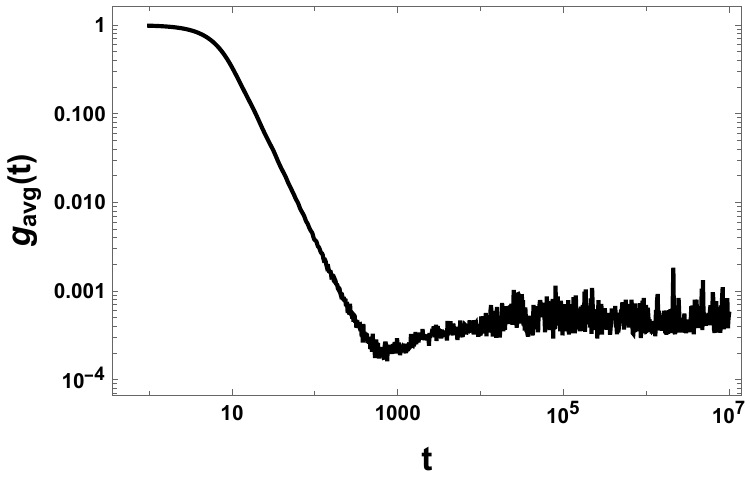}
         \caption{$\sigma_0^2=10$}
         \label{comp4}
     \end{subfigure}
     \caption{Ensemble-averaged spectral form factor as a function of the variances, for \(\beta=0\). For concreteness, the black hole variance is fixed at \(\sigma_0^2=10^{-4}\), while the variance at the cosmological brick wall is varied. The remaining parameters are \(\omega_{\text{cut}}=1.09\) and \(s_e/s_c=10^{-2.5}\).}
     \label{avg_SFF_22_COSMO}
\end{figure}

In Fig.~\ref{avg_SFF_22_COSMO} we show the corresponding SFF at \(\beta=0\). For small variance at the cosmological wall, the SFF exhibits a clear linear ramp. As the variance increases, the slope of the ramp progressively decreases until the ramp eventually disappears. This is consistent with the increasingly Poisson-like character of the LSD shown in Figs.~\ref{comp51} and~\ref{comp61}.

Finally, in Fig.~\ref{KRYLOV_SCHW_FIG_2_COSMO} we present the Krylov complexity for the TFD state at \(\beta=0\), constructed from the spectrum considered above. The behavior is similar to what we observed in Fig.~\ref{KRYLOV_SCHW_FIG}. For small variance, the complexity exhibits a pronounced peak, consistent with the presence of a dominant peak in the LSD or a ramp in the SFF. As the variance increases, this peak becomes progressively less prominent and eventually disappears, in agreement with the Poisson-like character of the LSD.
\begin{figure}[t!]
    \centering
    \includegraphics[width=0.6\textwidth]{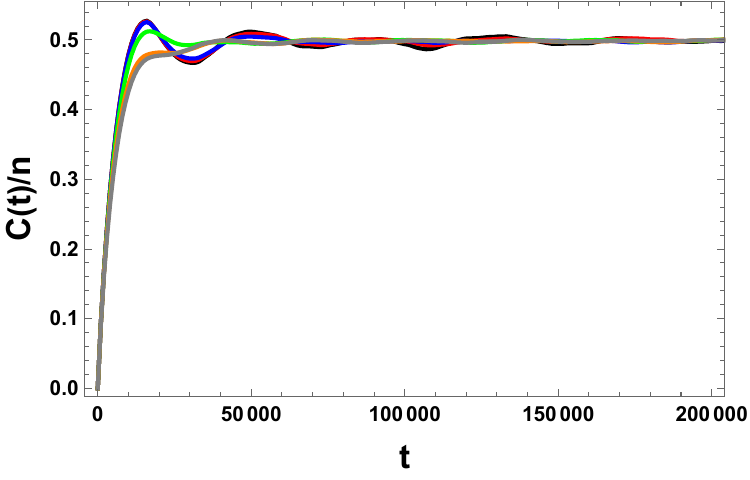}
    \caption{Krylov complexity of the normal modes when $\sigma_0^2= 0.1, ~ 0.2, ~ 0.3, ~ 0.8, ~ 2, ~ 10$ (black, red, blue, green, orange, gray), averaged over 30 realizations. We vary only the cosmological $\sigma_0$ and keep the black hole $\sigma_0^2$ to $10^{-4}$. The parameters are $\omega_{\text{cut}}=1.09$ and $s_e/s_c=10^{-2.5}$.}\LA{KRYLOV_SCHW_FIG_2_COSMO}
\end{figure}

One aspect that remains to be specified is the upper bound on the variances. Mathematically, Eqs.~\eqref{quantbh1}--\eqref{quantbh2} (or their one-horizon counterparts) can in principle be solved for arbitrarily large variances of \(\lambda_{1,2}\). From a physical perspective, however, the fluctuations should not be so large that they completely destroy the level correlations present in the original spectrum without fluctuations.
To ensure that the added fluctuations do not fully erase the underlying spectral correlations, the noise amplitude must remain small compared with the local level spacing. If the perturbed levels are
\begin{equation}
\widetilde{E}_l = E_l + \xi_l ,
\end{equation}
with \(\mathrm{Var}(\xi_l)=\sigma_0^2/l^2\), then the spacing fluctuations are broadened by \(\sqrt{2}\sigma_0/l\). A natural consistency condition for preserving the original correlations is therefore
\begin{equation}
\sqrt{2}\sigma_0/l < \Delta_{\mathrm{loc}} ,
\end{equation}
where \(\Delta_{\mathrm{loc}}\) denotes the local mean level spacing in the spectral window under consideration. For an approximately logarithmic spectrum, one has \(\Delta_{\mathrm{loc}} \sim 1/l\). Consequently, the most restrictive bound arises at the upper end of the chosen spectral window,
\begin{equation}
\sigma_0 < \frac{1}{\sqrt{2}} ,
\end{equation}
up to an order-one factor set by the coefficient \(A\) in \(E_l \sim A \log l\). In this regime, both the SFF and KC retain signatures of chaotic behavior. Hence, the effectively integrable regime lies well outside the physically relevant parameter range and is unlikely to be realized within the regime of validity of the brick wall EFT.

\section{Conclusions}\label{sec:conclusions}

In this paper, we have used the brick wall model to study quantum-chaotic signatures of asymptotically de~Sitter spacetimes. Specifically, we computed the normal modes of a massless scalar field in pure de~Sitter space and in the Schwarzschild--de~Sitter black hole, imposing Dirichlet-type boundary conditions on stretched horizons placed close to the corresponding geometric horizons. We then analyzed the single-particle spectra using three complementary diagnostics: the level-spacing distribution (LSD), the spectral form factor (SFF), and Krylov complexity (KC). Our main goal was to determine to what extent the lessons previously uncovered in AdS black holes survive in de~Sitter settings, and how they are modified in the genuinely two-horizon geometry of Schwarzschild--de~Sitter.

Our results show that the brick wall model continues to provide a remarkably simple but informative framework for accessing nontrivial spectral organization of de~Sitter horizons. In both pure de~Sitter space and the Schwarzschild--de~Sitter black hole, the normal modes exhibit a very slow logarithmic growth with the angular momentum quantum number \(l\), while remaining linear in the principal quantum number \(n\). This quasi-degeneracy along the angular direction is the same basic mechanism that underlies the area-law entropy in the brick wall model, and it appears to be a robust consequence of the large near-horizon redshift, or equivalently of the universal Rindler kinematics of non-extremal horizons, rather than a special feature of AdS black holes alone.

For pure de~Sitter space, the picture closely parallels the AdS story. With a fixed brick wall, the spectrum along the \(l\)-direction is highly structured and does not exhibit conventional Wigner--Dyson level repulsion. Nevertheless, it displays clear long-range signatures of nontrivial spectral correlations: the SFF develops a dip--ramp--plateau structure, while Krylov complexity exhibits the characteristic growth--peak--plateau behavior. Allowing the stretched horizon to fluctuate through Gaussian-distributed boundary conditions enriches this picture further. As the variance is increased, the spectrum interpolates from a sharply correlated regime toward a Poisson-like regime, passing through intermediate distributions that resemble the standard random-matrix universality classes, albeit with a small but nonzero value of \(p(0)\).\footnote{The small residual value of \(p(s=0)\) is not necessarily in tension with chaos. The Gaussian boundary fluctuations produce only effective Wigner--Dyson-like statistics, rather than an exact random-matrix ensemble, since they deform an underlying quasi-logarithmic spectrum in a structured way. Consequently, finite-size effects, binning, or residual inhomogeneity in the noisy spectrum can leave a small but nonzero weight near the origin. Similar behavior also appears in AdS brick wall spectra~\cite{Jeong:2024jjn}.} In the same regime, the ramp in the SFF and the peak in Krylov complexity gradually weaken and eventually disappear. Taken together, these results show that even when the nearest-neighbor statistics are not strictly Wigner--Dyson, the probe spectrum can still encode robust signatures of chaos.

The Schwarzschild--de~Sitter case is qualitatively richer and, in our view, contains the main conceptual message of this work. Because the static patch is bounded by a black hole horizon and a cosmological horizon, the semiclassical problem naturally involves two stretched horizons. In the WKB regime, where tunneling between the two classically allowed regions is exponentially suppressed, the quantization conditions factorize into two independent near-horizon problems. The full single-particle spectrum is therefore not a single sequence, but the union of two spectra associated with the two horizons. This provides a concrete geometric realization of spectral superposition in a gravitational setting.

This observation has important implications for the interpretation of chaos diagnostics. In particular, once two statistically independent or weakly coupled subsequences are superposed, the combined LSD can exhibit a pronounced nonzero value at \(s=0\) simply because levels from the two subsequences interleave, even when both spectra are Wigner--Dyson-like. Thus, the loss of strict level repulsion in the full spectrum does \emph{not} by itself imply the absence of chaos. Our results make this point explicit: although the LSD of the superposed spectrum exhibits structure reminiscent of Berry--Robnik-type behavior, the SFF still develops an approximately linear ramp, and Krylov complexity retains a clear growth--peak--plateau structure in the controlled small-variance regime. In this sense, Schwarzschild--de~Sitter furnishes a clean example in which short-range and long-range diagnostics must be interpreted with care: the combined spectrum is not purely Wigner--Dyson, yet it still carries clear signatures of nontrivial spectral correlations.

A natural objection is that this phenomenon may be nothing more than the familiar artifact of superposing spectra from different symmetry sectors. Indeed, whenever a system decomposes into independent symmetry blocks, failing to resolve those sectors generically removes level repulsion while leaving long-range correlations, such as the SFF ramp, largely intact. That mechanism is well known and, by itself, would not be conceptually new. The point here is more subtle. In the present setup, the superposition does not result from an arbitrary choice to combine sectors that should have been separated. Rather, it is an intrinsic consequence of the two-horizon geometry: in the semiclassical WKB regime, the suppression of tunneling dynamically gives rise to two approximately independent near-horizon subsectors whose union defines the physical single-particle spectrum. Schwarzschild--de~Sitter therefore provides a genuine gravitational realization of spectral superposition, rather than a trivial symmetry-mixing effect.

A second lesson concerns the role of stretched-horizon fluctuations. As in the AdS case, increasing the variance of the fluctuating boundary condition tends to drive the spectrum toward a more Poisson-like regime. In Schwarzschild--de~Sitter, this progressively weakens both the ramp in the SFF and the peak in Krylov complexity, with the SFF appearing more sensitive to noise than the Krylov diagnostic. At a mathematical level, this provides a useful interpolation between more and less correlated spectra. Physically, however, we believe this large-variance regime should be treated with caution. If gravitational horizons are indeed efficient scramblers on general grounds, then the regime of greatest interest is not one in which arbitrarily strong boundary fluctuations wash out all correlations, but rather the controlled probe regime in which those correlations remain visible. From this perspective, the key result is the persistence of the ramp and the Krylov peak over a substantial semiclassical window, even in the presence of two horizons and spectral superposition.

Taken together, our results suggest a clear picture. First, the qualitative probe-sector signatures of horizon chaos previously identified in AdS black holes extend nontrivially to de~Sitter space. Second, in a genuine two-horizon geometry such as Schwarzschild--de~Sitter, the physical spectrum is naturally a superposition of two near-horizon spectra, and this changes how standard spectral diagnostics must be interpreted. Most importantly, the appearance of \(p(s=0)\neq 0\) in the full spectrum does not, by itself, rule out chaos. The deeper lesson is that spectral ramps and Krylov peaks can survive the superposition, making them sharper probes of horizon chaos in multi-horizon systems.

There are several natural directions worth exploring. A first priority is to obtain sharper analytic control over the WKB quantization conditions in suitable limits, along the lines of~\cite{Krishnan:2023jqn}. Closed-form expressions for the spectrum would help clarify the origin of the structures observed in the LSD of the superposed spectrum and would also enable a more direct computation of the thermal entropy, allowing for a sharper comparison with the corresponding horizon entropy. A second important step is to go beyond the no-tunneling regime. In our analysis, tunneling between the two wells of the Schwarzschild--de~Sitter effective potential was consistently neglected within the WKB approximation. Solving the full problem numerically, including the exponentially small mixing between the two near-horizon sectors, would make it possible to track how the superposed spectrum is deformed once the two sets of modes begin to interact, and may reveal genuinely new spectral phenomena characteristic of two-horizon geometries. More broadly, it would be very interesting to consider more general boundary conditions at the stretched horizons, allowing the reflection coefficient to interpolate continuously between perfect reflection and full absorption. Such a deformation would provide a direct bridge between the normal modes studied here and the quasinormal modes of an open system. Understanding this interpolation could provide a more unified picture of how chaotic spectral signatures evolve from closed horizon systems to genuinely dissipative ones.

We hope that the results presented here help sharpen the role of the brick wall model as a probe of horizon dynamics beyond AdS, and that they provide a useful starting point for understanding how quantum-chaotic signatures are encoded in de~Sitter horizons and, more broadly, in multi-horizon spacetimes relevant to quantum gravity and cosmology.

\acknowledgments

We would like to thank José Barbón, Roberto Emparan, Keun-Young Kim, Chethan Krishnan, and Arnab Kundu for useful discussions. SD and JFP are supported by the Comunidad de Madrid `Atracción de Talento' program (ATCAM) grant 2020-T1/TIC-20495, the Spanish Research Agency via grants CEX2020-001007-S, PID2021-123017NB-I00 and PID2024-156043NB-I00, funded by MCIN/AEI/10.13039/501100011033, and ERDF `A way of making Europe.'
HSJ is supported by an appointment to the JRG Program at the APCTP through the Science and Technology Promotion Fund and Lottery Fund of the Korean Government. HSJ is also supported by the Korean Local Governments -- Gyeongsangbuk-do Province and Pohang City.
All authors contributed equally to this paper and should be considered as co-first authors.

%
\appendix

\section{WKB approximation method}\label{WKBapp}

In this appendix, we first show how to rewrite the radial Klein--Gordon equation as a Schr\"odinger equation with an effective potential, and then review how to solve it using the WKB approximation when the potential has one or two turning points, which are the cases of interest here.

The radial Klein--Gordon equation can be written in the following general form:
\begin{equation}
    a(r)\, \Phi''(r) + b(r)\, \Phi'(r) + c(r)\, \Phi(r) = 0,
\end{equation}
where \(\Phi(r)\) is the radial part of the Klein--Gordon field. We now introduce a new radial function \(\phi(r)\) such that
\begin{equation}
    \Phi(r) = g(r)\, \phi(r),
\end{equation}
where \(g(r)\) is an auxiliary function. Substituting this into the above equation gives
\begin{equation}
    a(r) g(r)\, \phi''(r)
    + \Big(2 a(r) g'(r) + b(r) g(r)\Big) \phi'(r)
    + \Big(a(r) g''(r) + b(r) g'(r) + c(r) g(r)\Big) \phi(r) = 0.
\end{equation}
We now impose that the coefficient of \(\phi'(r)\) vanishes, i.e.,
\begin{equation}
    2 a(r) g'(r) + b(r) g(r) = 0.
\end{equation}
Using this condition, we can express \(g'(r)\) and \(g''(r)\) in terms of \(g(r)\), which, after simplification, leads to
\begin{equation}\label{schrodinger}
    \phi''(r) + V(r)\, \phi(r) = 0,
\end{equation}
where the effective potential is given by
\begin{equation}
    V(r) = \frac{1}{4 a(r)^2}
    \Big[ b(r)^2 - 2 b(r)\, a'(r)
    + 2 a(r) \big(b'(r) - 2 c(r)\big) \Big].
\end{equation}
Equation~\eqref{schrodinger} is a Schr\"odinger equation with zero energy and effective potential \(V(r)\). Since Eq.~\eqref{schrodinger} cannot be solved exactly in our case, we now discuss how to solve it using the WKB approximation in the cases where the potential has one or two turning points, corresponding to the pure de~Sitter and Schwarzschild--de~Sitter cases, respectively.

\subsection{WKB with one turning point}\label{oneturning}

\begin{figure*}
    \centering
    \includegraphics[width=.52\textwidth]{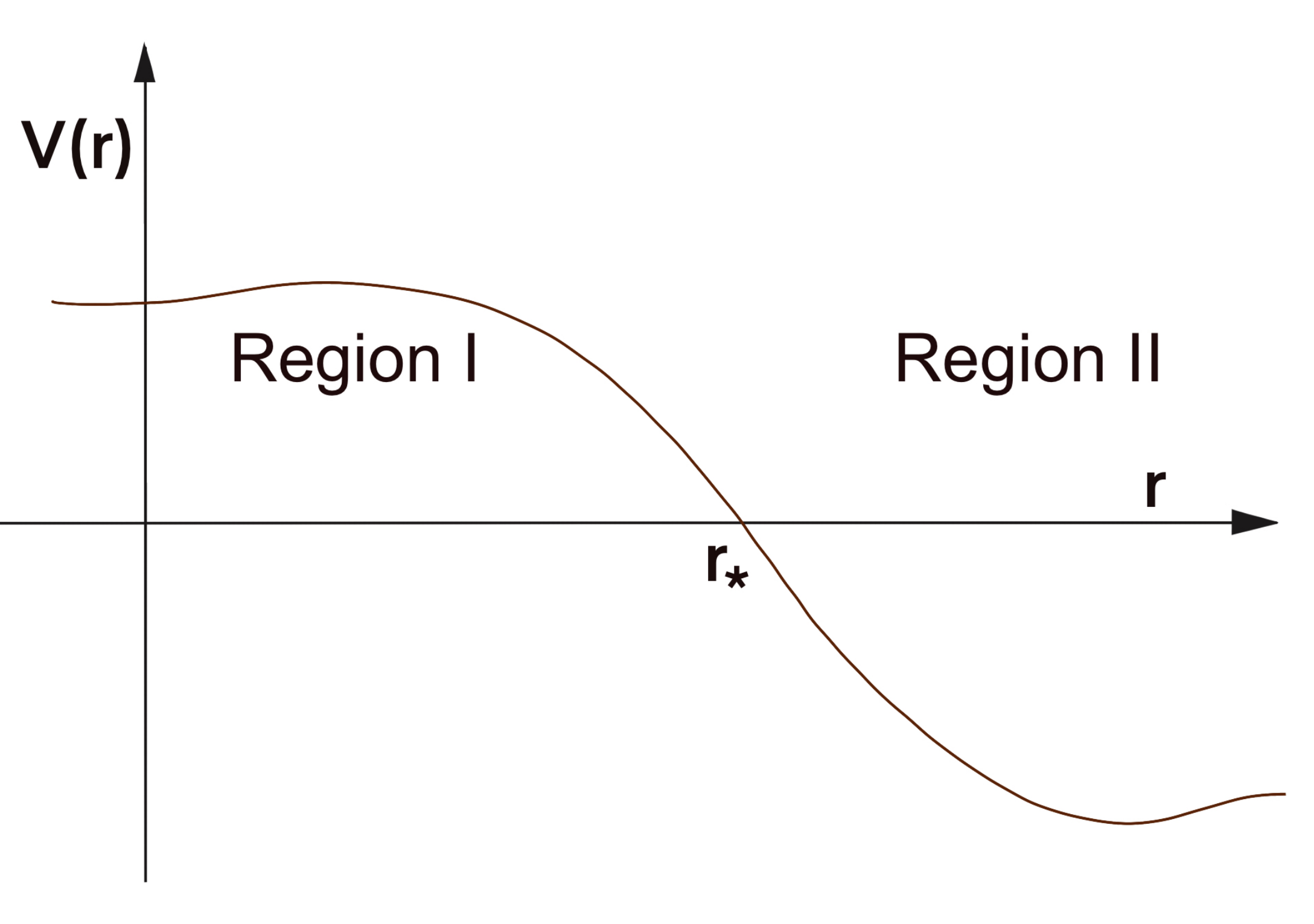}
   \caption{Schematic of a potential \(V(r)\) with a single turning point at \(r=r_*\). The region where \(V(r)>0\) corresponds to the classically forbidden region to the left of \(r_*\), whereas the region where \(V(r)<0\) to the right corresponds to the classically allowed region.}
    \label{WKB_pot1}
\end{figure*}
The structure of the potential in this case is shown in Fig.~\ref{WKB_pot1}, which has a single turning point at \(r=r_*\). The region to the left is classically forbidden, whereas the region to the right is classically allowed. The solutions in regions~I and~II can be written as
\begin{equation}\label{r1}
    \phi_{I}(r)
    = \frac{1}{|V(r)|^{1/4}}
      \left[ \tilde{a}\, e^{-\int_{r}^{r_*} \!\sqrt{V(r')}\,dr'}
      + \tilde{b}\, e^{\int_{r}^{r_*} \!\sqrt{V(r')}\,dr'} \right],
\end{equation}
and
\begin{equation}\label{r2}
    \phi_{II}(r)
    = \frac{1}{|V(r)|^{1/4}}
      \left[ \tilde{c}\, \cos\!\left( \int_{r_*}^{r} \!\sqrt{|V(r')|}\,dr' + \frac{\pi}{4} \right)
      + \tilde{d}\, \sin\!\left( \int_{r_*}^{r} \!\sqrt{|V(r')|}\,dr' + \frac{\pi}{4} \right) \right].
\end{equation}
Near the turning point \(r=r_*\), the potential can be approximated as
\begin{equation}
    V(r) \approx -|V'(r_*)|\, (r - r_*).
\end{equation}
The Schr\"odinger equation then becomes
\begin{equation}
    \phi''(r) + |V'(r_*)|\, (r - r_*)\, \phi(r) = 0,
\end{equation}
whose solution is sometimes called the \emph{patching} wave function and is denoted by \(\phi_p(r)\):
\begin{equation}\label{p0}
    \phi_{p}(r)
    = g_1\, \text{Ai}\!\left(-|V'(r_*)|^{1/3} (r - r_*)\right)
    + g_2\, \text{Bi}\!\left(|V'(r_*)|^{1/3} (r - r_*)\right).
\end{equation}
The asymptotic behaviors of the Airy functions are
\begin{align}
    \text{Ai}(-z) &\approx \frac{e^{-\frac{2}{3}(-z)^{3/2}}}{2\sqrt{\pi}(-z)^{1/4}}, \quad z \ll 0, \\
    \text{Bi}(-z) &\approx \frac{e^{\frac{2}{3}(-z)^{3/2}}}{\sqrt{\pi}(-z)^{1/4}}, \quad z \ll 0,
\end{align}
and
\begin{align}
    \text{Ai}(-z) &\approx \frac{1}{\sqrt{\pi}z^{1/4}} \sin\!\left(\frac{2}{3}z^{3/2} + \frac{\pi}{4}\right), \quad z \gg 0, \\
    \text{Bi}(-z) &\approx \frac{1}{\sqrt{\pi}z^{1/4}} \cos\!\left(\frac{2}{3}z^{3/2} + \frac{\pi}{4}\right), \quad z \gg 0.
\end{align}
Using these asymptotic expansions, for \(r\gg r_*\), Eq.~\eqref{p0} becomes
\begin{align} \label{Air1} \phi_{p}(r) \approx \frac{1}{\sqrt{\pi} |V'(r_*)|^{1/12} (r-r_*)^{1/4}} \Bigg[ g_1 & \sin\left( \frac{2}{3} |V'(r_*)|^{1/2} (r - r_*)^{3/2} + \frac{\pi}{4} \right) \nonumber \\ & + g_2 \cos\left( \frac{2}{3} |V'(r_*)|^{1/2} (r - r_*)^{3/2} + \frac{\pi}{4} \right) \Bigg]. \end{align}
Near the turning point, Eq.~\eqref{r2} can be approximated as
\begin{align} \label{R3} \phi_{II}(r) \approx \frac{1}{|V'(r_*)|^{1/4} (r-r_*)^{1/4}} \Bigg[ \tilde{d} & \sin\left( \frac{2}{3} |V'(r_*)|^{1/2} (r - r_*)^{3/2} + \frac{\pi}{4} \right) \nonumber \\ & + \tilde{c} \, \cos\left( \frac{2}{3} |V'(r_*)|^{1/2} (r - r_*)^{3/2} + \frac{\pi}{4} \right) \Bigg]. \end{align}
Matching Eqs.~\eqref{Air1} and~\eqref{R3}, we obtain
\begin{equation}\label{con1}
    g_1 = \sqrt{\pi}\, |V'(r_*)|^{-1/6}\, \tilde{d},
    \qquad
    g_2 = \sqrt{\pi}\, |V'(r_*)|^{-1/6}\, \tilde{c}.
\end{equation}
For \(r\ll r_*\), Eq.~\eqref{p0} can be approximated as
\begin{align}\label{Air2}
    \phi_{p}(r)
    &\approx \frac{1}{\sqrt{\pi}\,|V'(r_*)|^{1/12}(r_* - r)^{1/4}}
    \Big[ \tfrac{g_1}{2}\, e^{-\frac{2}{3}|V'(r_*)|^{1/2}(r_* - r)^{3/2}}
     + g_2\, e^{\frac{2}{3}|V'(r_*)|^{1/2}(r_* - r)^{3/2}} \Big].
\end{align}
Near the turning point, Eq.~\eqref{r1} can be approximated as
\begin{align}\label{RE2}
    \phi_{I}(r)
    &\approx \frac{1}{|V'(r_*)|^{1/4}(r_* - r)^{1/4}}
    \Big[ \tilde{a}\, e^{-\frac{2}{3}|V'(r_*)|^{1/2}(r_* - r)^{3/2}}
     + \tilde{b}\, e^{\frac{2}{3}|V'(r_*)|^{1/2}(r_* - r)^{3/2}} \Big].
\end{align}
Matching Eqs.~\eqref{Air2} and~\eqref{RE2}, we find
\begin{equation}\label{con2}
    g_1 = 2\sqrt{\pi}\, |V'(r_*)|^{-1/6}\, \tilde{a},
    \qquad
    g_2 = \sqrt{\pi}\, |V'(r_*)|^{-1/6}\, \tilde{b}.
\end{equation}
Finally, comparing Eqs.~\eqref{con1} and~\eqref{con2}, we arrive at the connection formula
\begin{equation}\label{connection_1}
    \tilde{c} = \tilde{b}, \qquad \tilde{d} = 2\tilde{a}.
\end{equation}

\subsection{WKB with two turning points}\label{twoturning}

\begin{figure*}
    \centering
    \includegraphics[width=.52\textwidth]{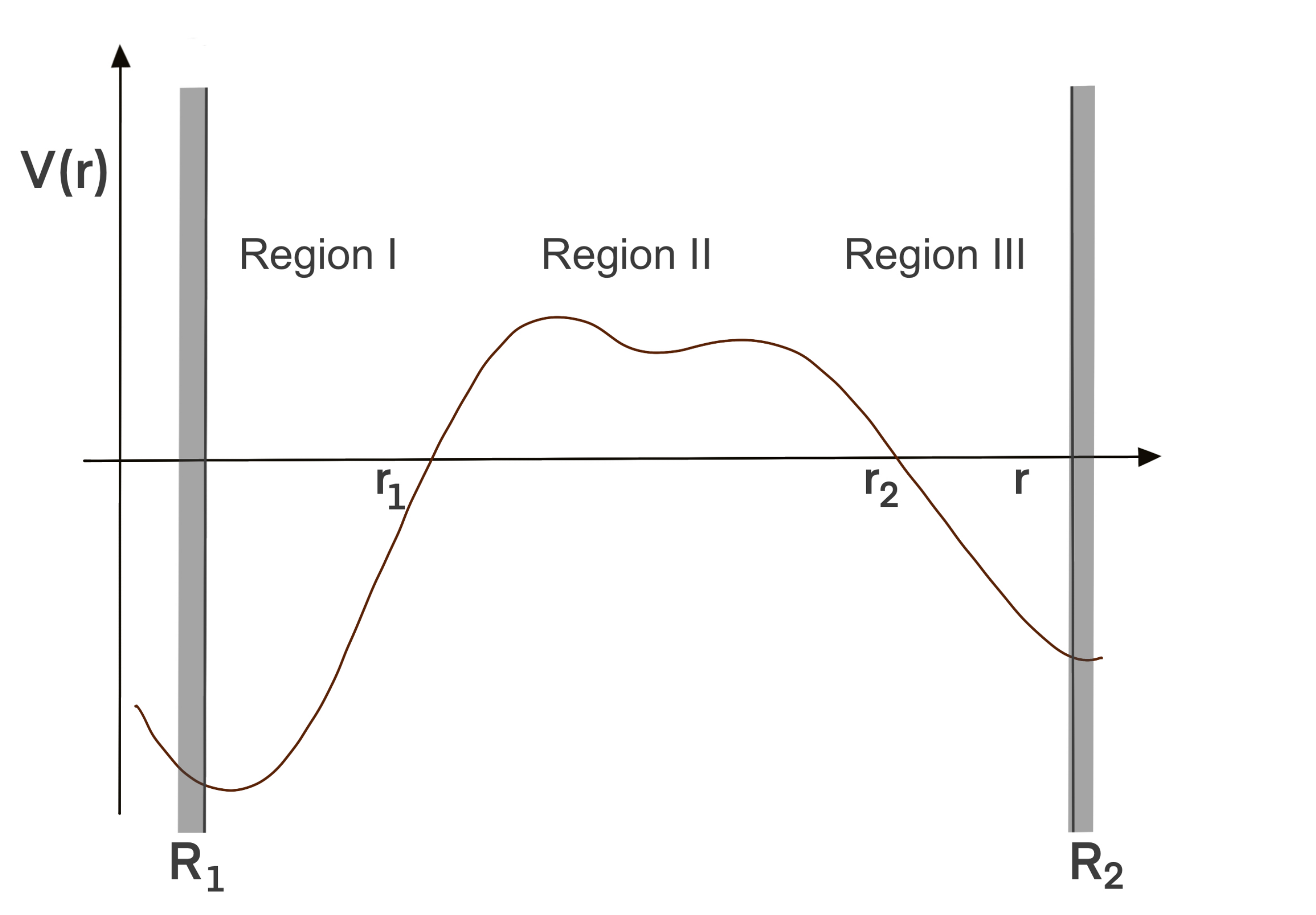}
   \caption{Schematic of the potential \(V(r)\) with two turning points at \(r=r_1\) and \(r=r_2\). Two infinite potential barriers (brick walls) are also placed at \(r=R_1\) and \(r=R_2\).}
    \label{WKB_pot2}
\end{figure*}
Let us now consider the Schr\"odinger equation with a potential \(V(r)\), as shown in Fig.~\ref{WKB_pot2}. As depicted in the figure, we place two infinite potential barriers at \(r=R_1\) and \(r=R_2\). Between them, the potential has two turning points at \(r=r_1\) and \(r=r_2\), with \(r_2>r_1\). The system is therefore divided into three regions, which we label Regions I, II, and III, corresponding to the allowed, forbidden, and allowed regions, respectively. We define
\begin{equation}
    p(r)=\sqrt{E-V(r)}, \qquad \kappa(r)= \sqrt{V(r)-E},
\end{equation}
both of which are real-valued. The solution in Region I is
\begin{equation} \label{reg1}
    \psi_{I}(r) = \frac{1}{\sqrt{p(r)}} \left[ a \, \cos \left( \int_{r}^{r_1} p(r') \, dr' + \frac{\pi}{4} \right) + b \, \sin \left( \int_{r}^{r_1} p(r') \, dr' + \frac{\pi}{4} \right) \right].
\end{equation}
Since Region II has turning points at both ends, the solution can be written in two equivalent forms. Anchored at \(r_1\), it reads
\begin{equation} \label{reg2}
    \psi_{II}(r) = \frac{1}{\sqrt{\kappa(r)}} \left[ A e^{-\int_{r_1}^{r} \kappa(r') dr'} + B e^{\int_{r_1}^{r} \kappa(r') dr'} \right],
\end{equation}
while, anchored at \(r_2\), it is
\begin{equation} \label{reg22}
    \psi_{II}(r) = \frac{1}{\sqrt{\kappa(r)}} \left[ \Tilde{A} e^{-\int_{r}^{r_2} \kappa(r') dr'} + \Tilde{B} e^{\int_{r}^{r_2} \kappa(r') dr'} \right].
\end{equation}
These two representations imply
\begin{equation}
    \Tilde{A}= B \, e^{S}, \quad \Tilde{B}= A \, e^{-S}, \quad \text{where } S=\int_{r_1}^{r_2} \kappa(r) \, dr .
\end{equation}
The solution in Region III is
\begin{equation} \label{reg3}
    \psi_{III}(r) =\frac{1}{\sqrt{p(r)}} \left[ c \, \cos \left( \int_{r_2}^{r} p(r') \, dr' + \frac{\pi}{4} \right) + d \, \sin \left( \int_{r_2}^{r} p(r') \, dr' + \frac{\pi}{4} \right) \right].
\end{equation}
To determine the relations among the constants appearing in these solutions, we now match them near the turning points using Airy-function approximations.

\vspace{0.2cm}
\noindent
\textbf{Matching near the first turning point.}
Near the first turning point, \(r=r_1\), the potential can be approximated as
\begin{equation}
    V(r)\approx E+|V'(r_1)| (r-r_1).
\end{equation}
The Schr\"odinger equation then becomes
\begin{equation}
    \psi''(r)-|V'(r_1)| (r-r_1) \psi(r)=0.
\end{equation}
The patching wave function near \(r=r_1\) is then
\begin{equation}\label{p1}
    \psi_{p_1}(r) = f_1 \, \text{Ai} \left(  V'(r_1)^{\frac{1}{3}} (r - r_1) \right) + f_2 \, \text{Bi} \left( V'(r_1)^{\frac{1}{3}} (r - r_1) \right).
\end{equation}
The asymptotic behaviors of the Airy functions are
\begin{align}
    \text{Ai}(z) & \sim \frac{e^{-\frac{2}{3}z^{3/2}}}{2 \sqrt{\pi} z^{1/4}}, \quad z \gg 0, \\
    \text{Bi}(z) &\sim \frac{e^{\frac{2}{3} z^{3/2}}}{\sqrt{\pi} z^{1/4}}, \quad z \gg 0,
\end{align}
and
\begin{align}
    \text{Ai}(z) &\sim \frac{1}{\sqrt{\pi} (-z)^{1/4}} \sin\left( \frac{2}{3} (-z)^{3/2} + \frac{\pi}{4} \right), \quad z \ll 0, \\
    \text{Bi}(z) &\sim \frac{1}{\sqrt{\pi} (-z)^{1/4}} \cos\left( \frac{2}{3} (-z)^{3/2} + \frac{\pi}{4} \right), \quad z \ll 0.
\end{align}
Thus, in the region \(r\ll r_1\), we have
\begin{align} \label{Airy1}
    \psi_{p_1}(r) \approx \frac{1}{\sqrt{\pi} V'(r_1)^{1/12} (r_1 - r)^{1/4}} 
    \Bigg[  f_1 & \sin\left( \frac{2}{3} V'(r_1)^{1/2} (r_1 - r)^{3/2} + \frac{\pi}{4} \right) \nonumber \\
    & + f_2 \cos\left( \frac{2}{3} V'(r_1)^{1/2} (r_1 - r)^{3/2} + \frac{\pi}{4} \right) \Bigg].
\end{align}
Near \(r=r_1\), Eq.~\eqref{reg1} can be approximated as
\begin{align} \label{reg11}
    \psi_{I}(r) \approx \frac{1}{V'(r_1)^{1/4}(r_1-r)^{1/4}} 
    \Bigg[  b\, & \sin\left( \frac{2}{3} V'(r_1)^{1/2} (r_1 - r)^{3/2} + \frac{\pi}{4} \right) \nonumber \\
    & + a \, \cos\left( \frac{2}{3} V'(r_1)^{1/2} (r_1 - r)^{3/2} + \frac{\pi}{4} \right) \Bigg].
\end{align}
Equating Eqs.~\eqref{Airy1} and~\eqref{reg11}, we obtain
\begin{equation}\label{connect1}
    f_1= \sqrt{\pi} V'(r_1)^{-\frac{1}{6}} \, b, \quad \quad f_2=\sqrt{\pi} V'(r_1)^{-\frac{1}{6}} \, a .
\end{equation}
Similarly, when \(r \gg r_1\), Eq.~\eqref{p1} can be approximated as
\begin{align} \label{Airy2}
    \psi_{p_1}(r) \approx \frac{1}{\sqrt{\pi} V'(r_1)^{1/12} (r - r_1)^{1/4}} 
    \Bigg[  \frac{f_1}{2} e^{-\frac{2}{3}(r-r_1)^{\frac{3}{2}}} + f_2 e^{\frac{2}{3}(r-r_1)^{\frac{3}{2}}} \Bigg].
\end{align}
Near \(r=r_1\), Eq.~\eqref{reg2} can be approximated as
\begin{equation}\label{REG2}
     \psi_{II}(r) \approx \frac{1}{V'(r_1)^{1/4}(r-r_1)^{1/4}} \left[ A \, e^{-\frac{2}{3}(r-r_1)^{\frac{3}{2}}} + B \, e^{\frac{2}{3}(r-r_1)^{\frac{3}{2}}} \right].
\end{equation}
Matching Eqs.~\eqref{Airy2} and~\eqref{REG2}, we obtain
\begin{equation}\label{connect2}
    f_1= 2\sqrt{\pi} V'(r_1)^{-\frac{1}{6}} \, A, \quad \quad f_2=\sqrt{\pi} V'(r_1)^{-\frac{1}{6}} \, B .
\end{equation}
Equating Eqs.~\eqref{connect1} and~\eqref{connect2}, we finally obtain
\begin{equation}\label{match1}
    a=B, \qquad b=2A .
\end{equation}

\noindent
\textbf{Matching near the second turning point.}
Near the second turning point, \(r=r_2\), the potential can be approximated as
\begin{equation}
    V(r)\approx E-|V'(r_2)| (r-r_2).
\end{equation}
The Schr\"odinger equation then becomes
\begin{equation}
    \psi''(r)+|V'(r_2)| (r-r_2) \psi(r)=0.
\end{equation}
Following the same procedure as for the first turning point, we obtain the relations
\begin{equation}\label{relation11}
    c= \tilde{B}=A e^{-S}=\frac{b}{2} e^{-S}, \qquad d=2 \tilde{A}=2 B e^S=2a e^S.
\end{equation}
%
%
\textbf{Finding the quantization condition.}
%
%
So far, we have not imposed the boundary conditions that the wave function vanishes at \(r=R_1\) and \(r=R_2\). We now impose them. The conditions \(\psi_I(R_1)=0\) and \(\psi_{III}(R_2)=0\) imply, respectively,
\begin{align}\label{EQ1}
    \left[ a \, \cos \left( \int_{R_1}^{r_1} p(r') \, dr' + \frac{\pi}{4} \right) + b \, \sin \left( \int_{R_1}^{r_1} p(r') \, dr' + \frac{\pi}{4} \right) \right]=0,
\end{align}
and
\begin{align}
     \left[ c \, \cos \left( \int_{r_2}^{R_2} p(r') \, dr' + \frac{\pi}{4} \right) + d \, \sin \left( \int_{r_2}^{R_2} p(r') \, dr' + \frac{\pi}{4} \right) \right]=0.
\end{align}
Here \(c\) and \(d\) can be written in terms of \(a\) and \(b\) using \eqref{relation11}, which gives
\begin{align}\label{EQ2}
    \left[ \frac{b \, e^{-S}}{2} \, \cos \left( \int_{r_2}^{R_2} p(r') \, dr' + \frac{\pi}{4} \right) + 2 a \, e^S \, \sin \left( \int_{r_2}^{R_2} p(r') \, dr' + \frac{\pi}{4} \right) \right]=0.
\end{align}
From these two equations, \eqref{EQ1} and \eqref{EQ2}, one can derive a single quantization condition,
\begin{equation}\label{quant}
    \tan \left( \int_{R_1}^{r_1} p(r') \, dr' + \frac{\pi}{4} \right) 
    \tan \left( \int_{r_2}^{R_2} p(r') \, dr' + \frac{\pi}{4} \right)
    = \frac{1}{4} e^{-2 S}.
\end{equation}
In a regime where \(S \gg 1\), the right-hand side of Eq.~\eqref{quant} can be neglected\footnote{If \(V(r)\) is sufficiently smooth, this corresponds to a regime in which the classically forbidden region between the two turning points is wide and/or the barrier is sufficiently high.} and the condition becomes
\begin{equation}
    \tan \left( \int_{R_1}^{r_1} p(r') \, dr'+ \frac{\pi}{4} \right) \tan \left( \int_{r_2}^{R_2} p(r') \, dr' + \frac{\pi}{4} \right) =0,
\end{equation}
which is equivalent to
\begin{equation}\label{quantization21}
    \tan \left( \int_{R_1}^{r_1} p(r') \, dr'+ \frac{\pi}{4} \right)=0 \implies \int_{R_1}^{r_1} \sqrt{E- V(r)} \, dx=\left(n-\frac{1}{4} \right)\pi,
\end{equation}
or
\begin{equation}\label{quantization22}
    \tan \left( \int_{r_2}^{R_2} p(r') \, dr' + \frac{\pi}{4} \right) =0 \implies \int_{r_2}^{R_2}\sqrt{E-V(r)} \, dr =\left(n-\frac{1}{4} \right)\pi,
\end{equation}
or both.

We can also consider more general boundary conditions, \(\Psi_I(R_1)=C_1\) and \(\Psi_{III}(R_2)=C_2\). If we make the following redefinitions of the constants:
\begin{equation}\label{quantization23}
    \begin{aligned}
        a = A\cos\lambda_1 ~,\qquad b = B\cos\lambda_1~,\\
        c = C\cos\lambda_2 ~,\qquad d = D\cos\lambda_2~,
    \end{aligned}
\end{equation}
together with
\begin{equation}\label{eqC1C2}
    \begin{aligned}
        C_1=-B\cos{\left(\int_{r_1}^{R_1} p(r') \, dr'+\frac{\pi}{4}\right)}\sin{\lambda_1}+A\sin{\left(\int_{r_1}^{R_1} p(r') \, dr'+\frac{\pi}{4}\right)}\cos{\lambda_1},\\
        C_2=-D\cos{\left(\int_{r_2}^{R_2} p(r') \, dr'+\frac{\pi}{4}\right)}\sin{\lambda_2}+C\sin{\left(\int_{r_2}^{R_2} p(r') \, dr'+\frac{\pi}{4}\right)}\cos{\lambda_2},
    \end{aligned}
\end{equation}
then, using the angle-sum formulas for sine and cosine, the boundary conditions can be rewritten as
\begin{equation}\label{EQ5}
     A \, \cos \left( \int_{R_1}^{r_1} p(r') \, dr' + \frac{\pi}{4} +\lambda_1\right) + B \, \sin \left( \int_{R_1}^{r_1} p(r') \, dr' + \frac{\pi}{4} + \lambda_1\right) =0,
\end{equation}
and
\begin{equation}\label{EQ6}
    C \, \cos \left( \int_{R_2}^{r_2} p(r') \, dr' + \frac{\pi}{4} +\lambda_2\right) + D \, \sin \left( \int_{R_2}^{r_2} p(r') \, dr' + \frac{\pi}{4} + \lambda_2\right)=0,
\end{equation}
Proceeding exactly as in the previous case, we obtain
\begin{equation}\label{quantization25}
    \tan \left( \int_{R_1}^{r_1} p(r') \, dr'+ \frac{\pi}{4} + \lambda_1 \right)=0 \implies \int_{R_1}^{r_1} \sqrt{E- V(r)} \, dx=\left(n-\frac{1}{4} \right)\pi - \lambda_1,
\end{equation}
or
\begin{equation}\label{quantization26}
    \tan \left( \int_{r_2}^{R_2} p(r') \, dr' + \frac{\pi}{4} + \lambda_2\right) =0 \implies \int_{r_2}^{R_2}\sqrt{E-V(r)} \, dr =\left(n-\frac{1}{4} \right)\pi - \lambda_2.
\end{equation}
%

\section{Normal modes of pure de Sitter using WKB}\label{WKBpuredS}

In this appendix, we apply the WKB approximation to the case of pure de~Sitter space. The main motivation is that, since the exact solution is known in this case, we can compare it with the WKB result and thereby assess the strengths and limitations of the approximation. This comparison will be useful in the Schwarzschild--de~Sitter case.

\begin{figure}[t!]
\begin{subfigure}{0.5\textwidth}
    \centering
    \includegraphics[width=\textwidth]{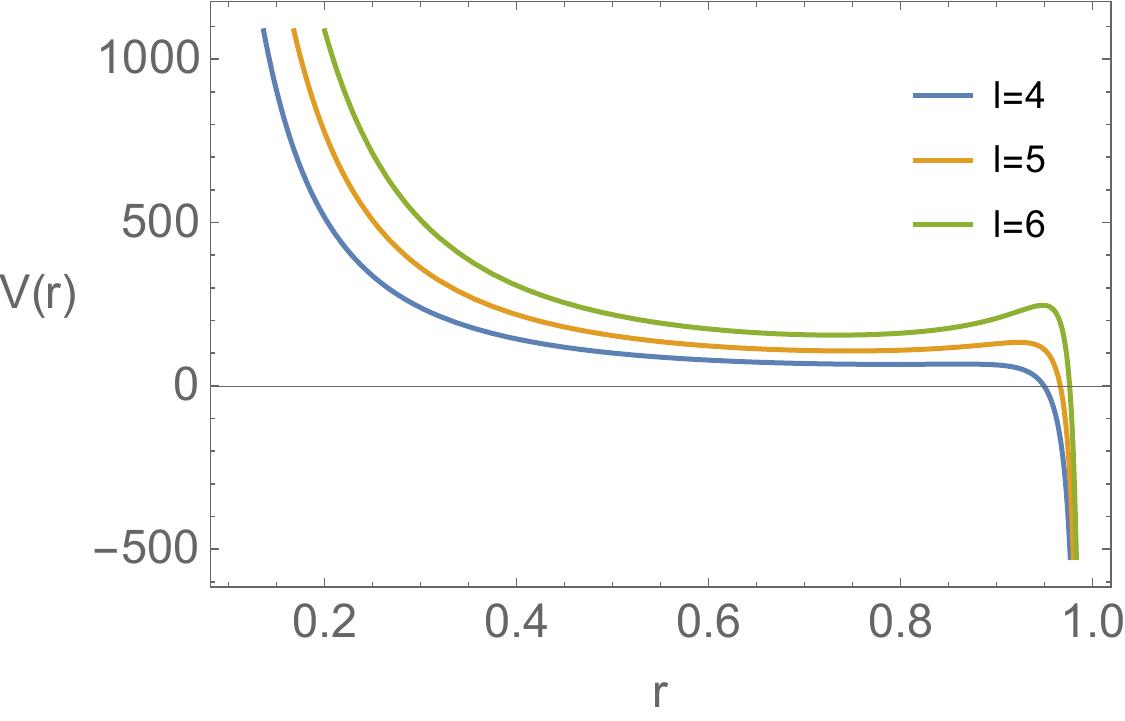}
    \end{subfigure}
    \hfill
    \begin{subfigure}{0.5\textwidth}
    \includegraphics[width=\textwidth]{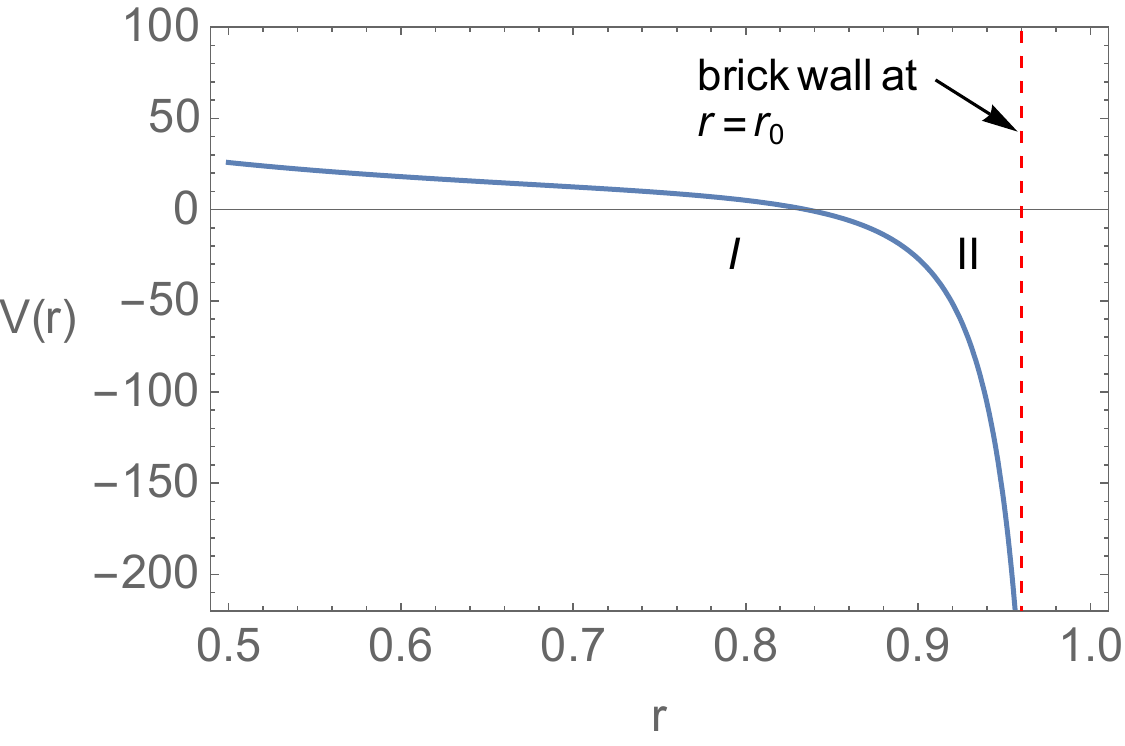}
    \end{subfigure}
    \caption{Effective potential in pure dS for \(d=3\) and \(\omega=1\). Left: \(V(r)\) as a function of the angular momentum \(l\). Right: \(V(r)\) for fixed \(l=2\), with the brick wall position indicated by the red vertical line. Since we consider a WKB problem with \(E=0\), the zero of \(V(r)\) defines the turning point: region~I (\(V>0\)) is classically forbidden, while region~II (\(V<0\)) is classically allowed.}
    \label{pot_empty_ds}
\end{figure}

\begin{figure}[t!]
    \centering
    \begin{subfigure}[t]{0.48\textwidth}
        \centering
        \includegraphics[width=\textwidth]{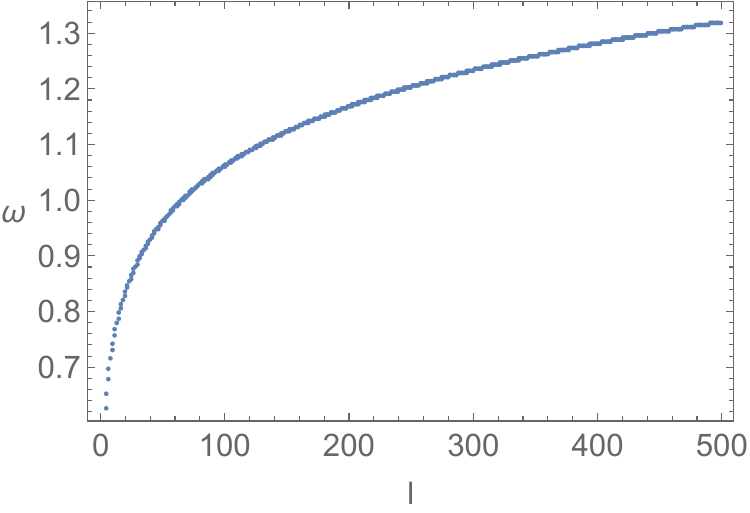}
        \caption{$\sigma_0^2=0$}
        \label{empty_ds_withoutnoise}
    \end{subfigure}
    \hfill
    \begin{subfigure}[t]{0.48\textwidth}
        \centering
        \includegraphics[width=\textwidth]{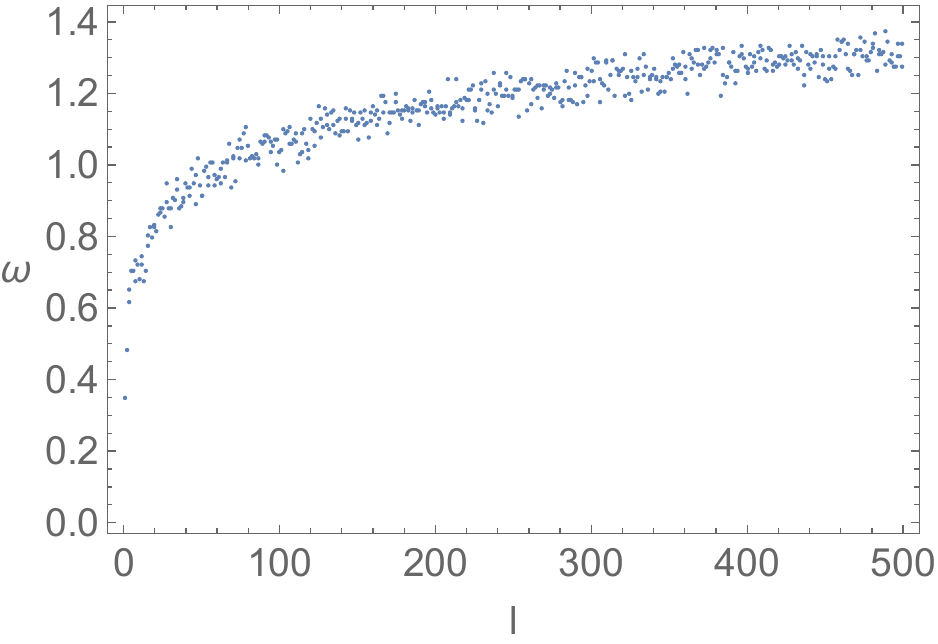}
        \caption{$\sigma_0^2=0.1$}
        \label{empty_ds_noise}
    \end{subfigure}
    \caption{WKB modes as functions of \(l\) for fixed \(n=6\), obtained by solving Eqs.~\eqref{quant01}-\eqref{quantnoise}. The modes are degenerate in the magnetic quantum number \(m\), which does not appear in the radial Klein--Gordon equation. The parameters chosen for the plots are \(r_C=1\) and \(\epsilon_0=10^{-15}\).}
    \label{empty_ds_wkb}
\end{figure}

Let us first rewrite the radial equation~\eqref{radial1} as
\begin{equation}
    a(r) \Phi''(r) + b(r) \Phi'(r) + c(r) \Phi(r) = 0,
\end{equation}
where
\begin{align}
    a(r) &= (r^2 - 1), \quad 
    b(r) = \left(\frac{1-d}{r} + r(1+d)\right), \quad 
    c(r) = \left(\frac{l(l+d-2)}{r^2} - \frac{\omega^2}{1-r^2}\right).
\end{align}
This can be recast as a Schr\"odinger equation, as explained in Appendix~\ref{WKBapp}, such that

\begin{equation}\label{scho0}
    \phi''(r) - V(r)\,\phi(r) = 0,
\end{equation}
where the effective potential \(V(r)\) is given by
\begin{align}\label{pot0}
    V(r) &= \frac{1}{4 a(r)^2} \big[ b(r)^2 - 2 b(r)\,a'(r) + 2 a(r)\big(b'(r) - 2c(r)\big) \big] \nonumber \\
    &= \frac{d^2 (r^2 - 1)^2 - 4d(l-1)(r^2 - 1) - 4l^2 (r^2 - 1) + 8l(r^2 - 1) - r^2(r^2 + 4\omega^2 + 6) + 3}{4r^2 (r^2 - 1)^2}.
\end{align}
Near \(r \to 0\),
\begin{equation}\label{r0_behavior}
    \lim_{r \to 0} V(r) = \frac{(d + 2l - 3)(d + 2l - 1)}{4r^2} + O(1),
\end{equation}
so the potential diverges to positive infinity. Near the cosmological horizon, \(r \to 1\),
\begin{equation}
    \lim_{r \to 1} V(r) = -\frac{\omega^2 + 1}{4(r - 1)^2}
    + \frac{-2d(l - 1) - 2(l - 2)l + \omega^2 - 1}{4(r - 1)} + \ldots,
\end{equation}
i.e. \(V(r)\) diverges to negative infinity there. Therefore, in between, \(V(r)\) develops a zero that serves as the turning point of the corresponding WKB problem (recall that Eq.~\eqref{scho0} is written with \(E = 0\)). A typical structure of this potential \(V(r)\) as a function of \(r\) is shown in Fig.~\ref{pot_empty_ds}.\footnote{
It is worth mentioning that, above a critical value of \(l\), \(V(r)\) develops a local maximum, which is reminiscent of the photon sphere in the geodesic picture. As a result, the potential admits two sets of bound states: one localized near the cosmological horizon and the other localized near the boundary. However, since our WKB analysis focuses on the \(E=0\) case, the method is suitable only for determining the modes localized near the horizon.
}
The figure shows two regions: region~I (forbidden) and region~II (allowed). The WKB solutions in these regions are
\begin{equation}\label{reg10}
    \phi_I(r) = \frac{1}{V(r)^{1/4}} 
    \left[ A_1 e^{-\int_r^{r_*} \!\sqrt{V(r')}\,dr'} 
    + B_1 e^{\int_r^{r_*} \!\sqrt{V(r')}\,dr'} \right],
\end{equation}
and
\begin{equation}\label{reg20}
    \phi_{II}(r) = \frac{1}{|V(r)|^{1/4}} 
    \left[ A_2 \cos\!\left( \int_{r_*}^{r} \!\sqrt{|V(r')|}\,dr' + \frac{\pi}{4} \right)
    + B_2 \sin\!\left( \int_{r_*}^{r} \!\sqrt{|V(r')|}\,dr' + \frac{\pi}{4} \right) \right].
\end{equation}
The WKB matching condition, Eq.~\eqref{connection_1}, gives
\begin{equation}
    A_2 = B_1, \qquad B_2 = 2A_1.
\end{equation}
Up to this point, we have not yet imposed the regularity condition near \(r \to 0\), where the potential diverges. The solution of Eq.~\eqref{scho0} with \(V(r)\) as in Eq.~\eqref{r0_behavior} (for \(d = 3\)) is
\begin{equation}
    \phi(r) = c_1\,r^{\frac{1}{2}(1 + a)} + c_2\,r^{\frac{1}{2}(1 - a)}, 
    \quad \text{where} \quad a = \sqrt{(d + 2l - 3)(d + 2l - 1)}.
\end{equation}
Normalizability requires \(c_2 = 0\), and hence the normalizable solution is
\begin{align}
    \phi(r) &= c_1 \sqrt{r}\,r^{\frac{a}{2}}
    = c_1 \sqrt{r}\, e^{\frac{a}{2} \log r}
    = c_1 \sqrt{r}\, e^{-\frac{a}{2} \!\left( \int_r^{\Lambda} \frac{dr'}{r'} - \log \Lambda \right)}
    \sim \frac{\tilde{c}_1}{|V(r')|^{1/4}} e^{-\int_r^{r_*} \sqrt{|V(r)|}\,dr'},
\end{align}
where all \(r\)-independent constants have been absorbed into \(\tilde{c}_1\). Comparing this with Eq.~\eqref{reg10}, we find
\begin{equation}
    B_1 = 0 \quad \implies \quad A_2 = 0.
\end{equation}
Imposing the Dirichlet boundary condition at \(r = r_0\) ($\Phi=0$) then gives
\begin{align}\label{quant0}
    \sin\!\left( \int_{r_*}^{r_0} \!\sqrt{|V(r)|}\,dr + \frac{\pi}{4} \right) &= 0 \nonumber \\[3pt]
    \implies \int_{r_*}^{r_0} \!\sqrt{|V(r)|}\,dr &= \left(n - \frac{1}{4}\right)\pi, 
    \quad n \in \mathbb{Z}.
\end{align}
This is the required quantization condition determining the allowed values of \(\omega\). It is not possible to obtain a closed-form expression for the integral in Eq.~\eqref{quant0} for the potential in Eq.~\eqref{pot0}. Following~\cite{Das:2024fwg}, we assume that the turning point \(r_*\) lies close to the horizon and expand \(V(r)\) as
\begin{equation}\label{taylor}
    \lim_{r \to r_C - \epsilon} V(r) \approx A_0 + \frac{A_1}{\epsilon} - \frac{A_2}{\epsilon^2},
\end{equation}
with
\begin{align}
    A_0 &= \frac{1}{16}\!\left[4\!\left(d^2 + 5d(l - 1) + 5l^2\right) - 40l - 3\omega^2 + 13\right], \nonumber \\
    A_1 &= \frac{1}{4}\!\left[2d(l - 1) + 2(l - 2)l - \omega^2 + 1\right], \quad 
    A_2 = \frac{1}{4}(\omega^2 + 1).
\end{align}
With this expansion, the integral on the left-hand side of~\eqref{quant0} can be evaluated to yield
\begin{equation}\label{quant01}
    \int_{r_*}^{r_0 = r_C - \epsilon_0} \!\sqrt{|V(r)|}\,dr =
    -\left(\frac{1}{2} \sqrt{A_2} \log \left(\frac{(a+1)^2+b^2}{(1-a)^2+b^2}\right)+\frac{A_1 \tan ^{-1}\left(\frac{d}{c}\right)}{2 \sqrt{A_0}}+\sqrt{A_2-\epsilon_0 (A_0 \epsilon_0+A_1)}\right),
\end{equation}
where
\begin{align*}
    a &=-\sqrt{\frac{A_2-\epsilon_0 (A_0 \epsilon_0+A_1)}{A_2}}, \quad b=\epsilon_0 \sqrt{\frac{A_0}{A_2}}, \\
    c &=\frac{2 A_0 \epsilon_0}{A_1}+1, \qquad d=\frac{2 \sqrt{A_0 (A_2-\epsilon_0 (A_0 \epsilon_0+A_1))}}{A_1}.
\end{align*}
Equation~\eqref{quant0} can be solved numerically using, for example, the \texttt{FindRoot} command in \textit{Mathematica}. In Fig.~\ref{empty_ds_withoutnoise}, we plot the normal modes as functions of the angular momentum \(l\).\footnote{This method is not very useful for computing modes along the \(n\) direction for small values of \(l\), since the Taylor expansion~\eqref{taylor} is not reliable in the regime of small \(l\) and large \(\omega\). Consequently, the integral appearing in Eq.~\eqref{quant0} must be evaluated numerically. Details are provided in Appendix~\ref{appenC}.} The resulting spectrum exhibits a very slow, logarithmic growth, a feature closely related to the quasi-degeneracy observed in earlier studies of AdS normal modes.

We now consider the situation in which, instead of imposing the Dirichlet boundary condition \(\Phi=0\), we allow for a more general boundary condition \(\Phi=\Phi_{0l}\), where \(\Phi_{0l}\) is a constant that depends only on \(l\). For an arbitrary choice of \(\Phi_{0l}\), the reflection coefficient is not equal to one, which leads to complex modes. To avoid this pathology, Refs.~\cite{Das:2024fwg,Jeong:2024jjn} choose \(\Phi_{0l}\) in a special way. Following the same logic, we choose here $\Phi_{0l}=\frac{B_2}{|V(r_0)|^{1/4}}\,\sin\lambda_l$, so that the quantization condition becomes\footnote{In general,
\[
\sin x = \sin y
\quad \Longleftrightarrow \quad
\begin{cases}
x = y + 2n\pi, \\[4pt]
x = \pi - y + 2n\pi,
\end{cases}
\qquad n \in \mathbb{Z}.
\]
Here we choose the first branch.}
\begin{align}\label{quantnoise}
\sin\!\left( \int_{r_*}^{r_0} \!\sqrt{|V(r)|}\,dr + \frac{\pi}{4} \right)
&= \sin\lambda_l , \nonumber \\[3pt]
\implies \int_{r_*}^{r_0} \!\sqrt{|V(r)|}\,dr
&= \lambda_l + \left(2n - \frac{1}{4}\right)\pi ,
\quad n \in \mathbb{Z}.
\end{align}
In Fig.~\ref{empty_ds_noise}, we present the normal modes as functions of \(l\), obtained by solving Eq.~\eqref{quantnoise} with \(\lambda_l\) drawn from a Gaussian distribution with zero mean and variance \(\sigma_0^2=0.1\).

\section{WKB without using the Taylor approximation to $V(r)$}\label{appenC}

In this appendix, we assess the validity of the near-horizon Taylor expansion of the WKB potential. In particular, we show that for sufficiently low values of \(l\) and large $n$, the Taylor expansion ceases to approximate the exact potential reliably, making it necessary to use exact numerical integration to determine the corresponding normal modes.

\begin{figure}[t!]
    \centering
    \includegraphics[width=0.5\linewidth]{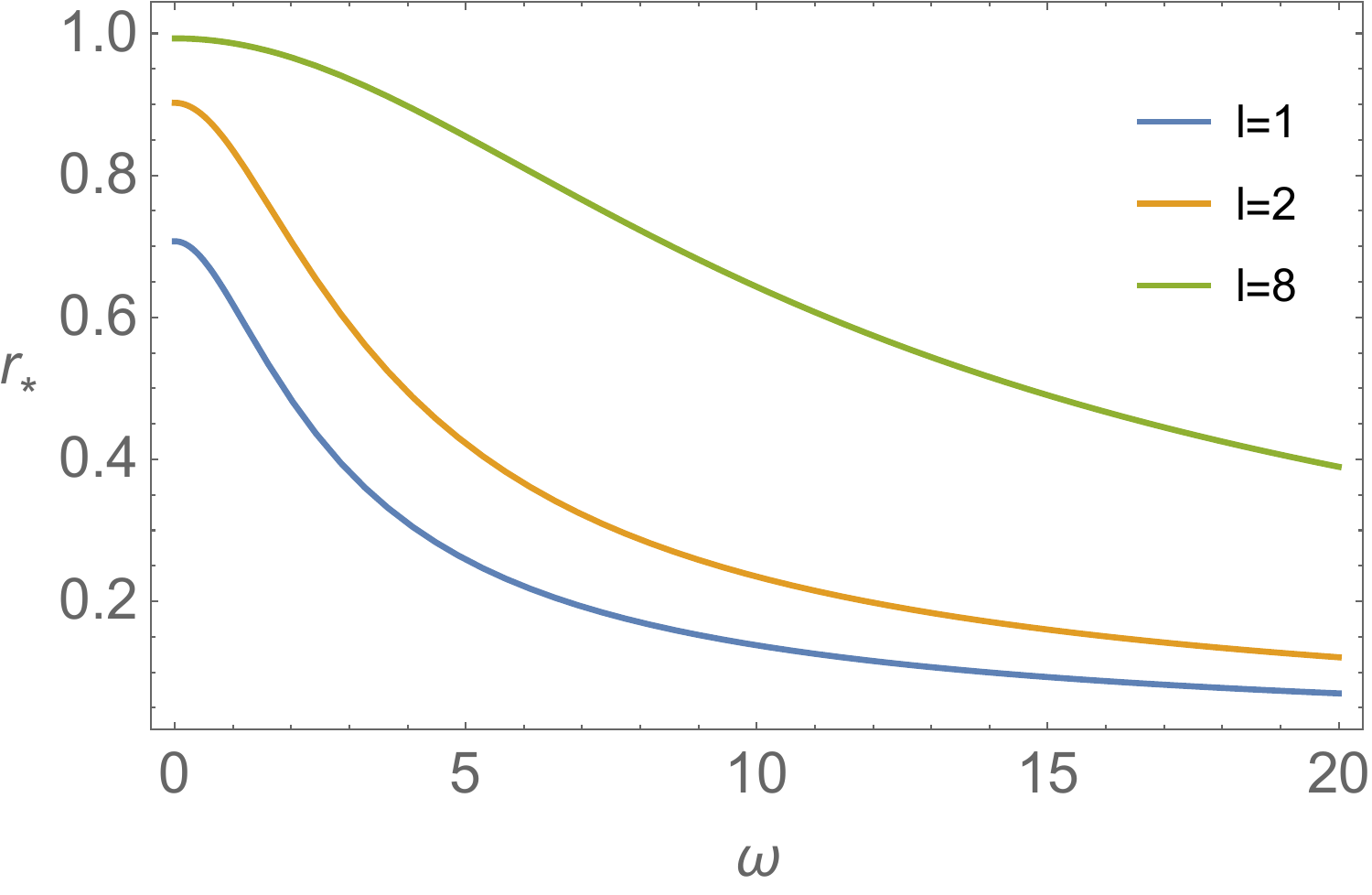}
    \caption{Turning point \(r_*\) as a function of \(\omega\) for different values of \(l\). The figure shows that, for moderately large \(l\) and sufficiently small \(\omega\), the turning point remains close to the horizon.}
    \label{turningpoint}
\end{figure}

\begin{figure}[t!]
    \centering\hspace{-0.7cm}
    \includegraphics[width=0.55\linewidth]{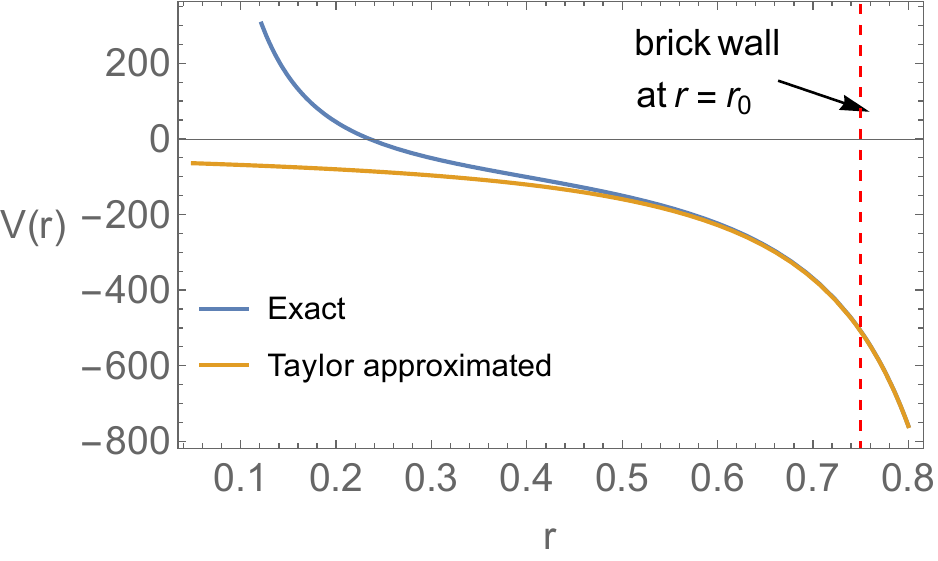}
    \caption{Comparison between the full potential~\eqref{full_pot} and its Taylor approximation~\eqref{approx_pot} for \(l=2\) and \(\omega=10\). Although the exact potential exhibits a turning point, the Taylor-approximated potential fails to capture this feature.}
    \label{taylor_failure}
\end{figure}

\begin{figure}[t!]
\begin{subfigure}{0.5\textwidth}
    \centering
    \includegraphics[width=\textwidth]{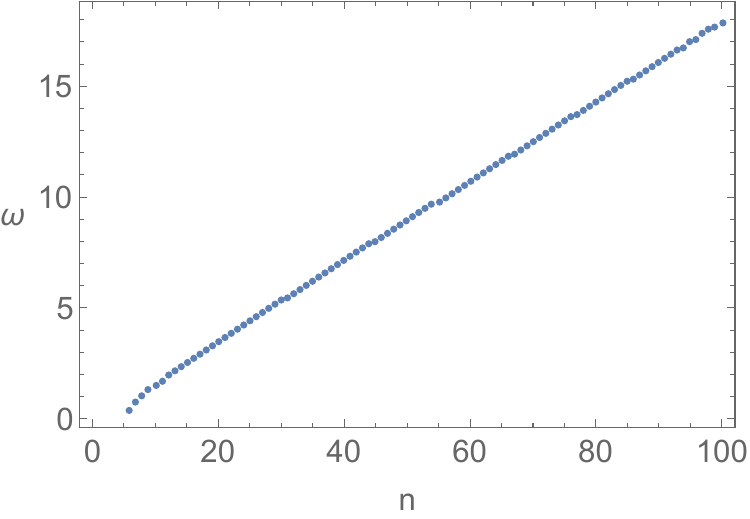}
\end{subfigure}
\hfill
\begin{subfigure}{0.5\textwidth}
    \centering
    \includegraphics[width=\textwidth]{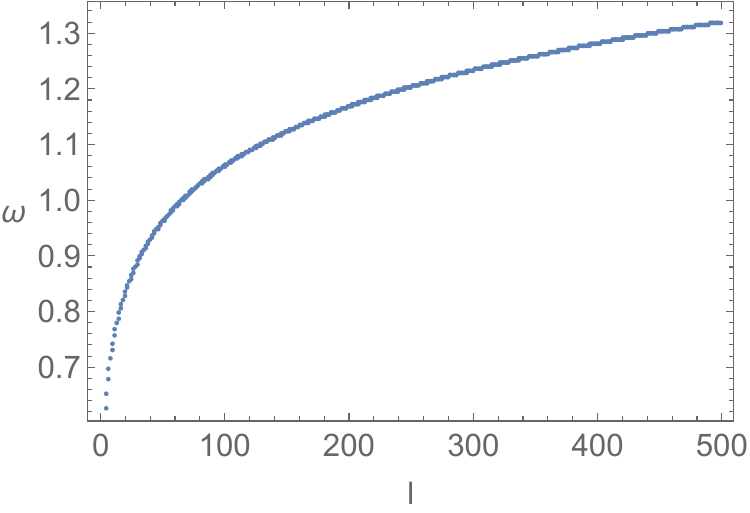}
\end{subfigure}
\caption{Exact WKB modes for brick wall position \(\epsilon_0=10^{-15}\), computed using the full potential without employing the Taylor approximation. In the left panel, \(l\) is fixed to \(1\), while in the right panel \(n=6\). The plots show an approximately linear growth along the \(n\) direction and a slow logarithmic growth along the \(l\) direction, as expected.}
\label{exact_modes_dS}
\end{figure}

For concreteness, we consider the case \(d=3\). The effective WKB potential is given by
\begin{equation}\label{full_pot}
    V(r)=\frac{-l^2 \left(r^2-1\right)-l r^2+l+r^2 \left(2 r^2-\omega ^2-3\right)}{r^2 \left(r^2-1\right)^2}\, .
\end{equation}
Near the horizon, its Taylor expansion takes the form
\begin{equation}\label{approx_pot}
    \lim_{r \rightarrow 1}V(r)
    \approx -\frac{\omega^{2}+1}{4(r-1)^{2}}
    +\frac{2l^{2}+2l-\omega^{2}-5}{4(r-1)}
    +\frac{20l^{2}+20l-3\omega^{2}-11}{16}\, .
\end{equation}
The full potential \(V(r)\) has a turning point located at
\begin{equation}
    r_{*}
    =\frac{1}{2}\sqrt{
    -\sqrt{\left(-l^2-l-\omega ^2-3\right)^2-8 \left(l^2+l\right)}
    +l^2+l+\omega ^2+3}\, .
\end{equation}
Figure~\ref{turningpoint} shows the behavior of \(r_*\) as a function of \(\omega\) for fixed \(l\). It is evident that, for fixed \(l\), as \(\omega\) increases, the turning point moves away from the horizon.
As a result, the Taylor expansion truncated to three terms fails to approximate the potential accurately over the interval from \(r_0\) to \(r_*\). This failure is illustrated in Fig.~\ref{taylor_failure}, where the Taylor-expanded potential is compared with the exact potential.
Consequently, for large values of \(\omega\), which correspond to higher values of the quantum number \(n\), the approximate integration formula~\eqref{quant01} is no longer reliable for computing the normal modes. In this regime, the integral appearing in Eq.~\eqref{quant0} must be evaluated numerically. We have performed this numerical integration and solved for the corresponding modes, which we refer to as ``exact WKB'' modes, in the sense that the full potential \(V(r)\) has been used without invoking the Taylor approximation. In Fig.~\ref{exact_modes_dS}, we present these exact WKB modes as functions of \(n\) and \(l\). The figure shows that the growth along the \(n\) direction is linear, which is significantly faster than the slow logarithmic growth observed along the \(l\) direction.

\begin{figure}[t!]
    \centering
    \includegraphics[width=0.58\linewidth]{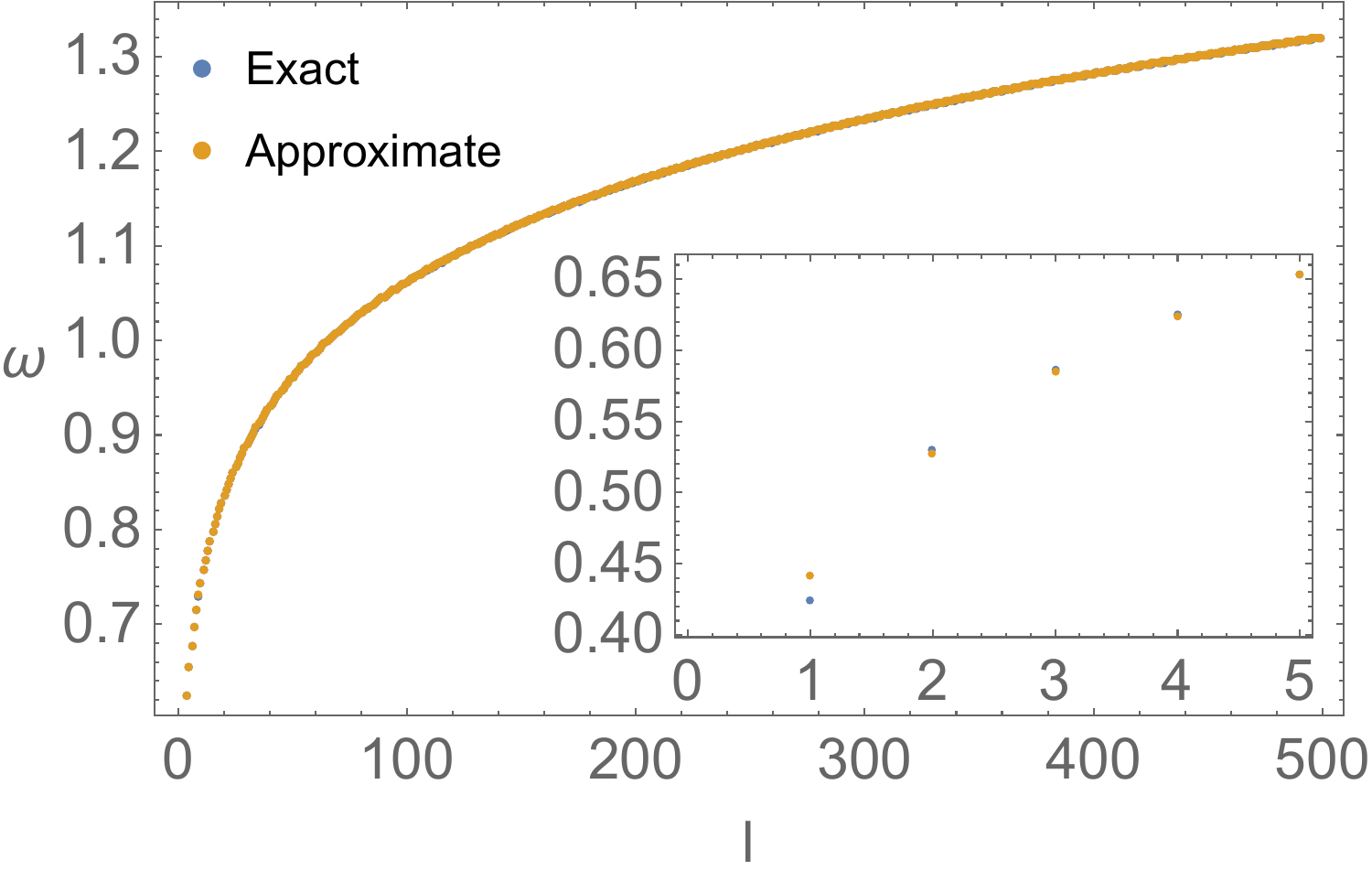}
    \caption{Comparison between the modes obtained using the approximate analytic formula~\eqref{quant01} and the exact WKB modes computed from the full potential. Apart from a very small discrepancy for the first few values of \(\omega\), the two sets of modes display excellent agreement.}
    \label{modes_comp_ds}
\end{figure}

For sufficiently small values of \(\omega\), corresponding to low \(n\) (but still large enough for the WKB approximation to be valid), the Taylor expansion provides a very good approximation to the potential. In this regime, one expects the quantization condition~\eqref{quant01} to yield reliable results for the modes along the \(l\) direction. To verify this, in Fig.~\ref{modes_comp_ds} we compare the exact WKB modes obtained using the full potential with those computed using the approximate formula~\eqref{quant01}.
Although there is a slight disagreement for the first few modes, the overall agreement is excellent. This demonstrates that the approximate integration formula~\eqref{quant01} can be safely used for modes along the \(l\) direction, provided that the cutoff \(\epsilon_0\) is sufficiently small.\footnote{These conditions ensure that \(\omega\) remains small, so that the turning point satisfies \(r_* \simeq r_C\).}

\section{More on the spectral analysis of Schwarzschild--de~Sitter}\label{appenD}

\begin{figure}[t!]
     \centering
     \begin{subfigure}[b]{0.45\textwidth}
         \centering
         \includegraphics[width=\textwidth]{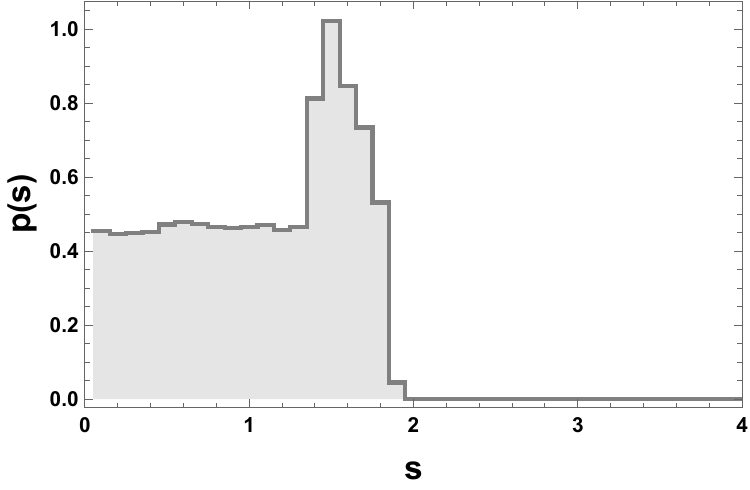}
         \caption{$\sigma_0^2=0.1$}
         \label{LSDcompE1}
     \end{subfigure}
     \hfill
     \begin{subfigure}[b]{0.45\textwidth}
         \centering
         \includegraphics[width=\textwidth]{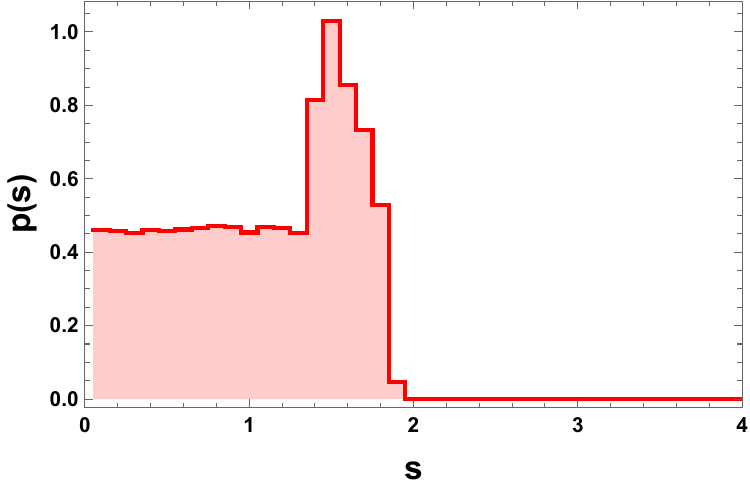}
         \caption{$\sigma_0^2=0.2$}
         \label{LSDcompE2}
     \end{subfigure}
     \hfill
     \begin{subfigure}[b]{0.45\textwidth}
         \centering
         \includegraphics[width=\textwidth]{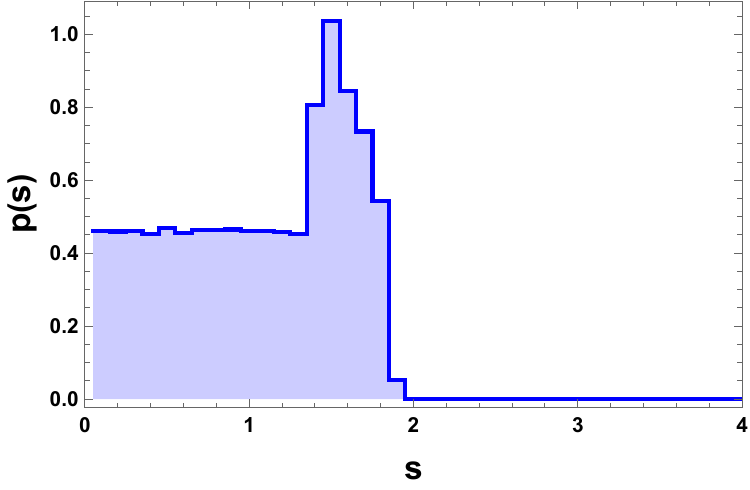}
         \caption{$\sigma_0^2=0.3$}
         \label{LSDcompE3}
     \end{subfigure}
     \hfill
     \begin{subfigure}[b]{0.45\textwidth}
         \centering
         \includegraphics[width=\textwidth]{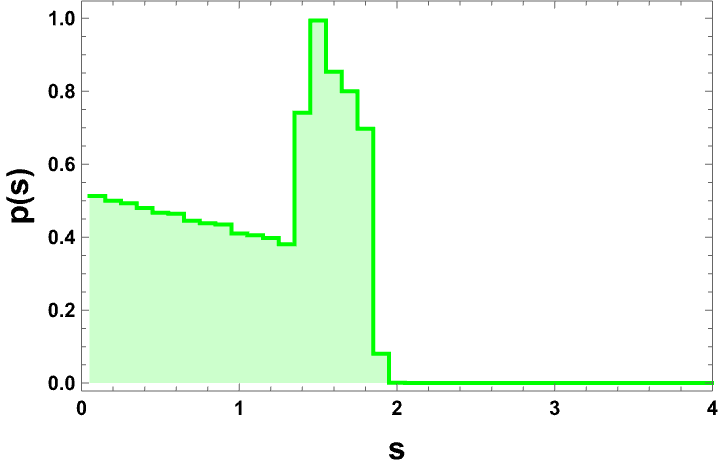}
         \caption{$\sigma_0^2=0.8$}
         \label{LSDcompE4}
     \end{subfigure}
     \hfill
     \begin{subfigure}[b]{0.45\textwidth}
         \centering
         \includegraphics[width=\textwidth]{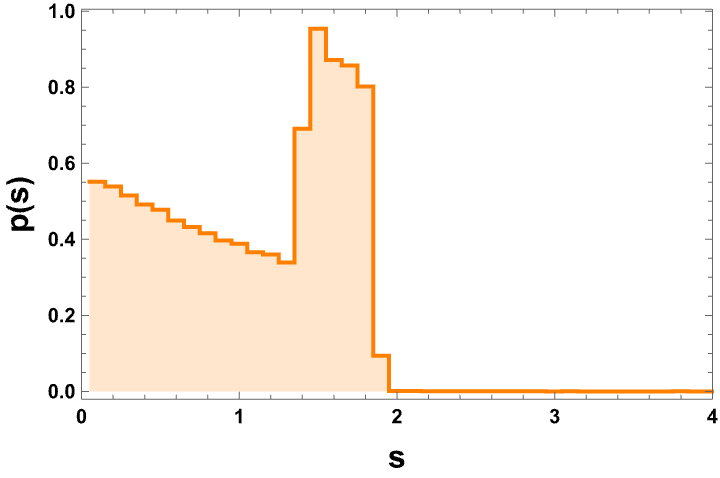}
         \caption{$\sigma_0^2=2$}
         \label{LSDcompE5}
     \end{subfigure}
     \hfill
     \begin{subfigure}[b]{0.45\textwidth}
         \centering
         \includegraphics[width=\textwidth]{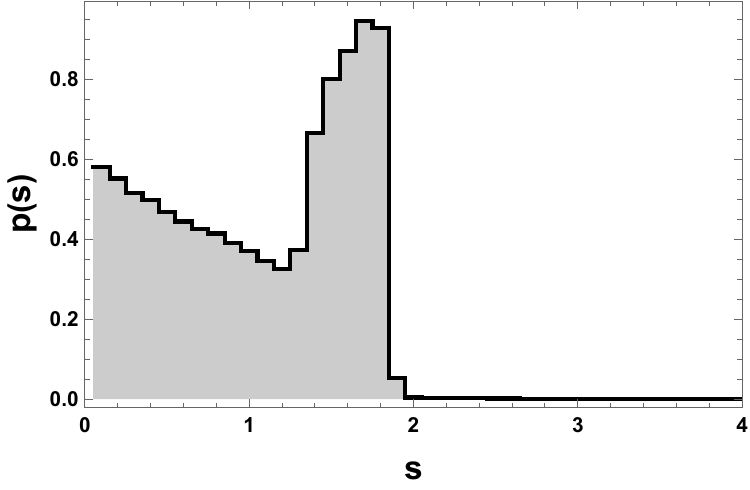}
         \caption{$\sigma_0^2=10$}
         \label{LSDcompE6}
     \end{subfigure}
\caption{Level-spacing distribution as a function of the black-hole brick wall variance, with cosmological variance fixed at \(\sigma_0^2=10^{-4}\), and parameters \(\omega_{\text{cut}}=1.09\) and \(s_e/s_c=10^{-2.5}\).}
     \label{avg_LSD_2_COSMO}
\end{figure}
\begin{figure}[t!]
     \centering
     \begin{subfigure}[b]{0.45\textwidth}
         \centering
         \includegraphics[width=\textwidth]{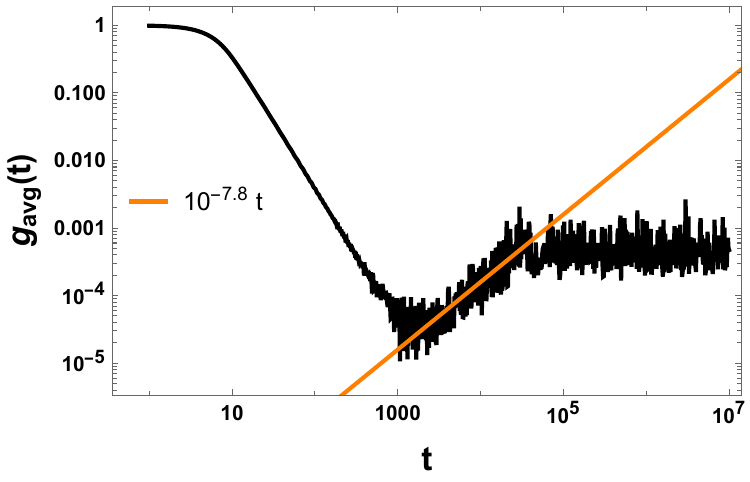}
         \caption{$\sigma_0^2=0.1$}
         \label{SFFcompE1}
     \end{subfigure}
     \hfill
     \begin{subfigure}[b]{0.45\textwidth}
         \centering
         \includegraphics[width=\textwidth]{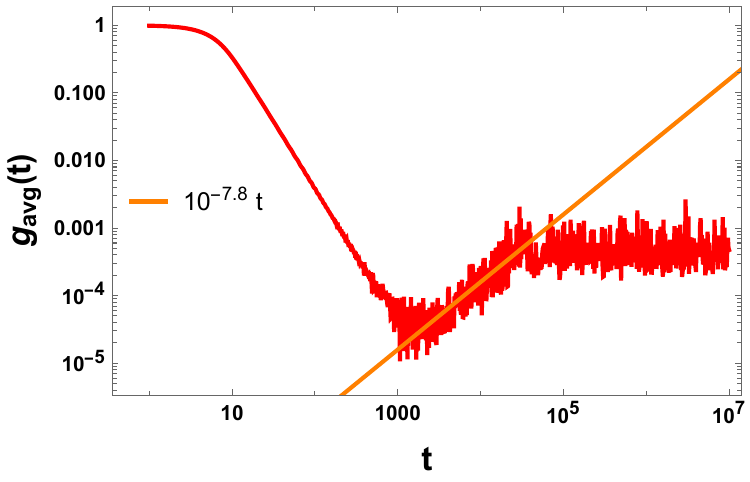}
         \caption{$\sigma_0^2=0.2$}
         \label{SFFcompE2}
     \end{subfigure}
     \hfill
     \begin{subfigure}[b]{0.45\textwidth}
         \centering
         \includegraphics[width=\textwidth]{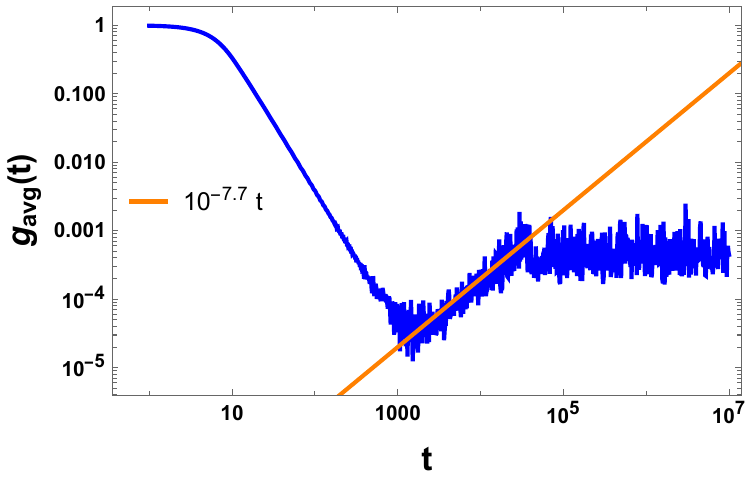}
         \caption{$\sigma_0^2=0.3$}
         \label{SFFcompE3}
     \end{subfigure}
     \hfill
     \begin{subfigure}[b]{0.45\textwidth}
         \centering
         \includegraphics[width=\textwidth]{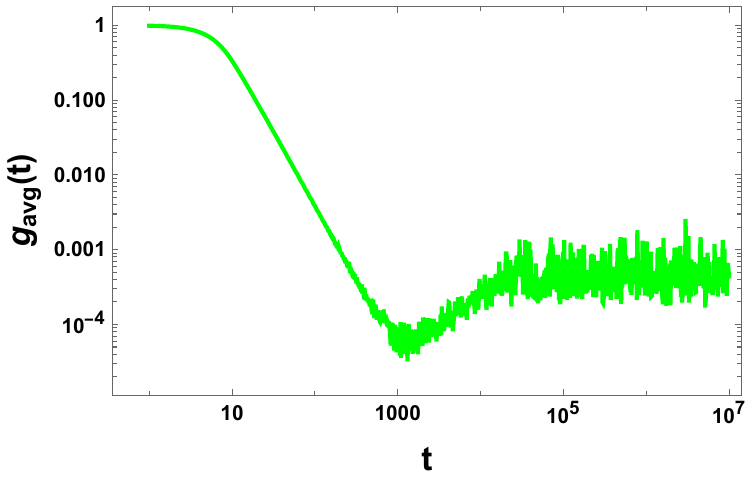}
         \caption{$\sigma_0^2=0.8$}
         \label{SFFcompE4}
     \end{subfigure}
     \hfill
     \begin{subfigure}[b]{0.45\textwidth}
         \centering
         \includegraphics[width=\textwidth]{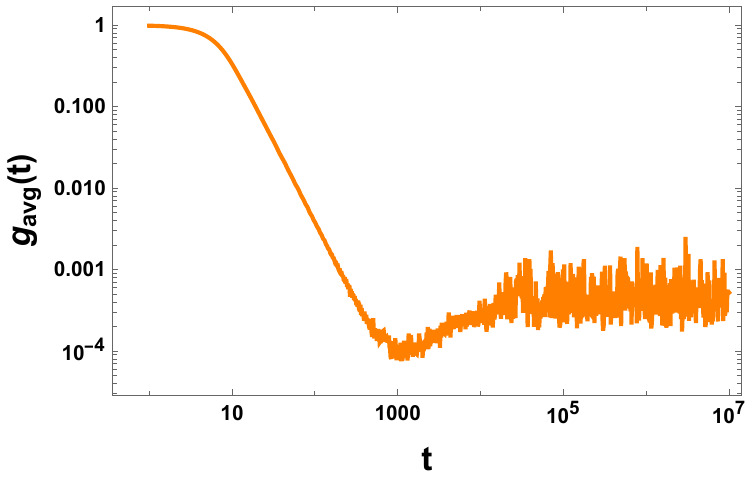}
         \caption{$\sigma_0^2=2$}
         \label{SFFcompE5}
     \end{subfigure}
     \hfill
     \begin{subfigure}[b]{0.45\textwidth}
         \centering
         \includegraphics[width=\textwidth]{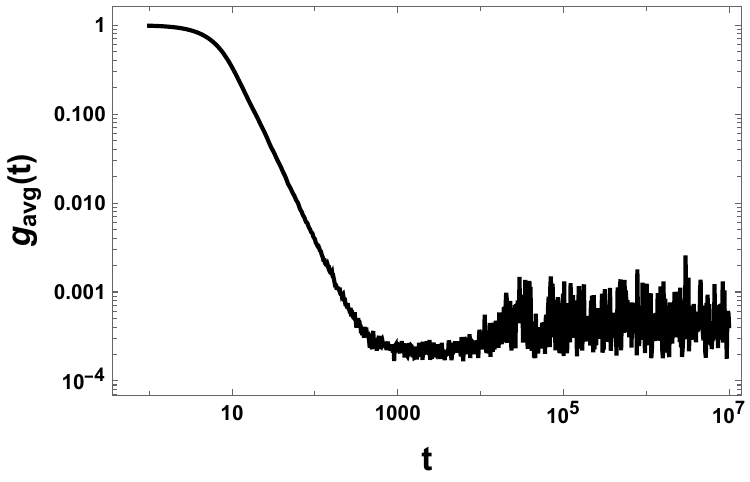}
         \caption{$\sigma_0^2=10$}
         \label{SFFcompE6}
     \end{subfigure}
     \caption{Ensemble-averaged SFF at \(\beta=0\), as a function of the black hole variance. The cosmological variance is fixed at \(\sigma_0^2=10^{-4}\), along with parameters \(\omega_{\text{cut}}=1.09\) and \(s_e/s_c=10^{-2.5}\).}
     \label{avg_SFF_2_EVENT}
\end{figure}

\paragraph{Varying the variance of the black hole brick wall.}
Here, we consider the opposite experiment to that discussed in the main text, namely, we fix the variance of the wall near the cosmological horizon and vary the variance of the wall near the event horizon. The corresponding LSD is shown in Fig.~\ref{avg_LSD_2_COSMO}. It appears that, in this case, the effect of \(\sigma_0\) is less pronounced than in the previous case shown in Fig.~\ref{avg_LSD_22_COSMO}. Empirically, this occurs because for the chosen parameter values the black hole spectrum lies above the cosmological spectrum (see, for example, Fig.~\ref{modes_BH}). Fig.~\ref{avg_SFF_2_EVENT} shows the corresponding SFF, which exhibits a linear ramp for small values of the variance, but loses it as the variance increases. Finally, in Fig.~\ref{KRYLOV_SCHW_FIG_2_EVENT}, we present the behavior of Krylov complexity, which does not show any drastic change as the variance increases. This suggests that the KC may be less sensitive to the addition of noise in the spectrum than the SFF.

\begin{figure}[t!]
    \centering
    \includegraphics[width=0.6\textwidth]{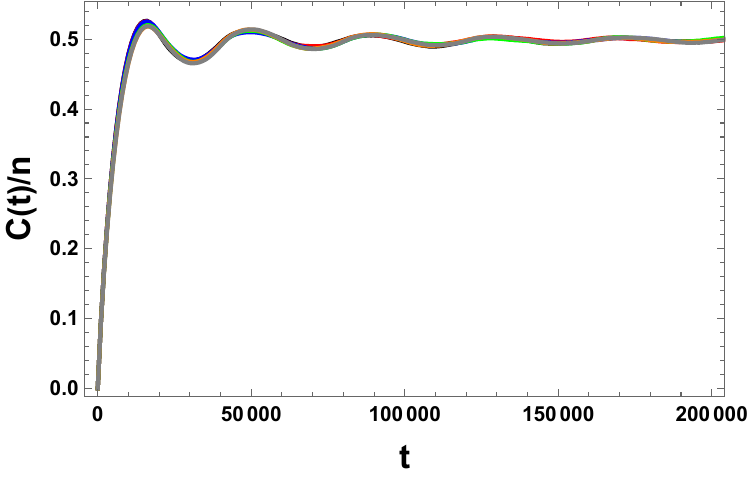}
    \caption{Ensemble-averaged KC for the TFD state at $\beta=0$ and black hole variance \(\sigma_0^2 = 0.1,\,0.2,\,0.3,\,0.8,\,2,\,10\) (black, red, blue, green, orange, and gray, respectively). The cosmological variance is fixed at \(\sigma_0^2=10^{-4}\), along with parameters \(\omega_{\text{cut}}=1.09\) and \(s_e/s_c=10^{-2.5}\).}\LA{KRYLOV_SCHW_FIG_2_EVENT}
\end{figure}

\begin{figure}[t!]
     \centering
     \begin{subfigure}[b]{0.45\textwidth}
         \centering
         \includegraphics[width=\textwidth]{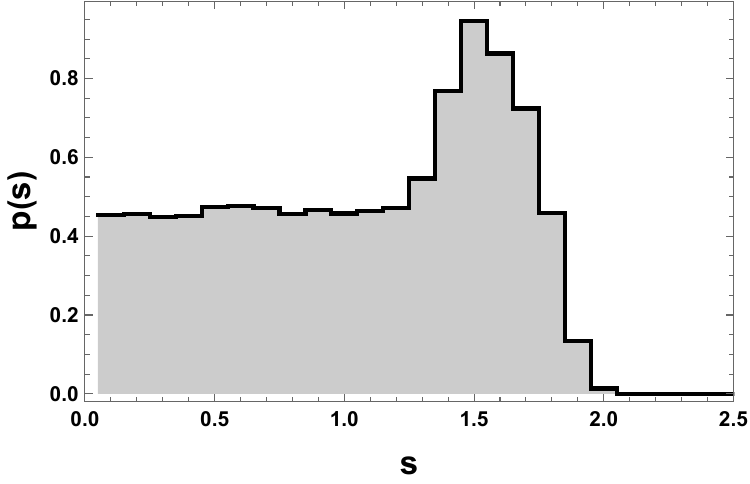}
         \caption{$\sigma_0^2=0.1$}
         \label{LSDcompEC1}
     \end{subfigure}
     \hfill
     \begin{subfigure}[b]{0.45\textwidth}
         \centering
         \includegraphics[width=\textwidth]{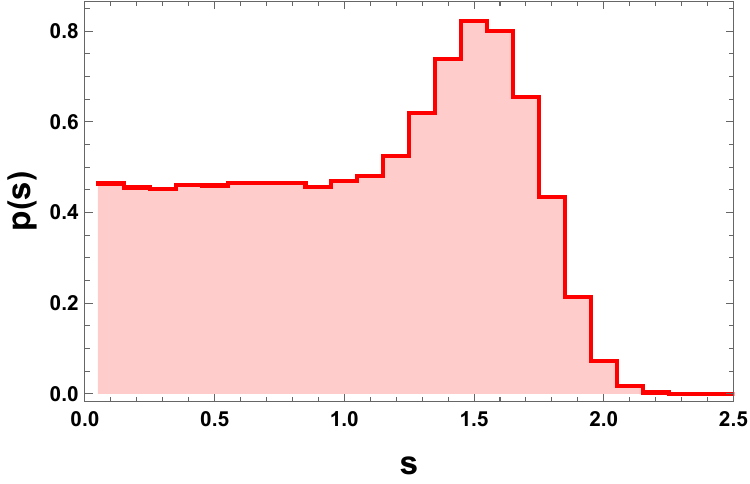}
         \caption{$\sigma_0^2=0.2$}
         \label{LSDcompEC2}
     \end{subfigure}
     \hfill
     \begin{subfigure}[b]{0.45\textwidth}
         \centering
         \includegraphics[width=\textwidth]{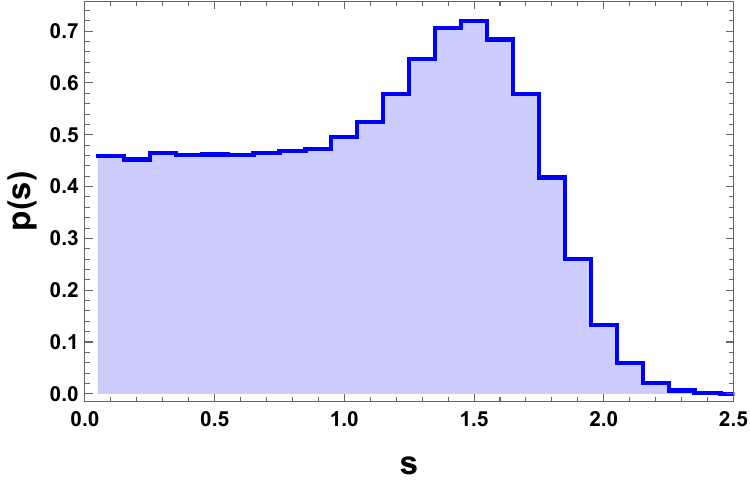}
         \caption{$\sigma_0^2=0.3$}
         \label{LSDcompEC3}
     \end{subfigure}
     \hfill
     \begin{subfigure}[b]{0.45\textwidth}
         \centering
         \includegraphics[width=\textwidth]{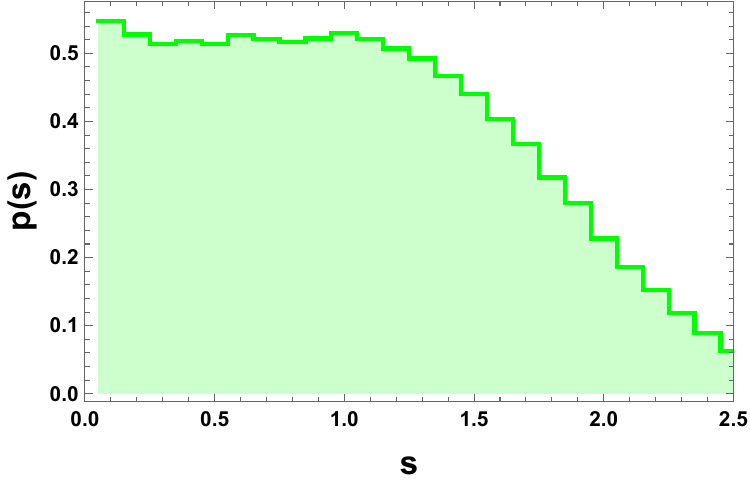}
         \caption{$\sigma_0^2=0.8$}
         \label{LSDcompEC4}
     \end{subfigure}
     \hfill
     \begin{subfigure}[b]{0.45\textwidth}
         \centering
         \includegraphics[width=\textwidth]{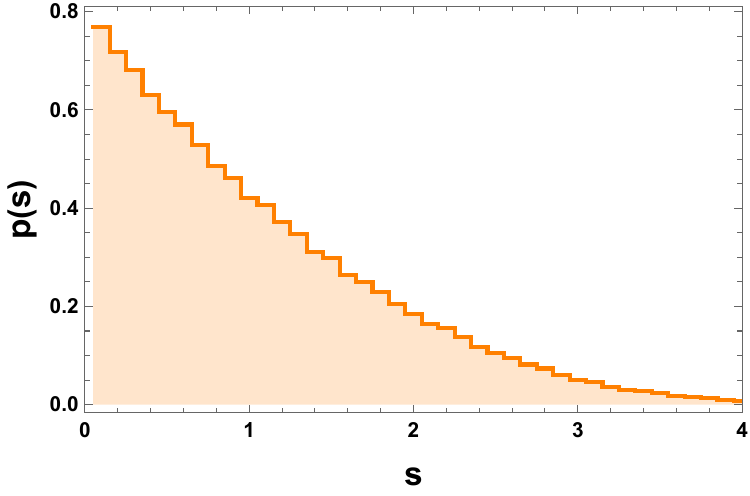}
         \caption{$\sigma_0^2=2$}
         \label{LSDcompEC5}
     \end{subfigure}
     \hfill
     \begin{subfigure}[b]{0.45\textwidth}
         \centering
         \includegraphics[width=\textwidth]{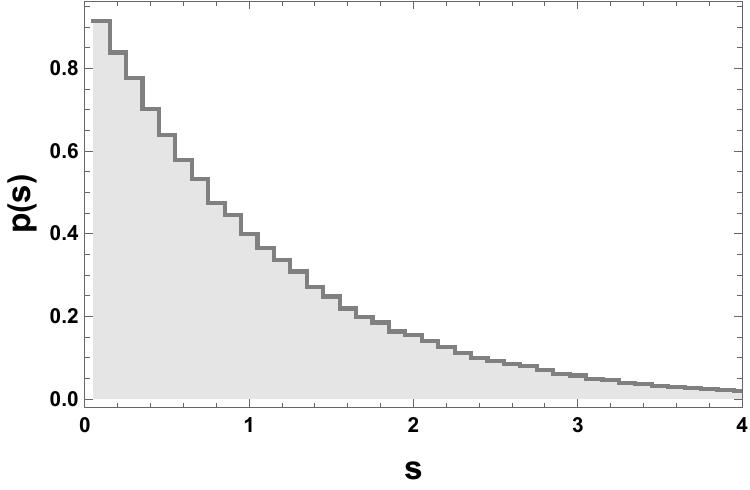}
         \caption{$\sigma_0^2=10$}
         \label{LSDcompEC6}
     \end{subfigure}
     \caption{LSD as a function of $\sigma_{BH}=\sigma_{C}=\sigma_0$, and parameters $\omega_{\text{cut}}=1.09$ and $s_e/s_c=10^{-2.5}$.}
     \label{avg_LSD_2_BOTH}
\end{figure}

\begin{figure}[t!]
     \centering
     \begin{subfigure}[b]{0.45\textwidth}
         \centering
         \includegraphics[width=\textwidth]{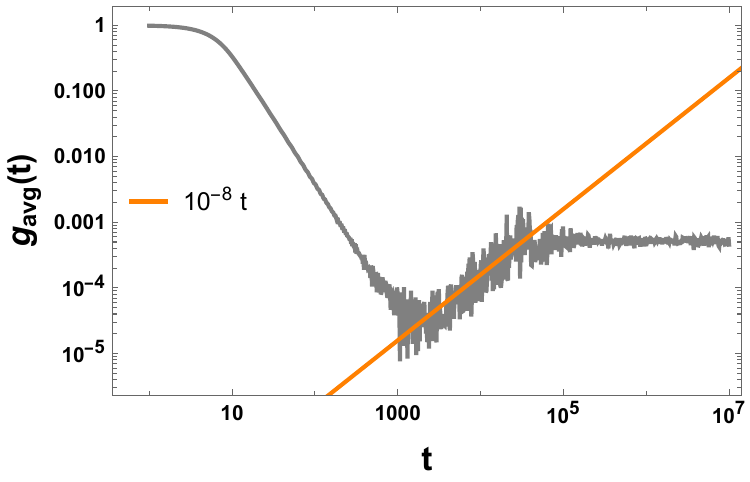}
         \caption{$\sigma_0^2=0.1$}
         \label{SFFcompEC1}
     \end{subfigure}
     \hfill
     \begin{subfigure}[b]{0.45\textwidth}
         \centering
         \includegraphics[width=\textwidth]{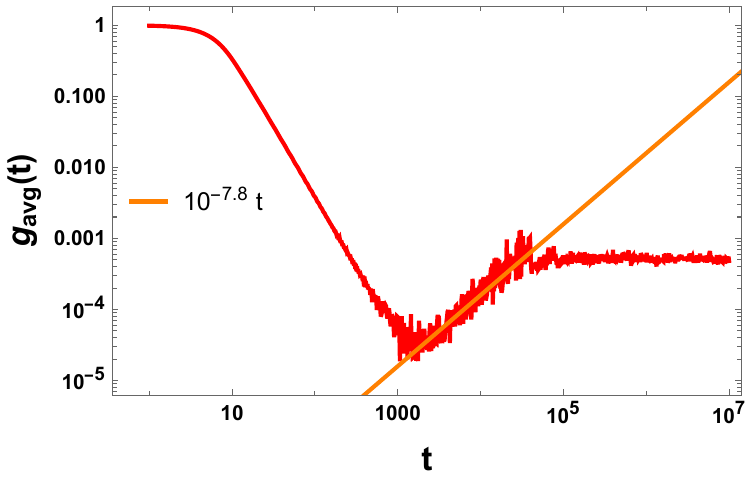}
         \caption{$\sigma_0^2=0.2$}
         \label{SFFcompEC2}
     \end{subfigure}
     \hfill
     \begin{subfigure}[b]{0.45\textwidth}
         \centering
         \includegraphics[width=\textwidth]{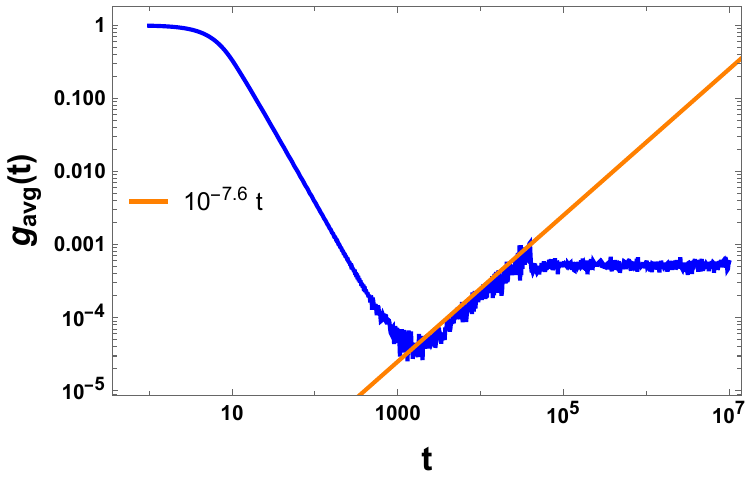}
         \caption{$\sigma_0^2=0.3$}
         \label{SFFcompEC3}
     \end{subfigure}
     \hfill
     \begin{subfigure}[b]{0.45\textwidth}
         \centering
         \includegraphics[width=\textwidth]{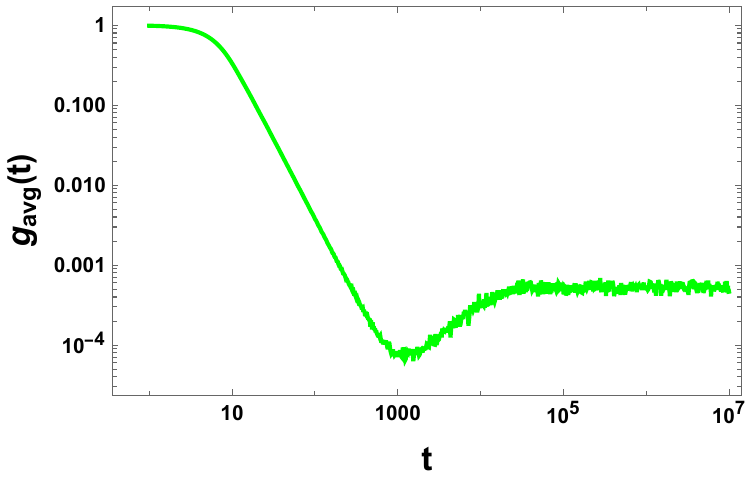}
         \caption{$\sigma_0^2=0.8$}
         \label{compEC4}
     \end{subfigure}
     \hfill
     \begin{subfigure}[b]{0.45\textwidth}
         \centering
         \includegraphics[width=\textwidth]{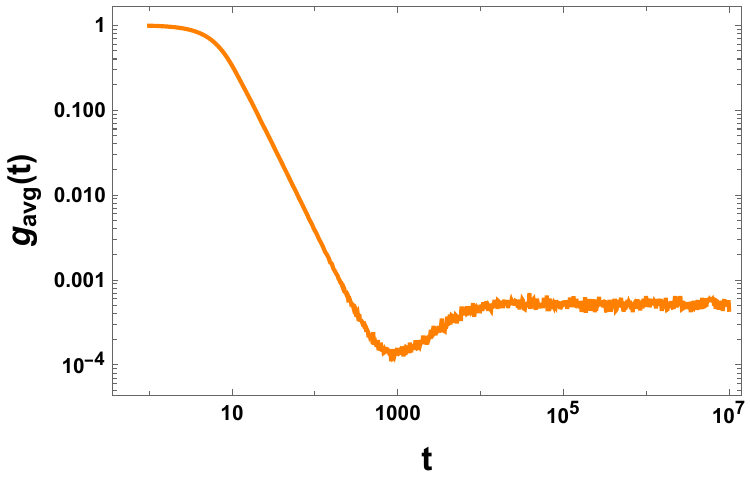}
         \caption{$\sigma_0^2=2$}
         \label{SFFcompEC5}
     \end{subfigure}
     \hfill
     \begin{subfigure}[b]{0.45\textwidth}
         \centering
         \includegraphics[width=\textwidth]{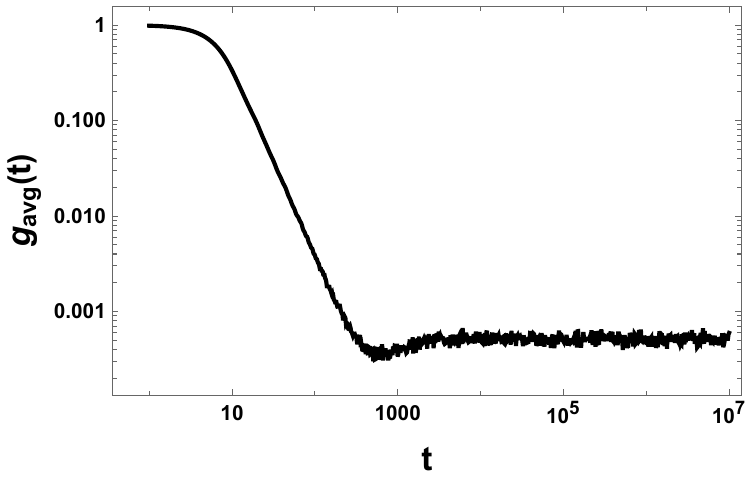}
         \caption{$\sigma_0^2=10$}
         \label{SFFcompEC6}
     \end{subfigure}
     \caption{Ensemble-averaged SFF at $\beta=0$ as a function of $\sigma_{BH}=\sigma_{C}=\sigma_0$, and parameters $\omega_{\text{cut}}=1.09$ and $s_e/s_c=10^{-2.5}$.}
     \label{avg_SFF_2_BOTH}
\end{figure}

\paragraph{The case $\sigma_{BH}=\sigma_{C}$.}
As a final experiment, we consider the special case in which both variances are taken to be equal, i.e. \(\sigma_{BH}=\sigma_{C}=\sigma_0\), and vary \(\sigma_0\). As in the main text, we present the response of the LSD, SFF, and KC to this variation. The results are shown in Figs.~\ref{avg_LSD_2_BOTH}, \ref{avg_SFF_2_BOTH}, and \ref{KRYLOV_SCHW_FIG_2_BOTH}. The overall behavior is qualitatively similar to that discussed in the main text: for small variance, the LSD retains a mixed profile with a nonzero value at the origin, the SFF exhibits a clear linear ramp, and the KC displays a pronounced peak. As the common variance increases, the LSD becomes progressively more Poisson-like, the ramp in the SFF gradually flattens and eventually disappears, and the peak in the KC becomes less pronounced. Taken together, these results are consistent with the interpretation that, for small variance, the spectrum retains a Berry--Robnik-type structure, whereas sufficiently large fluctuations progressively wash out the corresponding spectral correlations.

\begin{figure}[t!]
    \centering
    \includegraphics[width=0.6\textwidth]{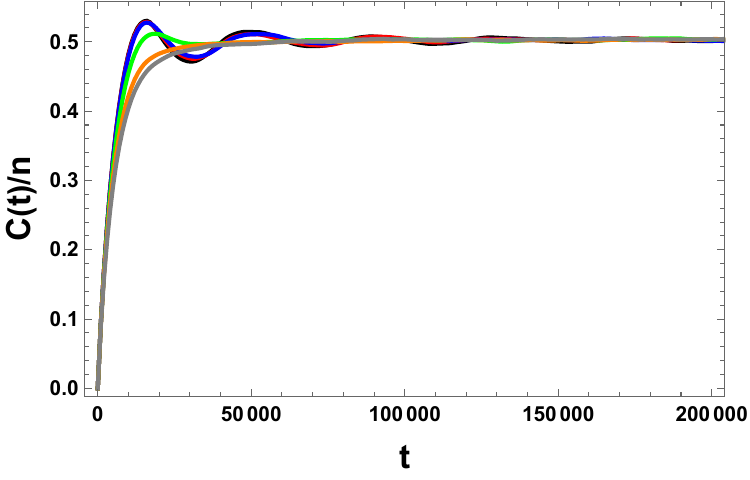}
    \caption{Ensemble-averaged KC for the TFD state at $\beta=0$ and $\sigma_{BH}=\sigma_{C}=\sigma_0^2= 0.1, ~ 0.2, ~ 0.3, ~ 0.8, ~ 2, ~ 10$ (black, red, blue, green, orange, gray), with $\omega_{\text{cut}}=1.09$ and $s_e/s_c=10^{-2.5}$.}\LA{KRYLOV_SCHW_FIG_2_BOTH}
\end{figure}

\bibliography{Refs}
\bibliographystyle{JHEP}

\end{document}